\documentclass{aa}  
\pdfoutput=1
\usepackage{graphicx,xcolor}
\usepackage{txfonts}
\usepackage{multirow}
\usepackage{lscape}
\usepackage{longtable}

\usepackage{tabularx}

\begin{document} 
\pdfoutput=1
\title{The EDIBLES survey III. \element[][][][2]{C}-DIBs and their profiles}

   \author{M.~Elyajouri
          \inst{1}
          \and
          R.~Lallement\inst{1}
          \and
          N.\,L.\,J.~Cox\inst{2}
          \and
          J.~Cami\inst{3,4}
          \and
          M.\,A.~Cordiner \inst{5,6}
          \and\\
          J.\,V.~Smoker\inst{7} 
          \and
          A.~Fahrang\inst{3,8}
          \and
          P.\,J.~Sarre\inst{9}
          \and
          H.~Linnartz\inst{10}
\fnmsep
          }
   \institute{GEPI, Observatoire de Paris, PSL University, CNRS,
Place Jules Janssen, 92190 Meudon, France\\
              \email{meriem.el-yajouri@obspm.fr}
         \and
             ACRI-ST, 260 route du Pin Montard, 06904, Sophia Antipolis, France
         \and
         Department of Physics and Astronomy and Centre for Planetary Science and Exploration (CPSX), The University of Western Ontario, London, ON N6A 3K7, 
Canada
\and
SETI Institute, 189 Bernardo Avenue, Suite 100, Mountain View, CA 94043, USA
\and
NASA Goddard Space Flight Center, 8800 Greenbelt Road, Greenbelt, MD 20771, USA
\and
Department of Physics, Catholic University of America, Washington, DC 20064, USA
\and
European Southern Observatory, Alonso de Cordova 3107, Vitacura,
  Santiago, Chile
  \and
  School of Astronomy, Institute for Research in Fundamental
  Sciences, 19395-5531 Tehran, Iran
\and
School of Chemistry, The University of Nottingham, University Park, Nottingham NG7 2RD, United Kingdom 
\and
Sackler Laboratory for Astrophysics, Leiden Observatory, Leiden
University, PO Box 9513, 2300 RA Leiden, The Netherlands
             }

   \date{Received 2018, ; accepted 2018}

  \abstract
   {An unambiguous identification of the carriers of the diffuse interstellar bands (DIBs) would provide important clues to the life cycle of interstellar matter. The so-called \element[][][][2]{C}-DIBs are a class of very weak bands that fall in the blue part of the optical spectrum and are associated with high column densities of the C$_{2}$ molecule. DIB profile structures constrain potential molecular carriers, but their measurement requires high signal-to-noise, high-resolution spectra and the use of sightlines without Doppler splitting, as typical for a  single-cloud situation.}
   {Spectra from the ESO Diffuse Interstellar Bands Large Exploration Survey (EDIBLES) conducted at the Very Large Telescope (ESO/Paranal) were explored to identify single-cloud and high C$_{2}$ column sightlines, extract the corresponding C$_{2}$-DIBs and study their strengths and profiles, and to investigate in detail any sub-structures.}
   {The target selection was made based on profile-fitting of the 3303 and 5895~\AA~\ion{Na}{i} doublets and the detection of C$_{2}$ lines. The C$_{2}$ (2-0) (8750–8849~\AA) Phillips system was fitted using a physical model of the host cloud. C$_{2}$ column densities, temperatures as well as gas densities were derived for each sightline.}
   {Eighteen known C$_{2}$-DIBs and eight strong non-C$_{2}$ DIBs were extracted towards eight targets, comprising seven single-cloud and one multi-cloud line-of-sights. Correlational studies revealed a tight association of the former group with the C$_{2}$ columns, whereas the non-C$_{2}$ DIBs are primarily correlated with reddening. We report three new weak diffuse band candidates at 4737.5, 5547.4 and 5769.8~\AA. We show for the first time that at least 14 C$_{2}$-DIBs exhibit spectral sub-structures which are consistent with unresolved rotational branches of molecular carriers. The variability of their peak separations among the bands for a given sightline implies that their carriers are different molecules with quite different sizes. We also illustrate how profiles of the same DIB vary among targets and as a function of physical parameters, and provide tables defining the sub-structures to be compared with future models and experimental results.
   }
   {}


  \maketitle

\section{Introduction}\label{sec:intro}
Diffuse interstellar bands (DIBs) are over 400 broad spectroscopic absorption features observed in stellar spectra in ultra-violet (UV), visible and infra-red (IR) ranges \citep[see][and references therein]{Sarre06}. Many of the DIB carriers are thought to be large carbonaceous molecules in the gaseous phase, but except for the very likely identification of C$_{60}^{+}$ \citep{Campbell15,Cordiner17,Spieler17,Walker17,Lallement18}, none of these bands has been definitely assigned to any given species. 
Identifying the DIB carriers and measuring their response to physical conditions and chemical compositions of interstellar clouds is of high importance. This would add to our rather limited knowledge of the molecular inventory of these clouds in which small molecules, like C3 \citep{Schmidt14} were long the largest species identified. It also would be key to understanding the life cycle of cosmic dust and carbonaceous compounds from dying star ejected envelopes to the birth places of new stars and planetary systems. In this respect, a sub-class of DIBs, the so-called \element[][][][2]{C}-DIBs, is particularly interesting because they are detected in interstellar clouds that are characterized by high columns of the \element[][][][2]{C} molecule  \citep{Thorburn03, Gala06, Kazmierczak2014}, i.e., environments more closely chemically related to the cool, dense molecular clouds where stars form, as opposed to the more diffuse, warmer clouds that pervade the Galaxy.

Sub-structures in DIB profiles are indicative of the molecular structure of carriers and open the way to comparisons with laboratory spectra. However, such measurements require high-performance spectrographs and to date their use has been restricted to relatively strong DIBs and a limited number of sightlines. In fact, disentangling the spectral components requires high spectral resolution, and, furthermore, is restricted to target stars for with only a small Doppler broadening along the line-of-sight --- the so-called single-cloud situation, to avoid blends and sub-structures arising from nearly overlapping bands in different clouds. Most of the strong and narrow optical DIBs have been found to have asymmetric profiles, and in the case of the strongest ones such as DIBs 5797 and 6614~\AA \, \citep{Sarre95,Ehrenfreund96,Gala02}, 5850, 6234 and 6270~\AA\, \citep[e.g.][]{Krelowski97,Gala02}, 6376, 6379 and 6196~\AA\: \citep{Walker01}, sub-structures have been clearly identified which may be consistent with unresolved rotational branch structures associated with electronic transitions of gas-phase molecules. Profiles of several \element[][][][2]{C}-DIBs and some other weak DIBs have been examined also by \cite{Gala02} and \cite{Slyk06}. Asymmetries were found, but in the case of the weak \element[][][][2]{C}-DIBs these were only seen for the 4963, 5418, 5512, 5541 and 5546~\AA\ bands. Distinct sub-structures were found towards HD\,147889 for the 4734, 5175, 5418, 5512, 5546 and 6729~\AA\ DIBs by \cite{Gala2008}.

In the case of the \element[][][][2]{C}-DIBs, the study of their profiles is particularly difficult. The weakness of the absorption requires a very high signal-to-noise ratio (S/N), a careful correction of telluric lines and  disentangling from weak stellar features. In addition, because they are narrow ($\sim 0.5$~\AA), the single-cloud condition\footnote{Note that the term single-cloud does not necessarily imply the existence of a unique homogeneous cloud but rather, a very narrow velocity dispersion of matter along the line-of-sight (much less than the instrumental resolution)} is even more stringent. Very high resolution spectra commonly reveal that an apparent single absorption can actually be decomposed into absorptions in several closely related clouds, reflecting the quasi-fractal nature of the dense interstellar matter \citep[][]{Welty01}. However, the clumps are generally fragments of the same complex, and often share a common environment and consequently possess very similar physical properties. Sightlines with high \element[][][][2]{C} columns are crossing a cloud core with a density $\gtrsim 100\, cm^{-3}$ (higher than the typical diffuse interstellar medium (ISM)). It is a relatively rare situation, and therefore, the probability of crossing two independent cloud cores with identical radial velocities and with interstellar columns on the same order is low, i.e., \element[][][][2]{C} sightlines with a single absorption offer a good probability of being dominated by a single structure. Detecting and studying the \element[][][][2]{C}-DIBs is one of the goals of the ESO/VLT EDIBLES survey. EDIBLES is devoted to a deep analysis of the DIB properties and is particularly suited for the study of weak DIBs and their sub-structures, thanks to the high resolution of the UVES spectrograph, and, especially, thanks to the high S/N \citep{edibles1}. In addition, the large spectral coverage allows the simultaneous study of many gaseous absorption lines across the entire blue to the near-infrared range, like \element[][][][+]{OH} that can be used as tracer for the cosmic ray ionization rate 
\citep{Bacalla2018}.

Here we present an analysis of individual \element[][][][2]{C}-DIBs in EDIBLES spectra recorded along seven single-cloud and one multi-cloud lines-of-sight. We additionally show the results of a technique allowing to better reveal spectral sub-structures, namely the stacking of spectra of different target stars after they have been Doppler shifted to the cloud reference frame. Stacking spectra to enhance the S/N of band profiles has proved to be efficient to reveal DIBs \citep[see, e.g.][]{Lan15}, however, this is the first use of stacking in a search for profile sub-structures.
In doing so, we have examined other absorption bands that may correspond to so far unreported \element[][][][2]{C}-DIBs.

In Sect.~\ref{sec:observations} we describe the observations, the telluric line removal and the selection of single-cloud observations. In Sect.~\ref{sec:C2} we describe the selection of the \element[][][][2]{C} sightlines and the analysis of the \element[][][][2]{C} lines.
Sect.~\ref{sec:C2-DIBs} presents the selected spectra in the intervening cloud frame, in the spectral regions of the \element[][][][2]{C}-DIBs classified by \cite{Thorburn03}. We also show the spectral regions containing our potentially new \element[][][][2]{C}-DIBs. 
In Sect.~\ref{sec:structures} we list the wavenumber intervals between the observed absorption sub-peaks, when measurable, and study their variability among the bands. 
In Sect.~\ref{sec:stack} we present our optimal profiles obtained by co-addition of the spectra of three targets that have relatively deep DIBs and very similar kinetic temperatures and show how these vary as a function of the kinetic temperature derived from the \element[][][][2]{C} analysis. We compare the profiles of each \element[][][][2]{C}-DIB towards several targets characterized by different \element[][][][2]{C} excitation temperatures in Sect.~\ref{sec:variab}. 
Finally, several correlational studies are presented in Sect.~\ref{sec:correl}. They reveal a distinction between \element[][][][2]{C} and non-\element[][][][2]{C} DIBs. We discuss the various results in Sect.~\ref{sec:discussion}.

\begin{figure*}[th!]
\centering
\includegraphics[height=10cm]{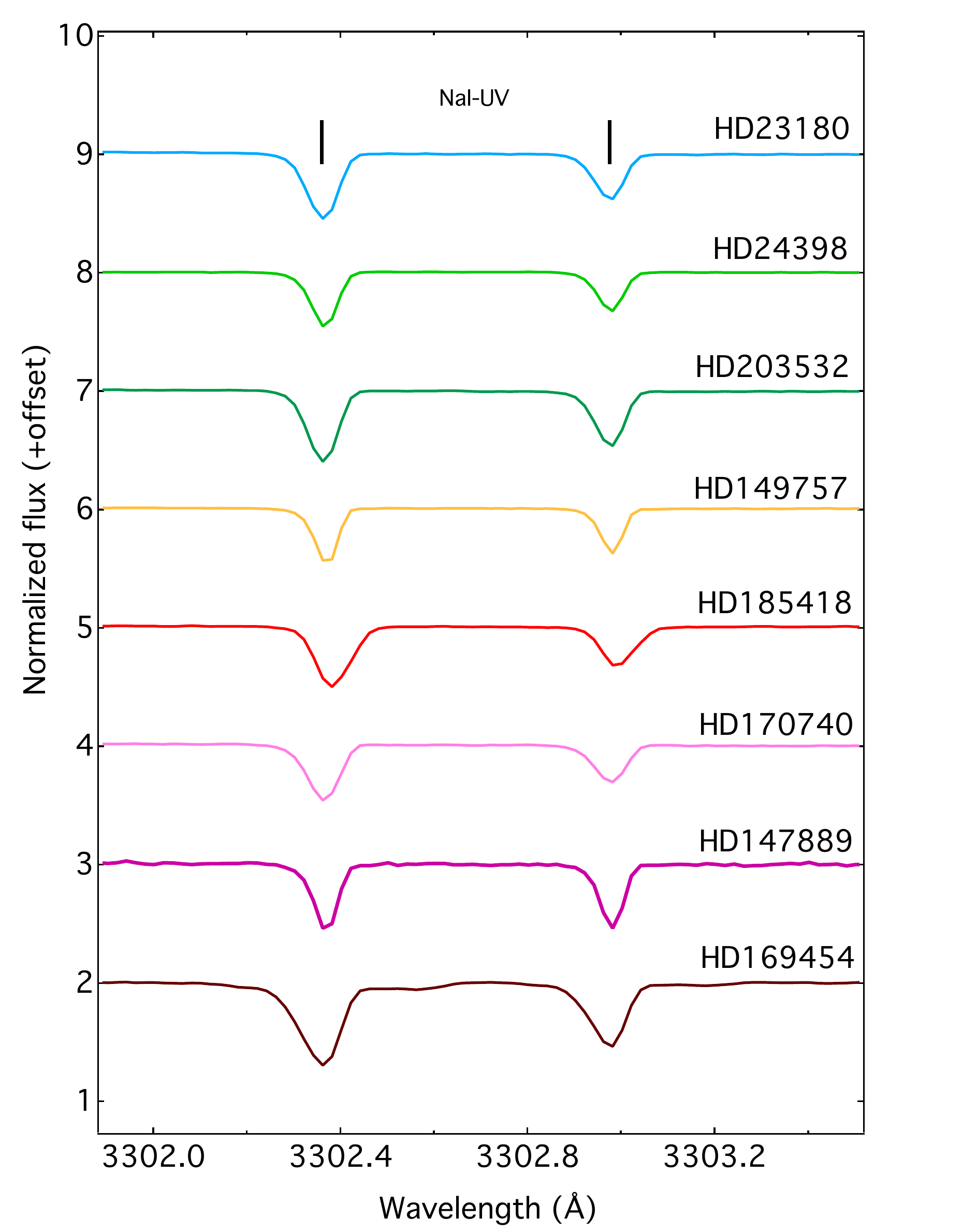}
\includegraphics[height=10cm]{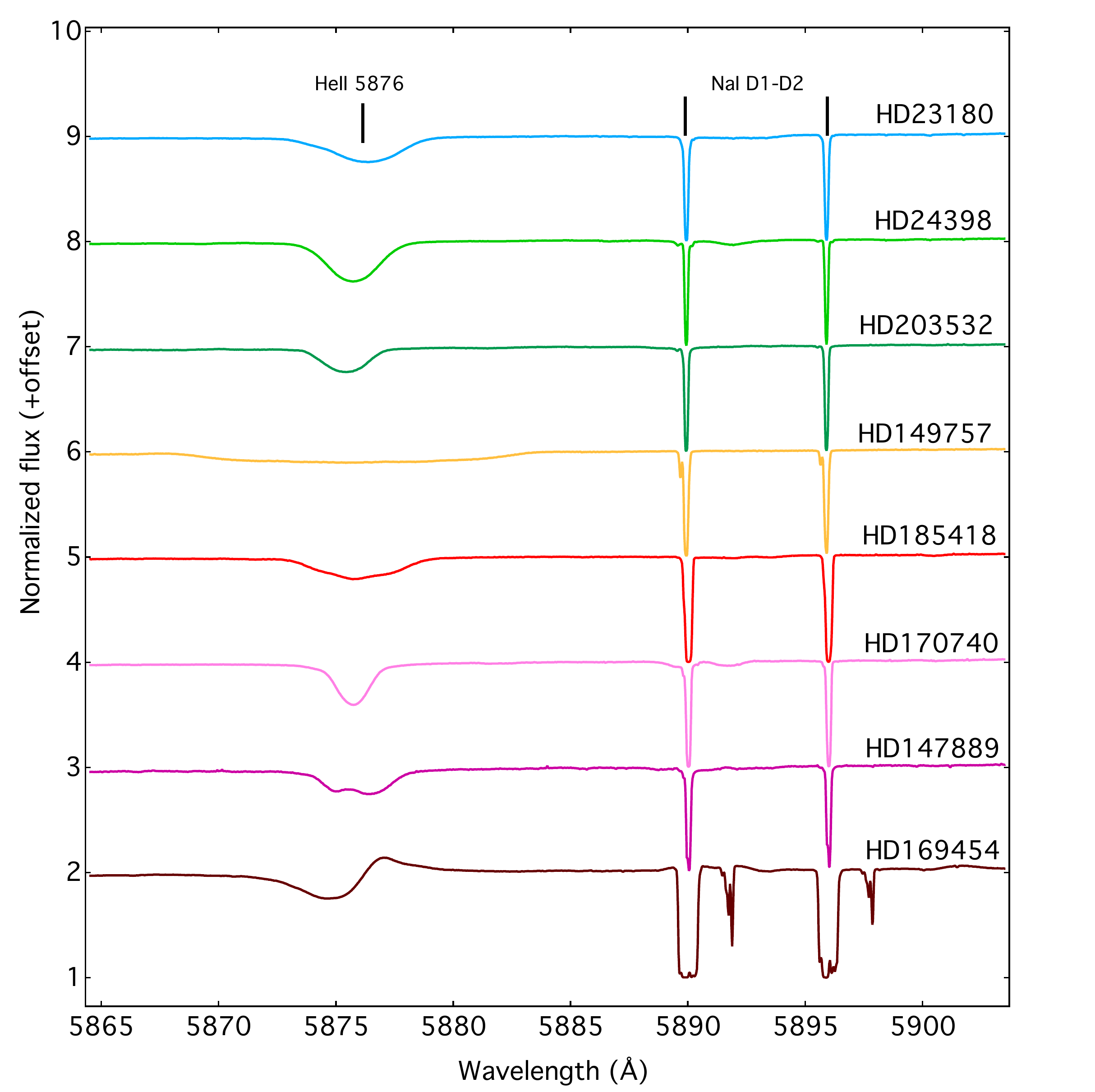}
\caption{\label{Fig:naIdoublet} \ion{Na}{i} doublet at 3302.368/3202.987 \AA\ (left), and
\ion{Na}{i} D1-D2 and the stellar \ion{He}{ii} (right) line are shown for the seven selected single-cloud line-of-sight (one main cloud component) and the multi-cloud target HD\,169454. The \ion{Na}{i} spectra have been shifted to their air rest wavelengths. The same velocity shift has been applied to the \ion{Na}{i} D1-D2 lines. The applied color coding is used  in all other figures as well.}
\end{figure*}

\section{Observations and single-cloud target selection}\label{sec:observations}
\begin{table*}[ht!]
\caption{Single-cloud lines-of-sight properties.}            
\label{table:sightline_properties} 
\centering                         
\begin{tabular}{l l  l l l l l l l}       
\hline                 
  &\multicolumn{2}{|c|}{Star} & \multicolumn{6}{c}{ISM}\\
\hline                 
Identifier & Spectral type  & $v$\,sin($i$) & E(B-V)\tablefootmark{a} & $R_V$ & Ref $R_V$\tablefootmark{b} & f(H$_2$)\tablefootmark{c}   & \ion{Na}{i}-UV velocity &5797/5780\tablefootmark{d} \\    
 & &  (km\,s$^{-1}$) & (mag) & & &  & (km\,s$^{-1}$) &\\ \hline
   HD 23180  & B1 III  &  100 &   0.28 & 3.11 & V04 & 0.51 &  13.8$\pm$0.3 &0.73 ($\zeta$)\\
   HD 24398  & B1 Ib  &  43   &  0.29 & 2.63 & W03 & 0.60    & 14.2$\pm$0.2 &0.53 ($\zeta$)\\ 
   HD 203532 & B3 IV   &   -   &  0.30 & 3.37 & V04 & 0.84    &15.2$\pm$0.2 & 0.48 ($\zeta$)\\ 
   HD 149757 & O9.2 IVnn  &  303&   0.32 & 2.55 & V04 & 0.63 &-14.2$\pm$0.5 & 0.43 ($\zeta$)\\
   HD 185418 & B0.5 IIIn &   -&  0.42 & 2.54 & V04 & 0.40    &-10.7$\pm$0.5 &  0.30 ($\sigma$) \\
   HD 170740 & B2 V   &  25  &  0.45 & 3.01 & V04 & 0.58    & -9.9$\pm$1.5 & 0.26 ($\sigma$)\\
   HD 147889 & B2 V   & 100 &  1.03 & 3.95 & V04 & 0.62    & -6.7$\pm$1.0 & 0.41 ($\zeta$) \\
   \hline
   HD 169454\tablefootmark{e} & B1 Ia &  49 & 1.03 & 3.37 & V04 & -  &-9.9$\pm$1.5 & 0.34 ($\sigma$)\\
   \hline
\end{tabular}
\tablefoot{
\tablefoottext{a}{E(B-V) is based on photometry and intrinsic colors.}
\tablefoottext{b}{From \citet{2004ApJ...616..912V} (V04) and \citet{2003AN....324..219W} (W03).}
\tablefoottext{c}{From \citet{Jenkins2009}.}
\tablefoottext{d}{The 5797/5780 DIB ratio is related to the effective UV radiation field
strength; UV exposed environments, $\sigma$-type, have ratios $<$ 0.35, while UV-shielded, $\zeta$-type environments, have ratios $\geq$~0.35 \citep[c.f.][]{Vos11}.}
\tablefoottext{e}{HD\,169454 is not a single-cloud line-of-sight but it is used for comparison (c.f. Sect.~\ref{sec:variab}).}
}
\end{table*}

\subsection{EDIBLES data\label{data}}
EDIBLES observations and data reductions are described in detail in \citet{edibles1}. The S/N values for most sightlines are 400--1000 in the red part, 300--400 in the UV and $\geq$ 300 in the near-IR. Two additional data treatments are necessary to realize the goals aimed for in this study.

First, we have corrected all individual exposures recorded with the 564-nm setting (including both the Red Upper and Red Lower detectors) for their telluric lines. The method follows the first of the two procedures described in \citet{edibles1}. It is appropriate for moderately weak lines and uses TAPAS synthetic spectra adapted to the observing site and season \citep{Bertaux14}. In addition to the telluric correction of the entire spectra, we have used the same method to remove the weak telluric lines that contaminate the spectral region comprising the C$_{2}$ Phillips system. During the correction process, we noticed an additional system of narrow telluric lines in the 5780-5815 \AA\ region that are lacking in the present TAPAS model. These very weak lines have been added very recently to the HITRAN database and correspond to the $b^{1} \Sigma_{g}^{+}$ (v=3)-X$^{3}\Sigma_{g}^{-}$ (v=0) band of O$_2$
\citep{HITRAN2017}. Pending their introduction into TAPAS, we have used a weakly reddened star and extracted the system of lines in this spectral area. We have used this empirical transmittance in the same way as the TAPAS one.

Second, we have stacked the individual multiple exposures of the same target, using the original spectra or the telluric-corrected spectra if necessary. The spectra were Doppler-corrected and interpolated to produce a single spectrum in the heliocentric frame. The spectra obtained with the different instrumental configurations have been concatenated. For overlapping spectral intervals we have not averaged the data; instead we have chosen the data obtained with the grating configuration providing the best S/N, i.e., the 437~nm setting was selected in the 346~nm/437~nm overlap region, and  the 560~nm-RedL was selected in the 564~nm-RedL/564~nm-RedU overlap region.

\subsection{Single-cloud sample selection}
Highly reddened lines-of-sight typically intersect more than one interstellar cloud \citep{Welty01}. Consequently, the resulting DIBs then represent some ill-defined average of the DIB properties along line-of-sight conditions, which complicates the interpretation. In the case of sub-structure studies, the existence of clouds at various radial velocities has the effect of smoothing and suppressing the structures. Therefore, we restrict our study of the \element[][][][2]{C}-DIBs to single-cloud sightlines, characterized by small Doppler broadening in atomic lines. 

In order to select single-cloud sightlines, we have fitted the \ion{Na}{i} 3302.368/3202.987~\AA\ UV doublet and the \ion{Na}{i} 5890/5896~\AA\ D2-D1 doublet. Due to the intrinsic weakness of the transitions for the UV doublet, there is no or very little saturation and it is easy to determine the number of intervening interstellar clouds with gas columns large enough to produce detectable DIBs. In the case of the 5890/5896~\AA\ doublet there is a strong saturation and line broadening for clouds with \element[][][][2]{C}, precluding a clear component separation. These transitions have been used to check the structure deduced from the UV doublet. We modeled the absorption lines of the doublets as convolved products of Voigt profiles in combination with a polynomial continuum, and we determined the radial velocities of the major components \citep[see, e.g.][]{Puspitarini12}. Here the shape of the continuum and the number of clouds are determined in a new and totally automated way. To do so, the noise standard deviation is initially measured in regions adjacent to the sodium lines. Model adjustments are then made iteratively for an increasing number of clouds until the residuals in the sodium area become equivalent to the standard deviation. The prior radial velocity of each additional cloud is derived from the location of the minimum in the residuals from the previous step of adjustment. 

All spectra and fit results were inspected visually to confirm the presence or absence of multiple velocity components. Fig.~\ref{Fig:naIdoublet} shows the lack of Doppler splitting in the interstellar \ion{Na}{i} lines for the single-cloud selected lines-of-sight which also have detectable \element[][][][2]{C} transitions and are subsequently kept in the present study. The selected EDIBLES target stars that satisfy all conditions are: HD\,24398, HD\,203532, HD\,149757, HD\,185418, HD\,170740 and HD\,147889. Note that, in the case of HD\,23180, \citet{Welty01} have found two distinct \ion{K}{i} lines at R $\sim$ 600\,000, at variance with HD\,24398 and HD\,149757 that show only one component at the same resolution. However, we keep HD\,23180 in the analysis since the Doppler splitting is very low and we assume that the two velocity components belong to two clumps of the same cloud.
Table~\ref{table:sightline_properties} lists their spectral type, radial velocity and rotation velocity as well as the line-of-sight reddening E(B-V), the total-to-selective extinction ratio R$_{V}$, the molecular H$_{2}$ fraction f(H$_{2}$), and the heliocentric radial velocity of the dominant absorbing cloud. Note that we also added to Fig.~\ref{Fig:naIdoublet}, to Table~\ref{table:sightline_properties} and to our study the target star HD\,169454. As shown in Fig.~\ref{Fig:naIdoublet}, this line-of-sight does not represent a single-cloud situation, instead there are two main clouds with radial velocities differing by $\simeq$5 km\,s$^{-1}$, and many more other smaller clouds. We have kept this target for comparison, and also because it is characterized by high reddening E(B-V), strong \element[][][][2]{C} lines and deep diffuse bands.
\begin{figure*}[th!]
\centering
\includegraphics[width=0.95\textwidth]{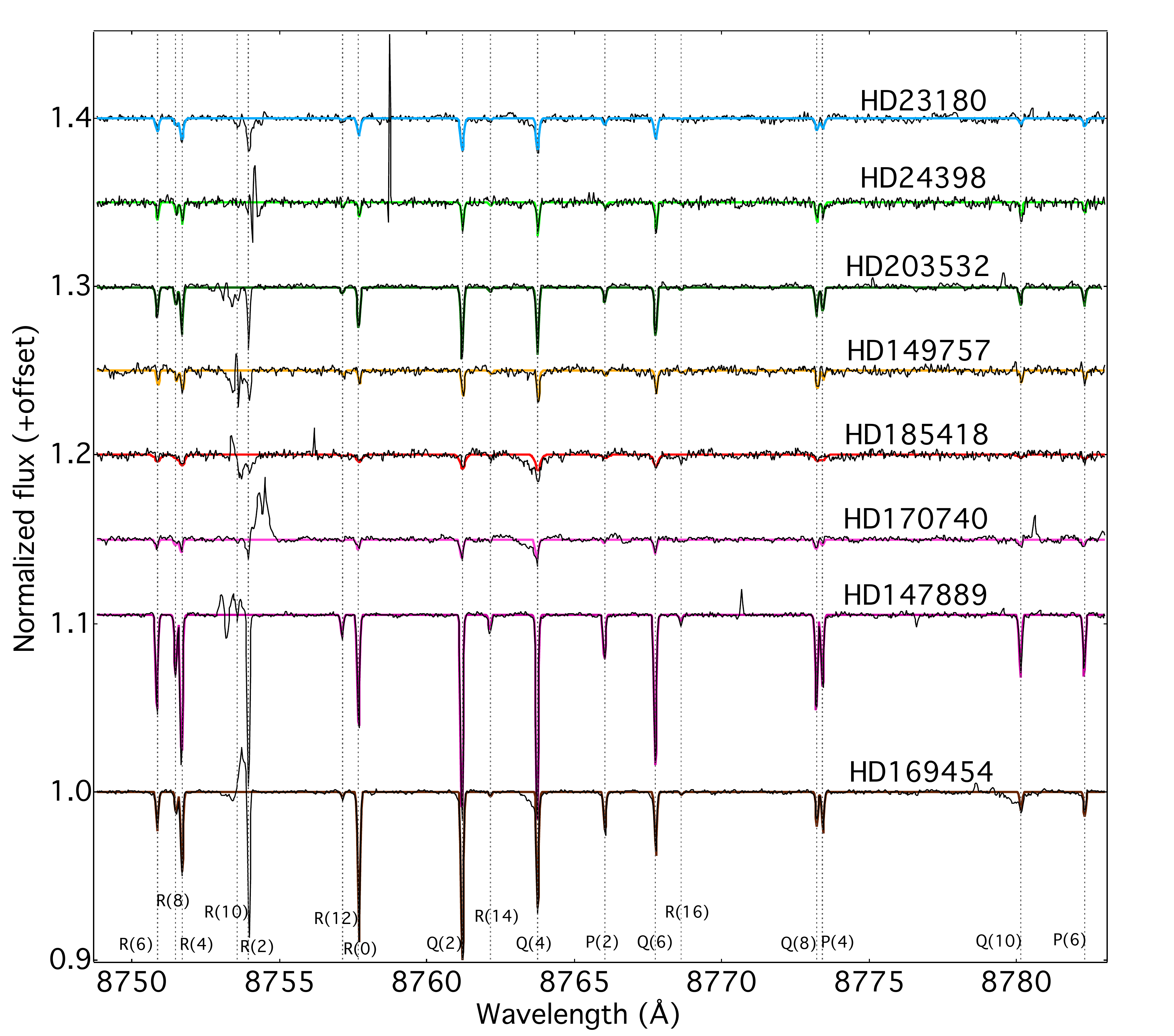}
\caption{\label{Fig:C2fits} 
 Observed Phillips (2-0) \element[][][][2]{C} bands and their best-fitting model profiles. The R$_{10}$ and R$_2$ lines at  8753.949 and  8753.578~\AA\: have not been included in the fit because of the presence of a CCD artifact near these wavelengths that could not be corrected by flat-fielding. In some cases, Q(4) line is blended with a suspected DIB (or stellar line).}
\end{figure*}
\section{Interstellar \element[][][][2]{C} lines: detection and analysis}\label{sec:C2}
As stated above, the second requirement for our target sample is that they additionally possess detectable \element[][][][2]{C} lines. All combined spectra have been visually inspected for the presence of the \element[][][][2]{C} (1-0), (2-0) and (3-0) Phillips bands. This requirement greatly reduces the number of appropriate sightlines. Fig.~\ref{Fig:C2fits} shows the (2-0) \element[][][][2]{C} Phillips lines for the resulting sample. The (2-0) band has been chosen for further analysis due to its high oscillator strength, lack of contamination and apparent absence of CCD fringing residuals that adversely affect analysis of the (1-0) band.

Column densities for \element[][][][2]{C} were derived using a custom fitting procedure written in Python. The (2-0) spectra in the range 8750--8785~\AA\ were normalized to remove the stellar continuum, and the \element[][][][2]{C} spectrum was fitted using a model consisting of Gaussian absorption profiles characterized by a single Doppler shift ($v$) and broadening parameter ($b$), convolved with the 2.8~km\,s$^{-1}$ UVES point spread function. The model opacity spectrum was generated as a function of column density ($N$), kinetic temperature ($T_{kin}$) and H + H$_2$ number density ($n$) using the excitation model of \citet{1982ApJ...258..533V}. Tabulated level populations were obtained from McCall's \element[][][][2]{C} Calculator (http://dib.uiuc.edu/c2/; see also \citealt{2006ARA&A..44..367S}) on a finely-sampled $T,n$ grid. Transition wavelengths and oscillator strengths were taken from \citet{2007ApJS..168...58S}. The set of variable model parameters ($v,b,N,T,n$), defined in Table~\ref{table:c2_properties}, was optimized using Levenberg-Marquardt $\chi^2$ minimization until the best fit to the observed spectrum was obtained. The resulting parameter values for each sightline are given in Table~\ref{table:c2_properties}, and the best-fitting spectral models are shown in Fig.~\ref{Fig:C2fits}. A complete description of all the EDIBLES \element[][][][2]{C} spectra and their properties will be presented in a separate study (Cordiner et al., in prep.). Parameter uncertainties ($\pm 1\sigma$) were generated using Monte Carlo noise resampling with 200 replications.
\begin{table*}[t!]
\caption{Gas kinetic temperature $T_\mathrm{kin}$, volume density $n$, 
column density $N$, Doppler $b$ parameter, radial velocity $v$ derived from Phillips (2-0) \element[][][][2]{C} system for the selected sightlines, and \element[][][][2]{C} column density normalized to extinction $N(C_2)$/E(B-V).}        
\label{table:c2_properties}     
\centering                       
{\renewcommand{\arraystretch}{1.5}
\begin{tabular}{l l l l l l l l l }       
\hline
&\multicolumn{6}{|c|}{This work} & \multicolumn{2}{c}{Kazmierczak et al.\tablefootmark{a}}\\
\hline
Identifier & $T_\mathrm{kin}$ & $n$ & log $N(C_2)$ & $b$ & $v$ & $N(C_2)$/E(B-V)  & $T_{02}$ & ref\\    
 & (K) & (cm$^{-3}$) & (cm$^{-2}$) & (km\,s$^{-1}$) & (km\,s$^{-1}$) &  & (K)  &\\ \hline 
   HD 23180  & 44.7$^{+5.9}_{-3.6}$  & 346.5$^{+68.3}_{-46.7}$ & 13.31 $\pm$ 0.02 & 1.95 $\pm$ 0.10 & 13.61 $\pm$ 0.06 & 7.3 $\times$~$10^{13}$   & 20$\pm$7& K09\\
   HD 24398  & 56.7$^{+10.9}_{-8.4}$ & 211.5$^{+51.9}_{-30.2}$ & 13.30 $\pm$ 0.02 &  0.93 $\pm$ 0.13 & 13.97 $\pm$ 0.06 & 6.9 $\times$~$10^{13}$   & -&\\
   HD 203532 & 40.7$^{+1.2}_{-1.3}$  & 282.1$^{+10.1}_{-10.2}$ & 13.58 $\pm$ 0.01 & 1.06 $\pm$ 0.04& 15.40 $\pm$ 0.01& 1.3 $\times$~$10^{14}$   &-&\\ 
   HD 149757 & 52.2$^{+10.9}_{-7.9}$ & 199.2$^{+40.6}_{-21.1}$ & 13.26 $\pm$ 0.02 &  0.90 $\pm$ 0.15&-14.34 $\pm$ 0.06 & 5.6 $\times$~$10^{13}$   &42$\pm$42& K10b\\
   HD 185418 & 66.8$^{+6.2}_{-8.2}$  & -                          & 13.1 $\pm$ 0.1 & 3.55 $\pm$ 0.30  &-10.73 $\pm$ 0.20 &  3.3 $\times$~$10^{13}$  &-&\\
   HD 170740 & 45.4$^{+7.5}_{-5.8}$  & 248.5$^{+42.9}_{-31.1}$ & 13.15 $\pm$ 0.02 &  1.83 $\pm$ 0.11  &-8.80 $\pm$ 0.07 & 3.1 $\times$~$10^{13}$  &14$\pm$5&K10b\\
   HD 147889 & 44.1$^{+0.4}_{-0.5}$  &  260.7$^{+3.2}_{-3.9}$ & 14.09 $\pm$ 0.01   & 0.98 $\pm$ 0.01 & -7.49 $\pm$ 0.01 & 1.2 $\times$~$10^{14}$  &49$\pm$7&K10a\\ \hline 
   HD 169454 & 18.8$^{+0.2}_{-0.2}$  & 356.1$^{+3.3}_{-2.4}$  & 13.85 $\pm$ 0.01  &  0.90 $\pm$ 0.01 &-8.92 $\pm$ 0.01 & 7.1 $\times$~$10^{13}$  &23$\pm$2&K10a\\ \hline 
\end{tabular}}
\tablefoot{
\tablefoottext{a}{Sources are K09: \citet{Kazmierczak2009}, K10a: \citet{Kazmierczak10a} and K10b: \citet{Kazmierczak10b}  }

}
\end{table*}

\section{Extractions of the \element[][][][2]{C}-DIBs}\label{sec:C2-DIBs}
\subsection{Extraction of individual \element[][][][2]{C}-DIBs}
\begin{figure*}
\centering
\includegraphics[width=.93\textwidth,clip=]{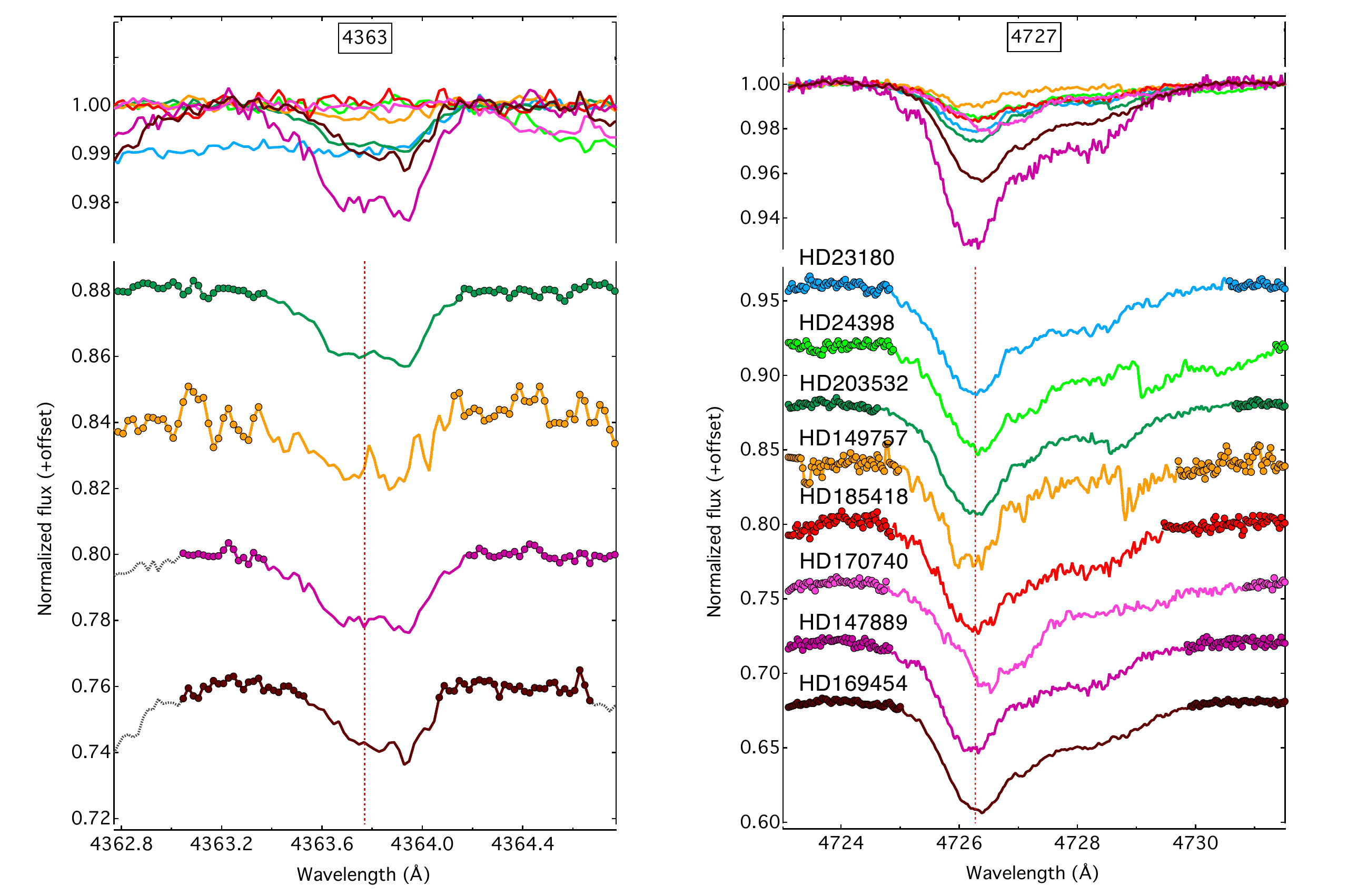}
\includegraphics[width=.93\textwidth,clip=]{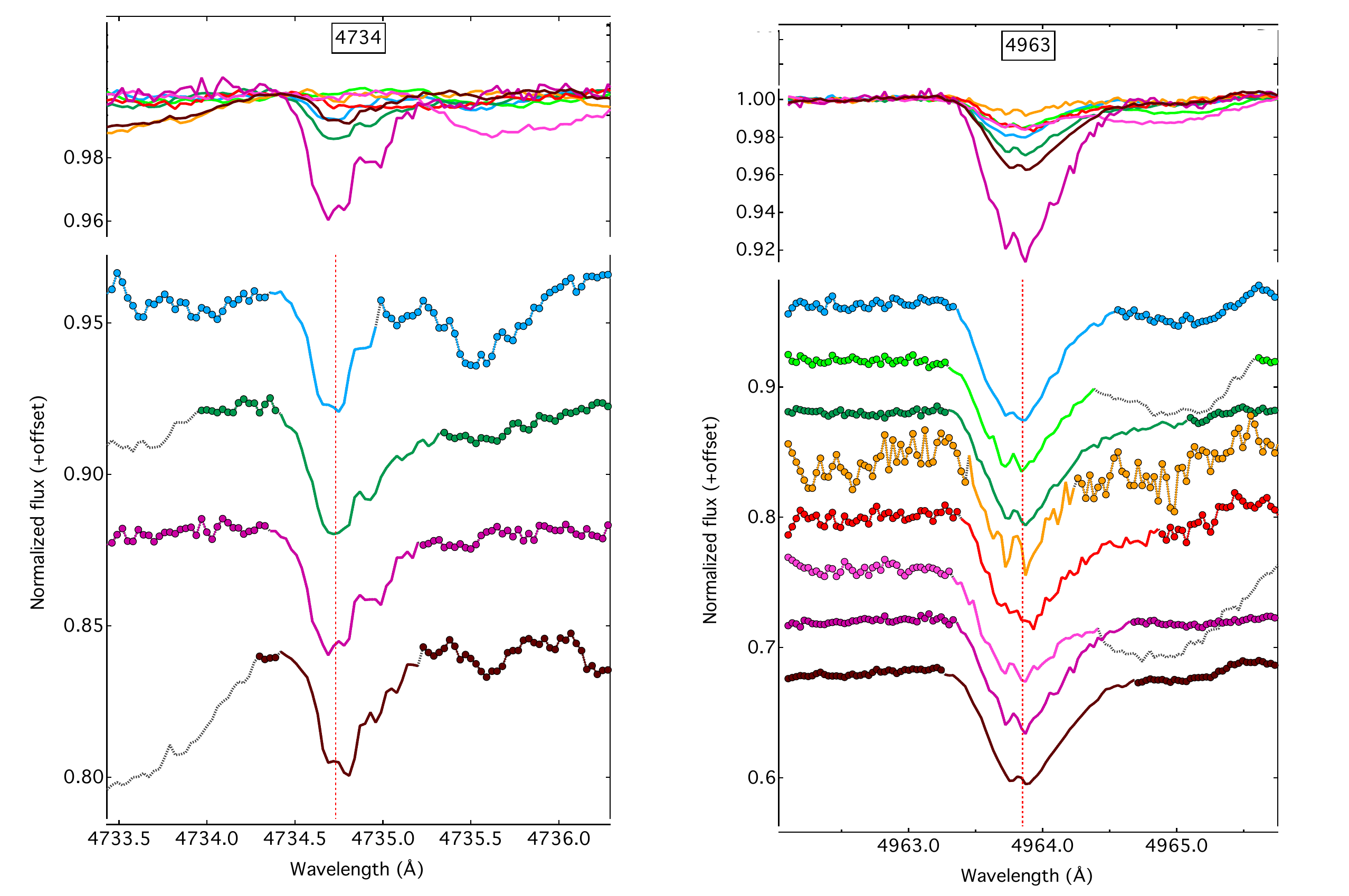}
 \caption{\label{dib0} The 4363, 4727, 4734, and 4963 C$_{2}$-DIBs recorded along seven single-cloud and one multi-cloud line-of-sight. Top in all four panels: overlaid C$_{2}$-DIB profiles at the same vertical scale for all stars. Bottom in all four panels: same C$_{2}$-DIB vertically displaced and depth-equalized profiles. The red dashed line shows the DIB location. See for color coding Fig.~\ref{Fig:naIdoublet}.}
\end{figure*}

\begin{figure*}
\centering

\includegraphics[width=.93\textwidth]{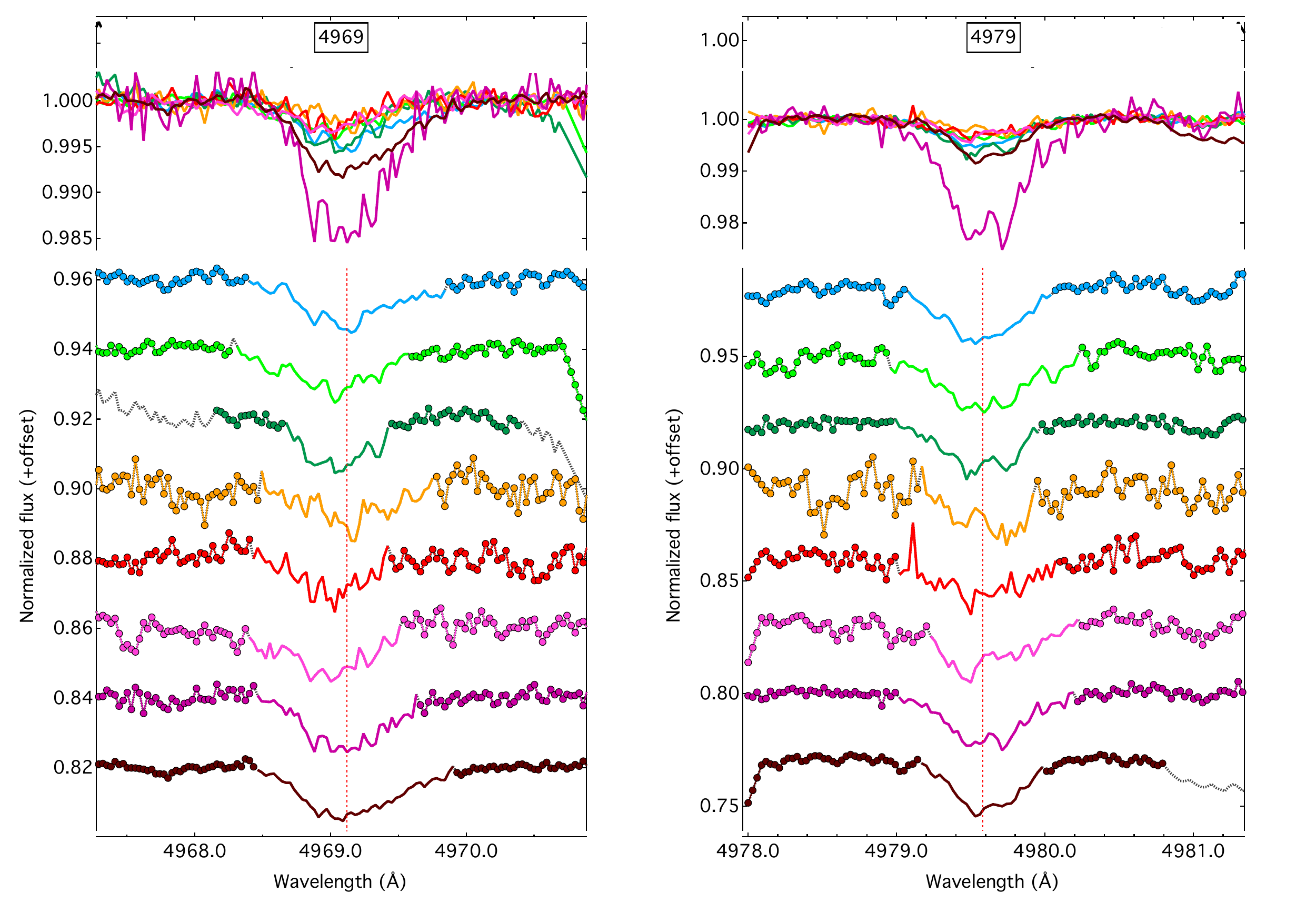}
\includegraphics[width=.93\textwidth]{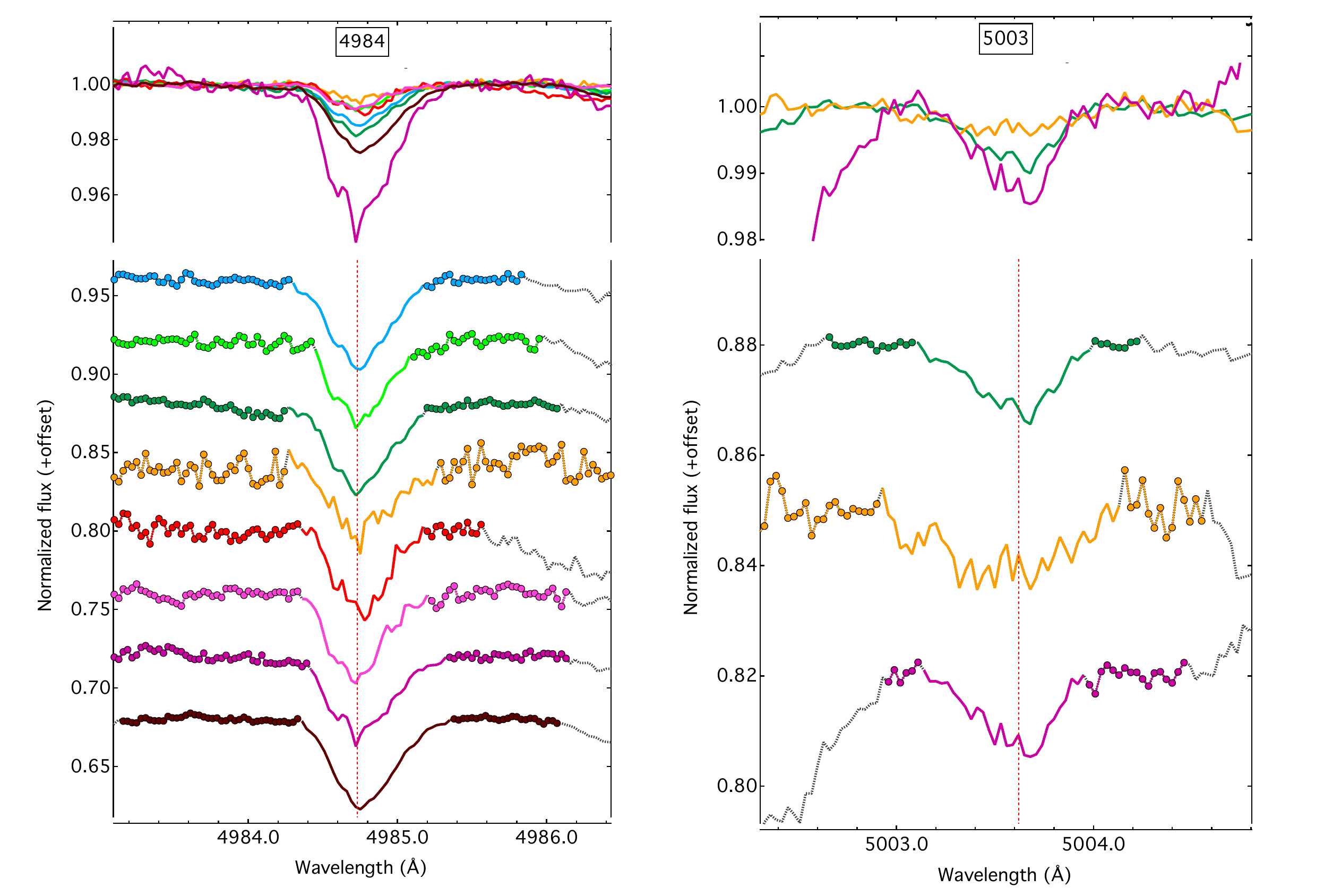}
\caption{\label{dib1} The 4969, 4979, 4984, and 5003 C$_{2}$-DIBs recorded along seven single-cloud and one multi-cloud line-of-sight. Top in all four panels: overlaid C$_{2}$-DIB profiles at the same vertical scale for all stars. Bottom in all four panels: same C$_{2}$-DIB vertically displaced and depth-equalized profiles. The red dashed line shows the DIB location. See for color coding Fig.~\ref{Fig:naIdoublet}. 5003 \AA\: band is blended with stellar lines.}
\end{figure*}

\begin{figure*}
\centering
\includegraphics[width=.93\textwidth]{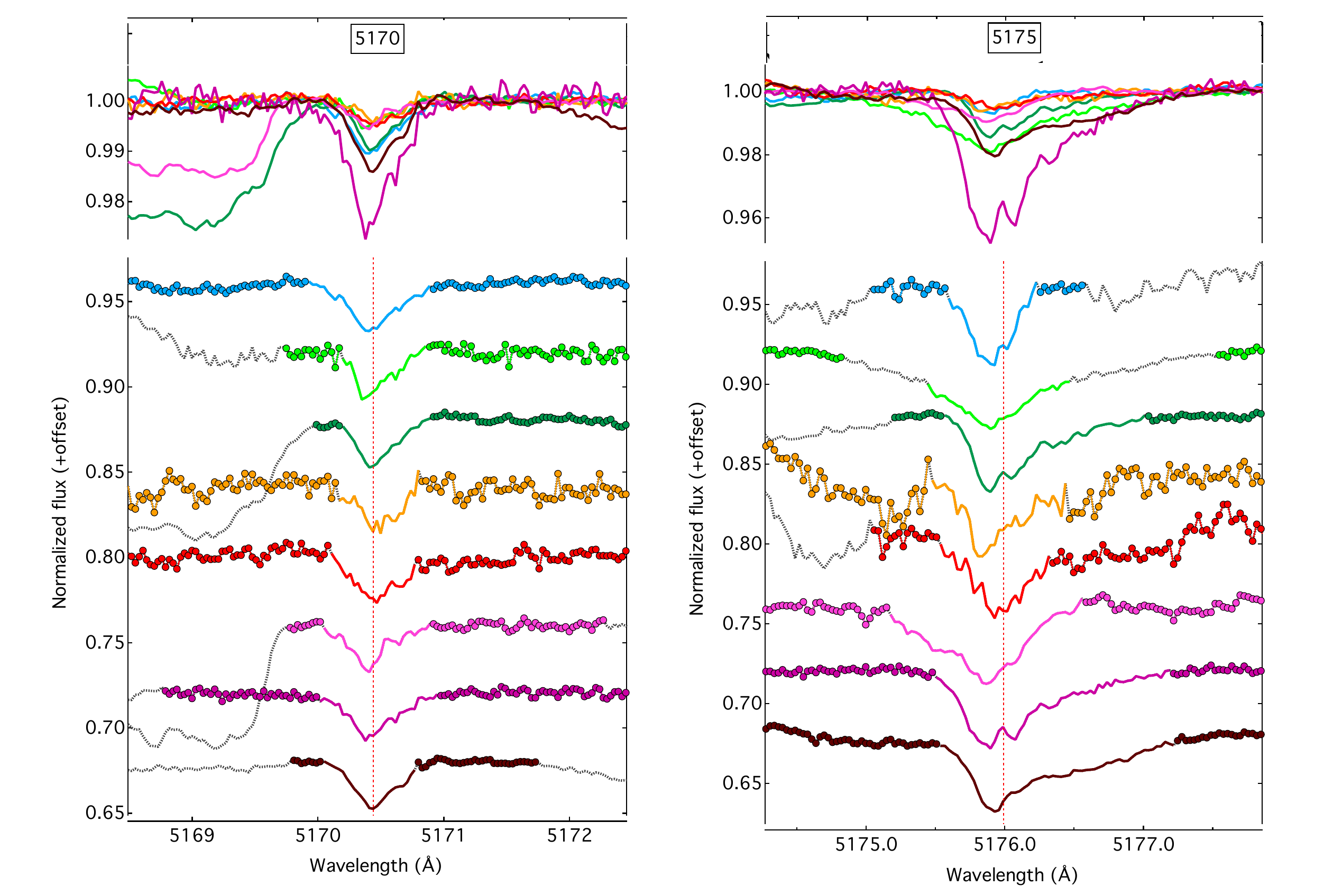}
\includegraphics[width=.93\textwidth]{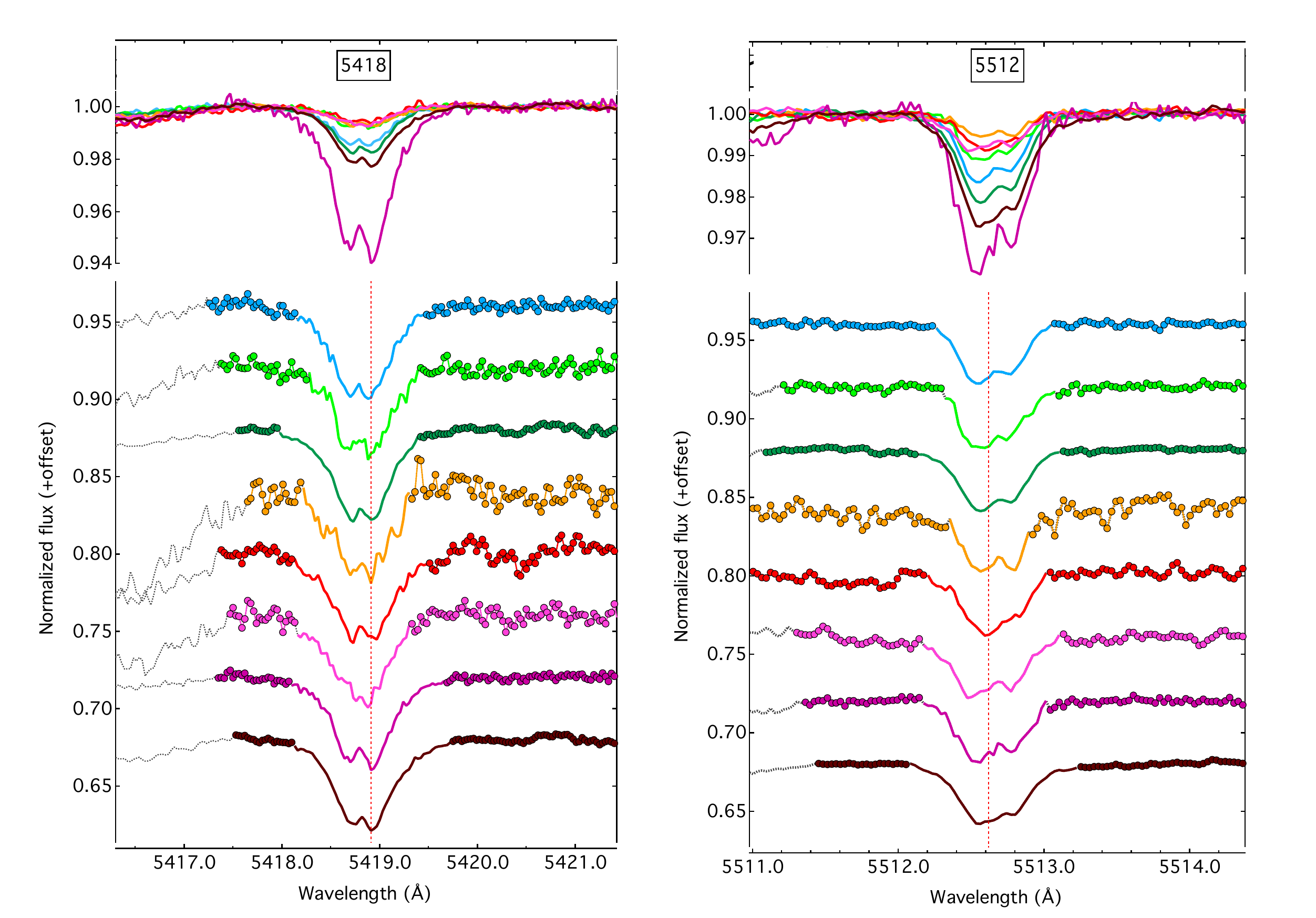}
\caption{\label{dib2} The 5170, 5175, 5418, and 5512 C$_{2}$-DIBs recorded along seven single-cloud and one multi-cloud line-of-sight. Top in all four panels: overlaid C$_{2}$-DIB profiles at the same vertical scale for all stars. Bottom in all four panels: same C$_{2}$-DIB vertically displaced and depth-equalized profiles. The red dashed line shows the DIB location. See for color coding Fig.~\ref{Fig:naIdoublet}.}
\end{figure*}

\begin{figure*}
\centering
\includegraphics[width=.93\textwidth]{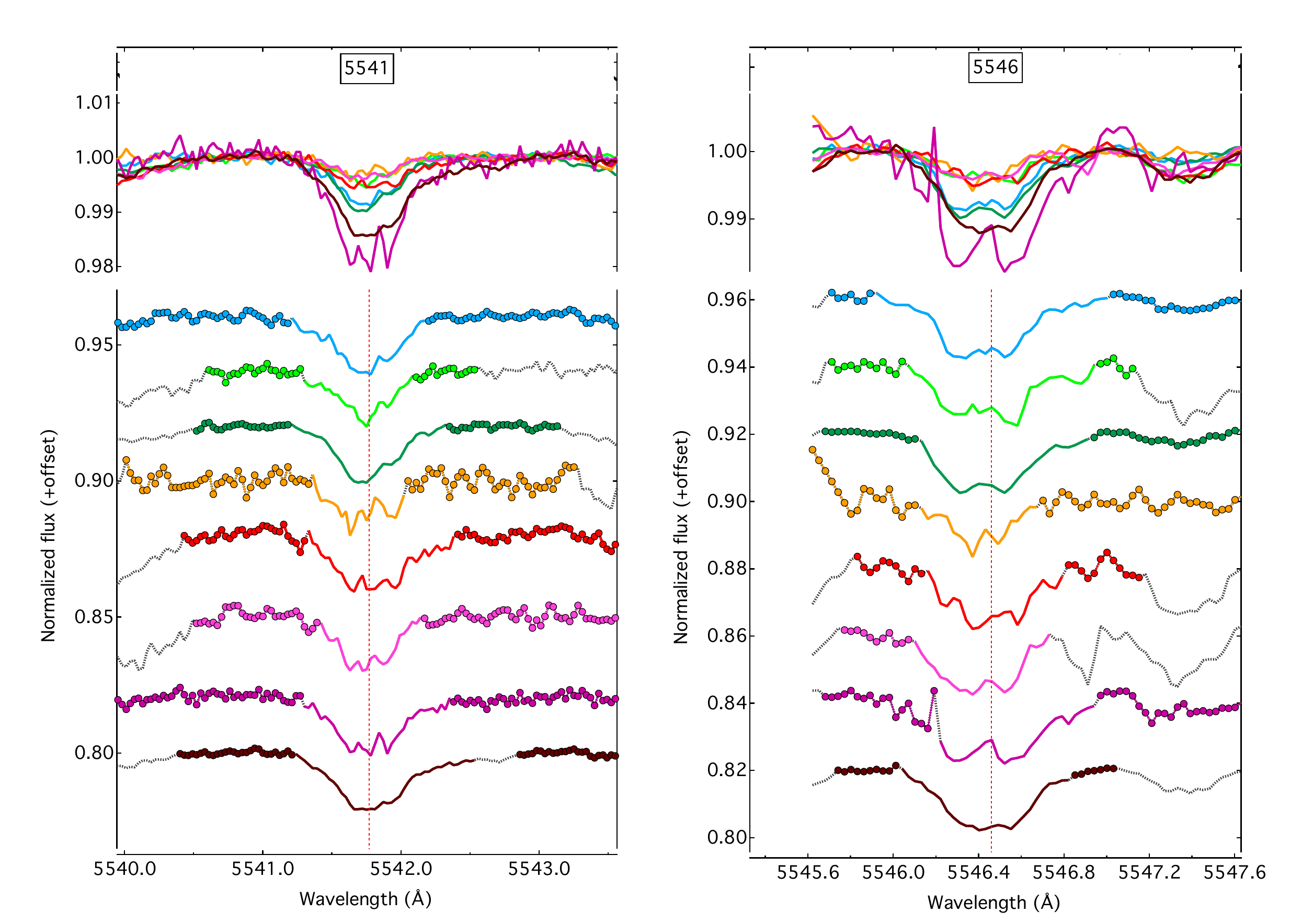}
\includegraphics[width=.93\textwidth]{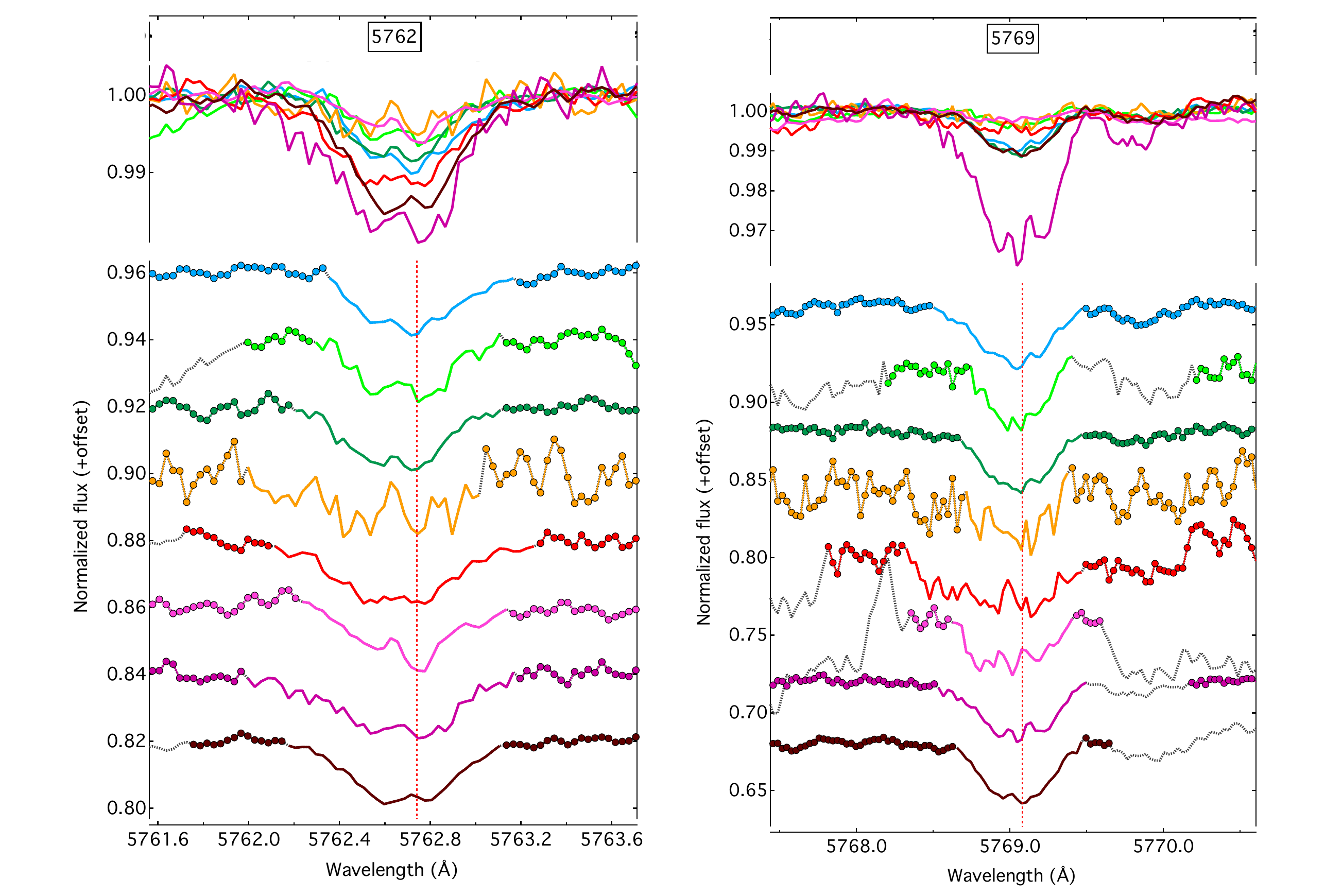}
\caption{\label{dib3}The 5541, 5546, 5762, and 5769 C$_{2}$-DIBs recorded along seven single-cloud and one multi-cloud line-of-sight. Top in all four panels: overlaid C$_{2}$-DIB profiles at the same vertical scale for all stars. Bottom in all four panels: same C$_{2}$-DIB vertically displaced and depth-equalized profiles. The red dashed line shows the DIB location. See for color coding Fig.~\ref{Fig:naIdoublet}.}
\end{figure*}

\begin{figure*}
\centering
\includegraphics[width=.93\textwidth]{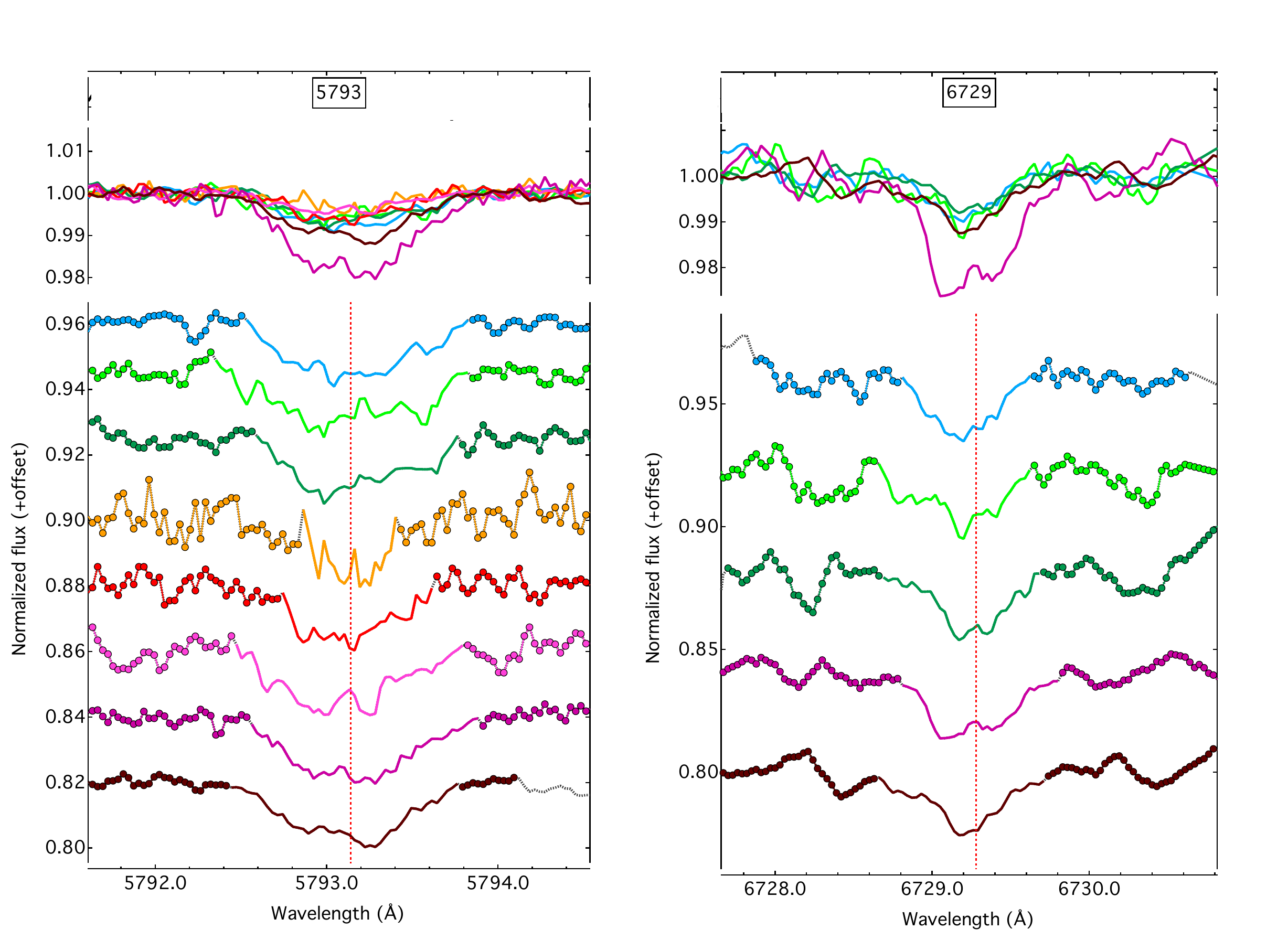}
\caption{\label{dib4} The 5793 and 6729 C$_{2}$-DIBs recorded along seven single-cloud and one multi-cloud line-of-sight. Top in all four panels: overlaid C$_{2}$-DIB profiles at the same vertical scale for all stars. Bottom in all four panels: same C$_{2}$-DIB vertically displaced and depth-equalized profiles. The red dashed line shows the DIB location. See for color coding Fig.~\ref{Fig:naIdoublet}.}
\end{figure*}

We have analyzed the spectral regions that correspond to the series of 18 DIBs classified as \element[][][][2]{C}-DIBs by \cite{Thorburn03}, using the DIB wavelength centers listed in this work. For each sightline and each DIB, we have fitted a third order polynomial continuum around the DIB, normalized the spectrum based on this continuum, then determined the DIB equivalent width (EW) as the area between the spectrum and the normalized continuum. Figs.~\ref{dib0}, \ref{dib1}, \ref{dib2}, \ref{dib3}, and~\ref{dib4} show the normalized spectra around each \element[][][][2]{C}-DIB for the seven single-cloud targets and HD\,169454. These figures also show our choice of the spectral areas on both sides of the band used for continuum-fitting as well as for the spectral interval used for the EW measurements. As stated in the introduction, it is essential here to identify weak and narrow overlapping stellar lines. In lieu of observing additional, lightly reddened stars of similar spectral types to be used as spectral standards, we used the eight targets together and the fact that stellar line widths must correlate with the rotations (at variance with DIBs). To do so we have used the full spectrum to estimate the rotational broadening for each target. We also used the rotation velocities $v$sin($i$) listed  in Table~\ref{table:sightline_properties} when available from the CDS-Simbad website. For example, visual inspection of the stellar \ion{He}{ii} 5876~\AA\: line in Fig.~\ref{Fig:naIdoublet} shows that HD\,149757, HD\,185418 are fast rotators while HD\,23180, HD\,24398, HD\,203532, HD\,170740 have narrower photospheric lines; in these latter cases, we checked even more carefully for the presence of narrow stellar lines in the DIB spectral regions that could mimic DIBs or DIB sub-structures (see the cases of 4727, 4963, 5541, 5546, 5762, 6729; Table~3 from \citealt{Slyk06}). 
In the case of the 4734, 5170, 5175, and 6729~\AA\: DIBs, we found that the continuum is contaminated by stellar lines that fall just outside the DIB and, as shown in Figs.~\ref{dib0}, \ref{dib1}, \ref{dib2}, \ref{dib3} and~\ref{dib4} these lines have been excluded from the fitted spectral interval during the continuum-fitting. In the case of the 4727~\AA\: DIB, it has a double structure suggestive of a blend with a stellar feature on its red side. However, the width of this structure appears to be independent of the stellar rotation. Rather, its shape is perfectly correlated with the main absorption. Therefore, we concluded that this absorption is part of the DIB or an additional, blended DIB, and we measured the total equivalent width of both structures, as shown in Fig.~\ref{dib0}. The situation for the 5003~\AA\: DIB is quite different. There we find a blend with an additional absorption, however, this absorption is absent in three other sightlines. We conclude that this feature must be of stellar origin and we keep only the three clean bands.

We used the residuals of continuum fitting to estimate the standard deviation in each DIB spectral interval and for each target. This standard deviation was used in a Monte-Carlo noise resampling with 100 replications and we estimated in this way the uncertainty on each DIB EW. All EWs and their uncertainties are listed in Table~\ref{table:2}. 
In exactly the same way, we measured the EWs and associated  uncertainties for eight strong non-\element[][][][2]{C} DIBs, and the results are also listed in Table~\ref{table:2}. Their EWs are used in Sect.~\ref{sec:correl}.

\begin{figure}
\centering
\includegraphics[width=0.87\columnwidth]{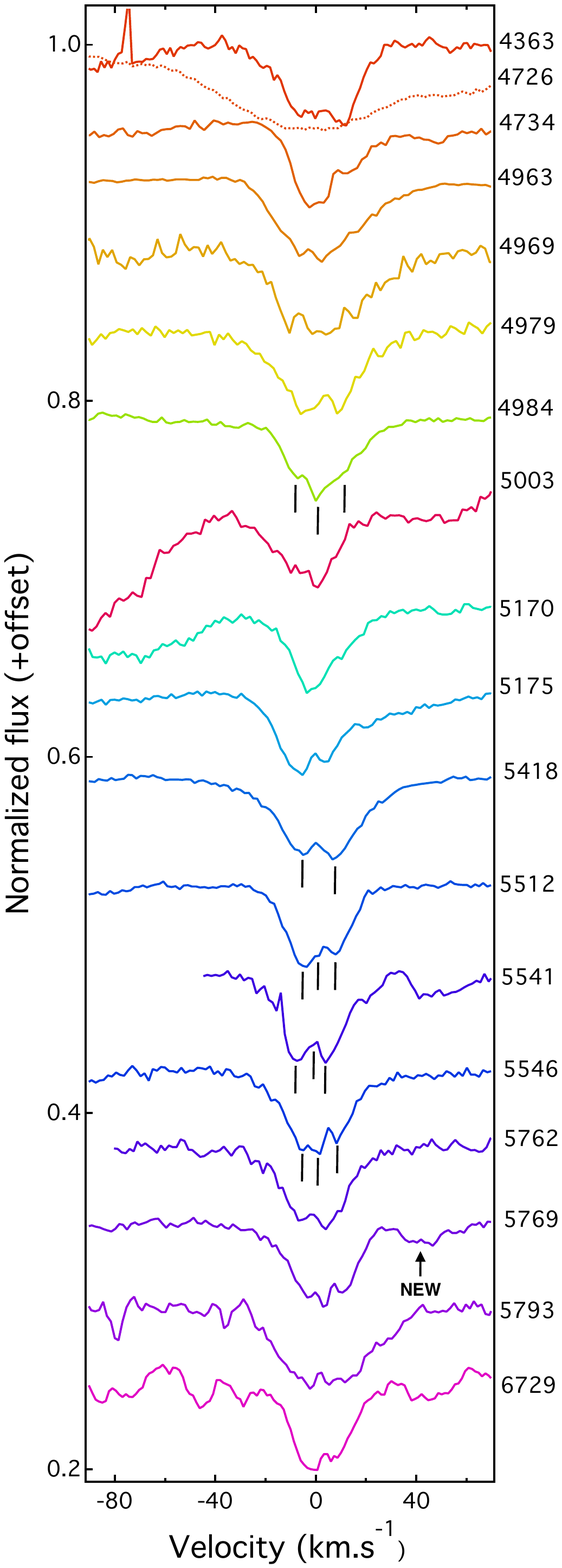}
\caption{Extracted average C$_{2}$-DIB profiles (i.e., from stacking profiles of HD\,23180, HD\,203532, and HD\,147889) normalized to the same depth of 4963~\AA\ DIB. Tick marks refer to the location of sub-peaks of Table \ref{table:3}.}\label{all}
\end{figure}

\begin{table*}
\caption{DIB equivalent widths.} \label{table:2}     
\centering                         
\begin{tabular}{l l l l l l l l l}        
\hline\hline                 
DIB & 23180 & 24398 & 203532 &149757 & 185418 & 170740 &  147889 & 169454\\   
(\AA) &\multicolumn{8}{c}{(m\AA)}  \\ \cline{4-6}
\hline 
\multicolumn{9}{c}{\element[][][][2]{C}-DIBs}  \\
\hline
4363 & -\tablefootmark{a} & <0.3 & 4.3$\pm$0.2 & 1.3$\pm$0.4 & < 0.4 & < 0.2 & 11.1$\pm$0.4 &4.4$\pm$ 0.2 \\
4727\tablefootmark{a} & 46.7$\pm$0.9 & 38.3$\pm$0.7 &55.3$\pm$1.1&17.0$\pm$1.0 & 35.8$\pm$1.0 & 45.3$\pm$1.0 &155.4$\pm$2.9& 95.0$\pm$1.1 \\
4734  & 2.3$\pm$0.8 & <0.5 & 6.5$\pm$0.7 & <0.6& <4 &<0.5 & 15.2$\pm$0.9 & 3.2$\pm$0.4\\
4963 & 12.0$\pm$0.8 & 9.6$\pm$0.3 & 21.2$\pm$0.4 & 3.8$\pm$0.6 & 11.6$\pm$0.7 & 11.2$\pm$0.3 &51.9$\pm$0.9 & 24.8$\pm$1.0\\
4969 & 3.8$\pm$0.3& 2.5$\pm$0.2 & 2.8$\pm$0.2&1.4$\pm$0.5&1.5$\pm$0.3& 2.2$\pm$0.3 & 9.7$\pm$0.8& 6.4$\pm$0.3\\
4979 &3.0$\pm$0.3&2.5$\pm$0.3&3.6$\pm$0.3&1.3$\pm$0.4&1.9$\pm$0.3&2.0$\pm$0.3 &  14.1$\pm$0.9& 3.9$\pm$0.3 \\
4984 &6.6$\pm$0.3 &3.6$\pm$0.3&8.2$\pm$0.4& 2.4$\pm$0.5 &4.0$\pm$0.3& 3.6$\pm$0.2 & 24.4$\pm$1.1 &  11.2$\pm$0.3  \\
5003 &-\tablefootmark{a} &-\tablefootmark{a}&4.1$\pm$0.2 &2.6$\pm$0.5& -\tablefootmark{a} & -\tablefootmark{a} &  5.6$\pm$0.5 & -\tablefootmark{a}\\
5170 &4.5$\pm$0.4 &1.6$\pm$0.2 & 3.2$\pm$0.5 & 1.3$\pm$0.5 & 1.9$\pm$0.3 &1.9$\pm$0.3 &11.1$\pm$0.8   &  5.3$\pm$0.2\\ 
5175 &2.3$\pm$0.2&13.5$\pm$0.4&7.9$\pm$0.3&2.8$\pm$0.8&2.3$\pm$0.5&6.2$\pm$0.5 & 31.7$\pm$1.0&16.1$\pm$0.8\\ 
5418 & 10.3$\pm$0.4 & 5.1$\pm$0.3 & 12.3$\pm$0.3 & 3.7$\pm$0.4 & 5.4$\pm$0.4 & 5.1$\pm$0.3 & 40.8$\pm$0.9 & 16.4$\pm$0.3  \\ 
5512 &7.5$\pm$0.3&5.0$\pm$0.2&10.6$\pm$0.3&2.2$\pm$0.4&3.8$\pm$0.4& 4.5$\pm$0.3 & 18.2$\pm$0.5& 14.6$\pm$0.3 \\
5541 &4.5$\pm$0.3&2.1$\pm$0.2&4.7$\pm$0.2&1.4$\pm$0.3&3.3$\pm$0.3&1.7$\pm$0.2 & 10.9$\pm$0.8 &  9.3$\pm$0.3 \\
5546 &4.0$\pm$0.4&2.0$\pm$0.3&4.7$\pm$0.3&1.5$\pm$0.4&1.9$\pm$0.3&1.7$\pm$0.1&8.2$\pm$0.8& 5.7$\pm$0.3\\
5762 &4.6$\pm$0.3&2.7$\pm$0.2&4.1$\pm$0.3&2.8$\pm$0.5&7.2$\pm$0.5& 2.3$\pm$0.3& 11.6$\pm$1.0 &8.1$\pm$0.3\\
5769 &4.8$\pm$0.4&1.5$\pm$0.3&5.2$\pm$0.3&<2&<4.5&1.2$\pm$0.2& 20.6$\pm$0.8  & 5.4$\pm$0.3\\
5793\tablefootmark{b} &6.3$\pm$0.7 &5.3$\pm$0.6&5.3$\pm$0.4&1.7$\pm$0.4&3.9$\pm$0.5&3.3$\pm$0.4&15.8$\pm$1.0& 8.7$\pm$0.3\\
6729 &4.2$\pm$0.6&5.1$\pm$1.5&3.5$\pm$0.7&<5&<4&<4 & 15.5$\pm$1.5 & 6.6$\pm$1\\
\hline  
\multicolumn{9}{c}{New \element[][][][2]{C}-DIB candidates}  \\
\hline
4737.5 &0.8$\pm$0.3&<0.4&1.2$\pm$0.2&<0.5&<0.5& blend\tablefootmark{a} &3.0$\pm$0.7 & 0.5$\pm$0.2\\
5547.4 &0.7$\pm$0.2&0.9$\pm$0.2&0.8$\pm$0.1&<0.7&1.3$\pm$0.2 &0.9$\pm$0.1&2.5$\pm$0.6&2.0$\pm$0.3\\
5769.8 &1.0$\pm$0.2&0.7$\pm$0.2&0.7$\pm$0.2&<1.3&<1.5& 0.7$\pm$0.1&1.9$\pm$0.7&1.8$\pm$0.2\\
\hline  
\multicolumn{9}{c}{Selected strong DIBs}\\
\hline
5850 & 31.1$\pm$0.6&20.4$\pm$0.7&28.0$\pm$0.3&10.9$\pm$0.6&27.6$\pm$0.6&21.0$\pm$0.4&64.6$\pm$1.0&61.0$\pm$0.7\\
5797 & 58.6$\pm$1.0 & 51.4$\pm$0.6& 53.1$\pm$1.1& 28.3$\pm$0.9 & 80.3$\pm$1.2& 61.2$\pm$0.9& 143.5$\pm$2.5&158.4$\pm$2.3\\
5780 & 80.0$\pm$0.9& 97.2$\pm$0.8 &108.6$\pm$0.9 & 65.6$\pm$1.3& 266.1$\pm$1.8& 237.0$\pm$1.6& 346.9$\pm$2.9& 463.9$\pm$2.2 \\
6196 & 12.6$\pm$0.6&13.6$\pm$0.7&12.7$\pm$0.6&8.5$\pm$0.7&33.7$\pm$2.1&23.7$\pm$0.9&34.8$\pm$1.6&52.4$\pm$0.6\\
6234&5.9$\pm$0.7& 5.9$\pm$0.5 &4.7$\pm$0.3&2.2$\pm$1.0 & 9.1$\pm$0.7&11.6$\pm$0.4& 14.4$\pm$1.4&15.1$\pm$1.2\\
6376&7.8$\pm$0.6&7.7$\pm$0.6&7.7$\pm$0.7&5.0$\pm$0.7&17.3$\pm$1.2&10.4$\pm$0.9&34.1$\pm$2.4&20.0$\pm$1.5\\
6379&37.6$\pm$0.6&57.4$\pm$0.7&43.2$\pm$0.4&15.2$\pm$0.7&71.5$\pm$0.7&55.6$\pm$0.7&82.4$\pm$1.8&93.3$\pm$1.4\\
6614 & 45.3$\pm$0.7 & 60.2$\pm$0.6 & 57.7$\pm$0.6&41.4$\pm$1.0 & 166.3$\pm$0.9&120.0$\pm$0.6&160.6$\pm$2.1&182.6$\pm$1.1\\
\hline
\end{tabular}
\tablefoot{
\tablefoottext{a}{blended with stellar lines}
\tablefoottext{b}{possible telluric residuals}
}
\end{table*}
Inspection of Figs.~\ref{dib0}-\ref{dib4} reveals clearly that sub-structures do exist for at least 14 DIBs.  The shapes look very similar to those of the strong bands in previous work which have been studied in depth and found to be compatible with rotational contours of large molecules. There are strong differences from one band to another, with two or three individual branches, depending on the band, accompanied in some cases by a broad red wing. Fig.~\ref{all} summarizes the observed variety of DIB profiles.

\subsection{Potential additional weak \element[][][][2]{C}-DIBs}
\begin{figure*}
\centering
 \includegraphics[width=.33\textwidth]{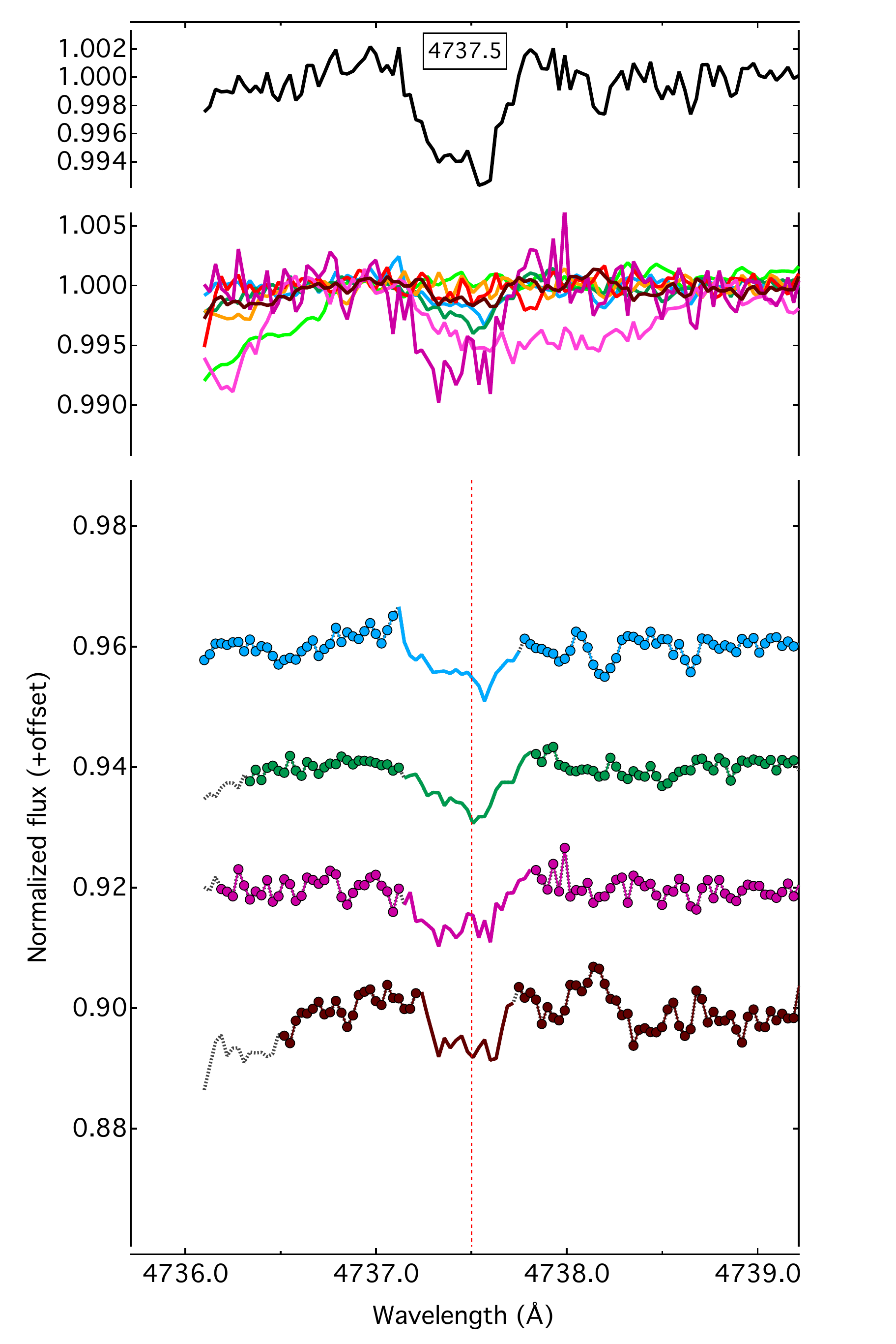}
 \includegraphics[width=.33\textwidth]{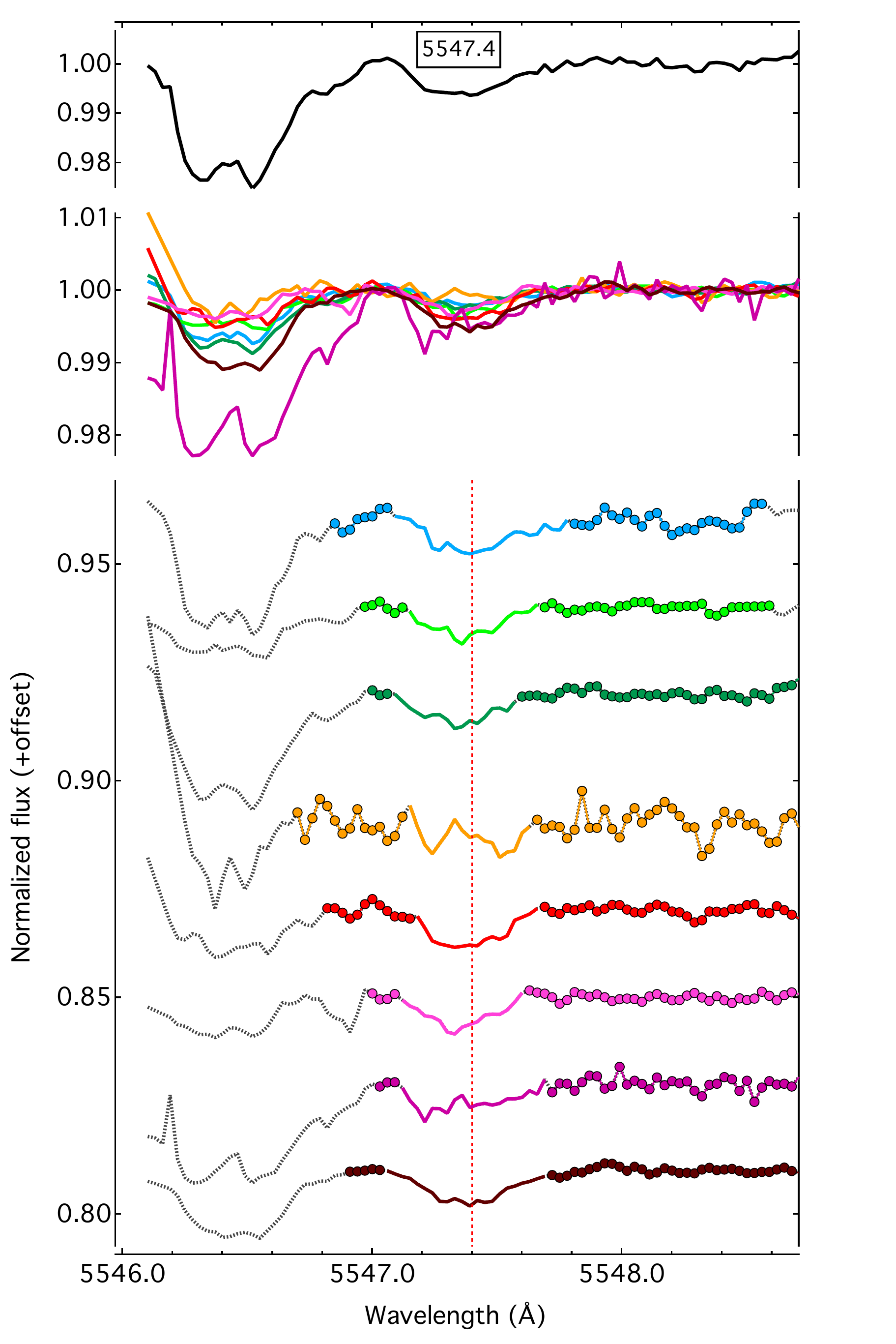}
  \includegraphics[width=.33\textwidth]{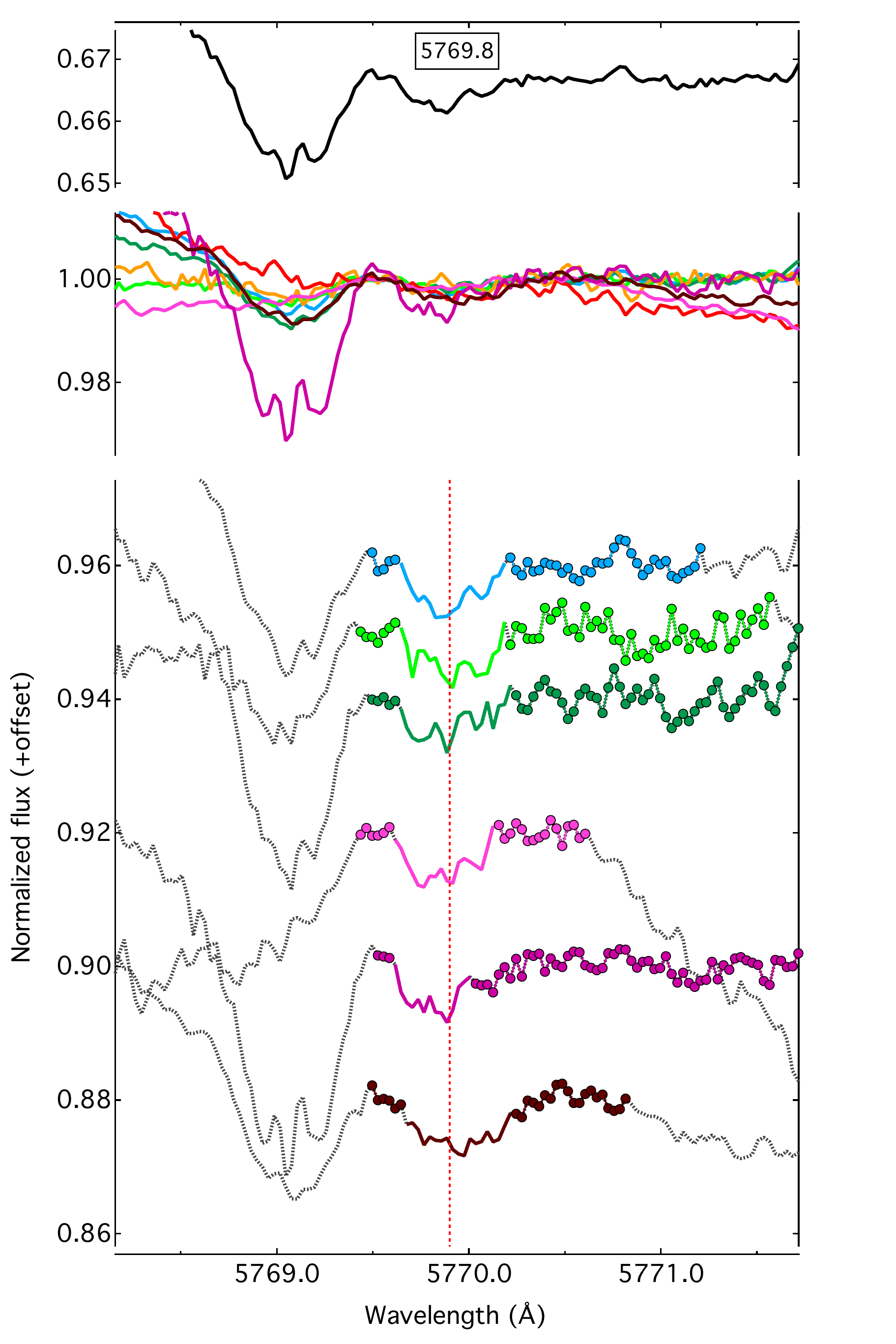}
\caption{\label{newdibs} Potential new detection of weak C$_{2}$-DIBs at 4737.5, 5547.4 and 5769.8~\AA\: (they appear toward sigthlines where the other C$_{2}$-DIBs are well-detected). The red dashed line shows the DIB location. Top graphs (black) in all three panels: extracted average normalized profile of C2-DIBs created from stacking three single-cloud targets: HD\,23180, HD\,203532 and HD\,147889, shown below. Middle in all three panels: overlaid C$_{2}$-DIB profiles at the same vertical scale for all stars. Bottom in all three panels: same C$_{2}$-DIB vertically displaced and depth-equalized profiles. The red dashed line shows the DIB location. The wavelengths are in the interstellar rest frame. See for color coding Fig.~\ref{Fig:naIdoublet}.}
\end{figure*}

In the course of this study we noticed three weak absorption features resembling DIBs falling at the same wavelength when spectra are shifted to the cloud frame, i.e. being very likely of interstellar origin. Fig.~\ref{newdibs} displays the spectra of the targets that show clearly these features. The relative strengths are roughly similar to those found for the other \element[][][][2]{C}-DIBs, which reinforces their identification as new DIBs. These potentially new bands are centered at 4737.5, 5547.4 and 5769.8~\AA, respectively. Their weakness prevents identification of sub-structures in their profiles. Two of them are found to lie 1~\AA\ to longer wavelength of a strong \element[][][][2]{C}-DIB. Further measurements will help in confirming these new bands. We excluded them from the correlational studies that are discussed in Sect.~\ref{sec:correl} due to large uncertaintiesin their EWs. For this reason, it is not possible to ascertain their \element[][][][2]{C}-DIB nature with confidence. 

\section{Variability of profiles among the eighteen C$_{2}$-DIBs}\label{sec:structures}

The main result of this study is the appearance of sub-structures with two or three sub-peaks in at least 14 \element[][][][2]{C}-DIBs, the remaining four being too weak or too contaminated (at 5003, 5762, 5793 and 6729~\AA ) to draw a definite conclusion. 
The two and three sub-peak profile structures are consistent with partially resolved rotational PQR-type branch structures and as such hint for a molecular nature of the involved carriers. It is recognised that it is not proven that the substructures are due to rotational branches and that the shortest wavelength component is an R feature, but for ease of discussion we have chosen this interpretation and to label the components as P (located at the longest wavelengths), Q and R (shortest wavelengths). When treating the profiles as molecular rotational contours, we assume that the upper state rotational constants are smaller than in the ground state. Selection rules can then result in different types of transitions resulting in either a two-peak profile similar to the 5797 or 6196~\AA\: DIB profiles that then represent unresolved 
P and R branches or a three-peak profile that additionally displays a central Q branch. 
With these conventions, we have determined the peak positions of the 
sub-peaks (in wavenumber space; $\nu_P$, $\nu_Q$ and $\nu_R$) and we have extracted the wavenumber intervals ($\Delta_{\nu_{PQ}}$ = $\nu_{Q}$ - $\nu_{P}$ , $\Delta_{\nu_{QR}}$ = $\nu_{R}$ - $\nu_{Q}$ and $\Delta_{\nu_{PR}}$= $\nu_{R}$ - $\nu_{P}$) between them in all relevant cases. This was done based on visual selection of regions possessing a well-defined minimum or secondary minimum, and subsequent fitting of the bottom region of each sub-structure with a second order polynomial to extract the minimum location. The \element[][][][2]{C}-DIB sub-structure peak separations $\Delta_{\nu_{PQ}}$, $\Delta_{\nu_{QR}}$ and $\Delta_{\nu_{PR}}$ are listed in Table~\ref{table:3}. These values are given in cm$^{-1}$, as this unit offer a wavelength independent value that allows a direct comparison between the different \element[][][][2]{C}-DIBs. We did not attempt to fit the profiles with the product of individual components, as the choice of the number of components is quite subjective, especially in the case of the presence of red wings or for the weakest bands. 
According to the results of Table~\ref{table:3} and as can be seen for all DIBs in Fig.~\ref{all}, the peak relative amplitudes and peak sub-structure separations are quite different among the DIBs. The separation between the P and R branch scales roughly with $B\times T$, where B is the rotational constant and T the carrier's rotational temperature (but see \citealt{Huang2015}). Therefore for the same cloud and within the assumption of the same physical conditions, this observed variability implies that the sizes of the carriers for the different \element[][][][2]{C}-DIBs are most likely quite different.
\section{Prominent sub-structures and average spectral profiles}\label{sec:stack}
\begin{figure*}
    \centering
    \includegraphics[width=0.8\textwidth]{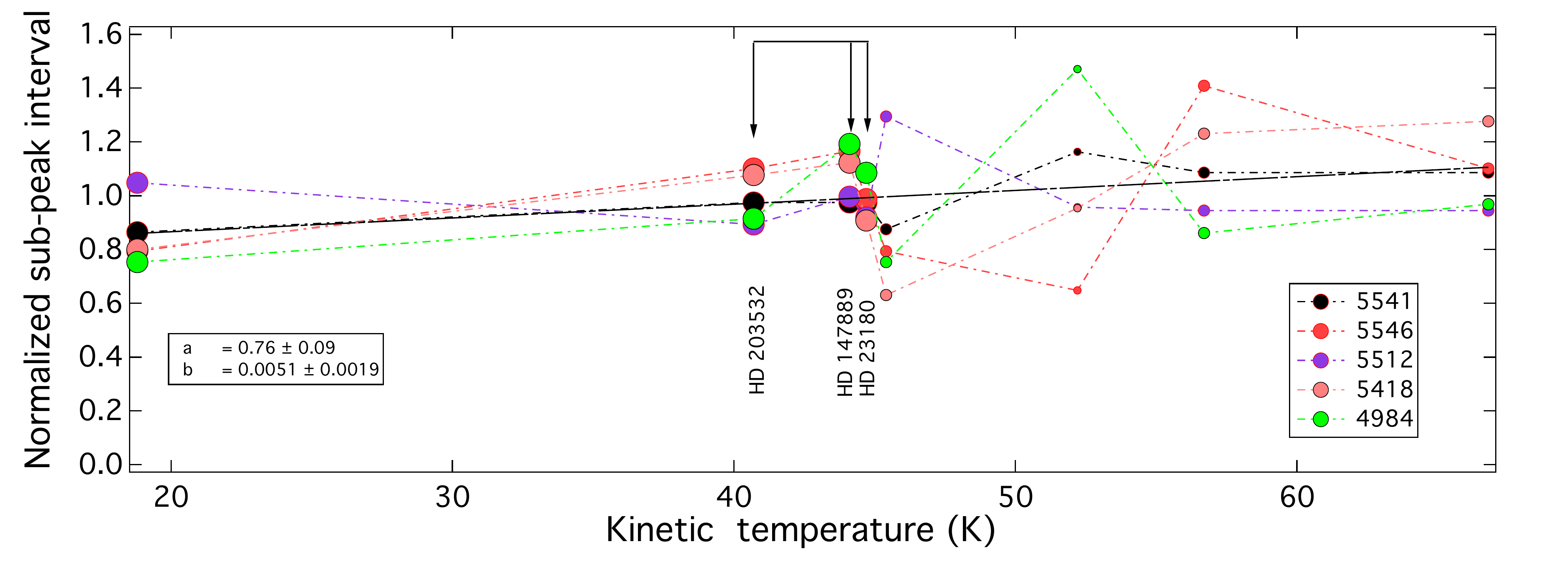}
    \caption{Normalized spectral intervals between the sub-peaks in the profiles of several C$_{2}$-DIBs as a function of the C$_{2}$-based kinetic temperature (see text) of each target. Uncertainties on the intervals can be deduced visually from the drawn profiles in Figs.~\ref{dib0}, \ref{dib1}, \ref{dib2}, \ref{dib3}, and~\ref{dib4}, and are on the order of the data point dispersion. The dashed line is a linear fit to the average over the five different DIBs. The size of the marker indicates the reliability of the sub-peak interval determination, again visually determined from the quoted figures. The three targets used for the co-addition of the DIB profiles are marked by arrows.}
    \label{subpeaks}
\end{figure*}

The generally low signal strengths do not allow clear identifications of sub-structures for all combinations of DIBs and targets. A classical method used to increase the S/N and enhance weak features is the co-addition of profiles. Some caution is needed here since different sightlines may correspond to different shapes of the same DIB, in this case co-adding sub-structures is not physical. This could be the case if the carrier internal population distributions were markedly different from one sightline to another in response to different excitations. Indeed, such peak-to-peak variations have been observed and used to place constraints on the rotational excitation temperatures of the 6614 and 6196~\AA\: DIB carriers \citep{Cami04, Kazmierczak2009}. In order to test the co-addition feasibility, we have used the sub-peak intervals of Table~\ref{table:3} to estimate upper limits on the profile variability. We discarded all cases for which peak determinations were too uncertain, which resulted in the selection of only five bands. For each type of interval, we normalized it by simply dividing by the mean value of the sub-peak intervals over all eight targets. Fig.~\ref{subpeaks} displays the resulting normalized intervals between the sub-peaks, here as a function of the kinetic temperature $T_{kin}$ for the corresponding target, derived from the C$_{2}$ analysis.  
Fig.~\ref{subpeaks} shows that the variations from one sightline to the other for the same band are significantly smaller than the dispersion among the normalized intervals of the various DIBs for the same target, although a weak increase with temperature is discernible. It is difficult to draw firm conclusions from this, given the uncertainties in our sub-peak wavelengths. Three stars, namely HD\,23180, HD\,203532 and HD\,147889  have the deepest absorption bands and are characterized by a good S/N, thus have relatively clear sub-structures. Fig.~\ref{subpeaks} confirms that for these three targets (marked by arrows) the sub-peak interval variability is the smallest. For this reason, we performed a co-addition of the profiles over these three targets only. Before the co-addition, the normalized DIB profiles were shifted to a common interstellar rest frame and elevated at a power in such a way that their depths are the same for the three targets. The average profiles are displayed in Fig.~\ref{all}. Although they can not be considered as pure intrinsic profiles (as stated above, rotational contours may change from one target to the other), these average profiles are free from telluric or stellar contamination and we believe they are our most accurate representation of each DIB structure. Such information complements well that contained in the series of individual measurements. On the other hand, it is often difficult to preclude profile variation and for this reason we discuss the DIBs towards each of the three selected stars separately in the next section. 

\section{Search for profile variability among the sightlines}\label{sec:variab}
With the goal of going one step further in the validation of the average profiles, we investigated the potential variations of DIB profiles for the three selected single-cloud targets. In particular, measurements of the \element[][][][2]{C} molecule rotational temperatures $T_{02}$ defined as the average excitation temperature of the two (resp. three and four) lowest rotational levels of the ground electronic state have been performed by \cite{Kazmierczak2009} and \cite{Kazmierczak10a} for two of these single-cloud sightlines. The temperatures are different: $T_{02}$ = 20~K for HD\,23180 and 49~K for HD\,147889. 

\begin{figure*}
\centering
\includegraphics[width=.3\textwidth]{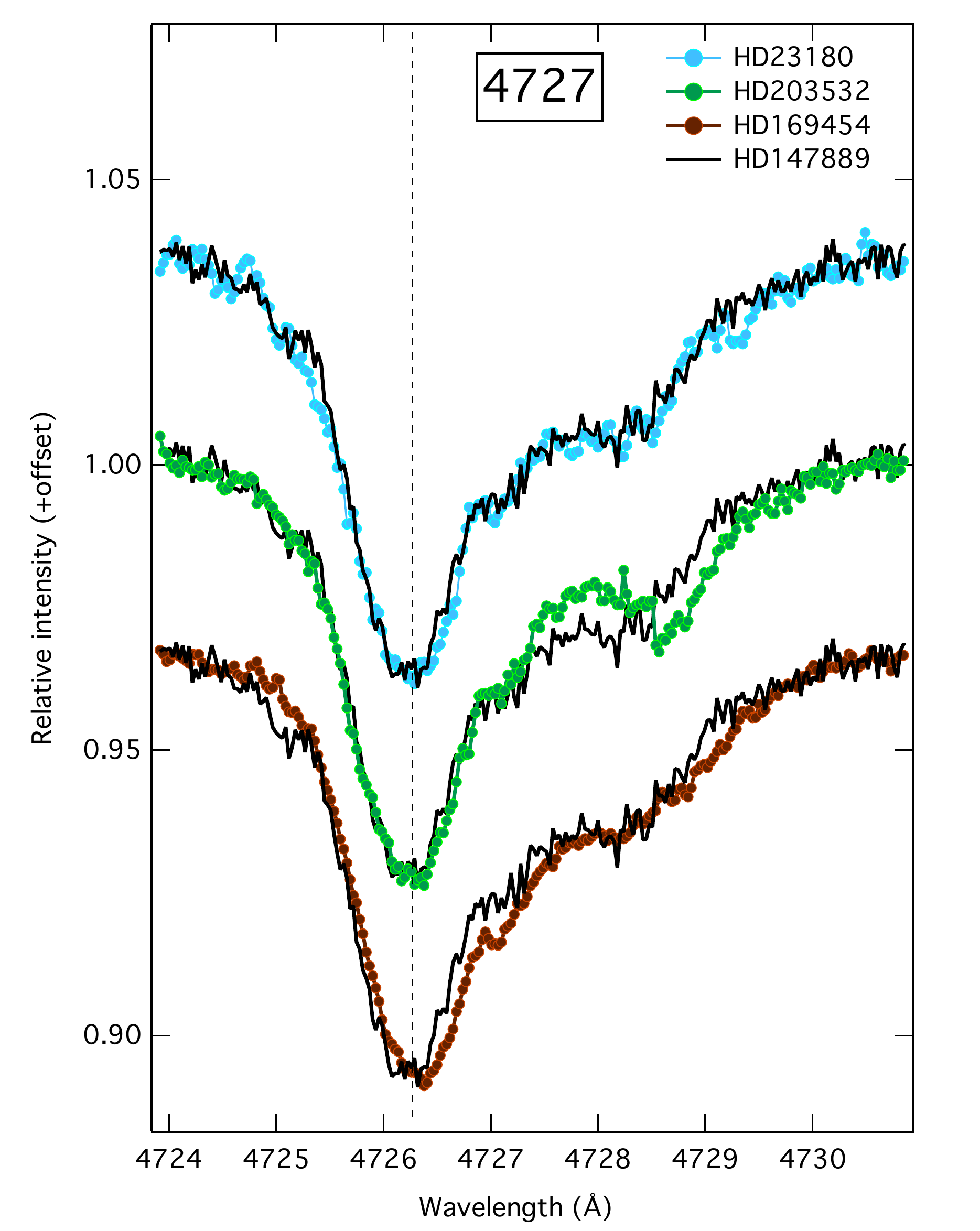}
\includegraphics[width=.3\textwidth]{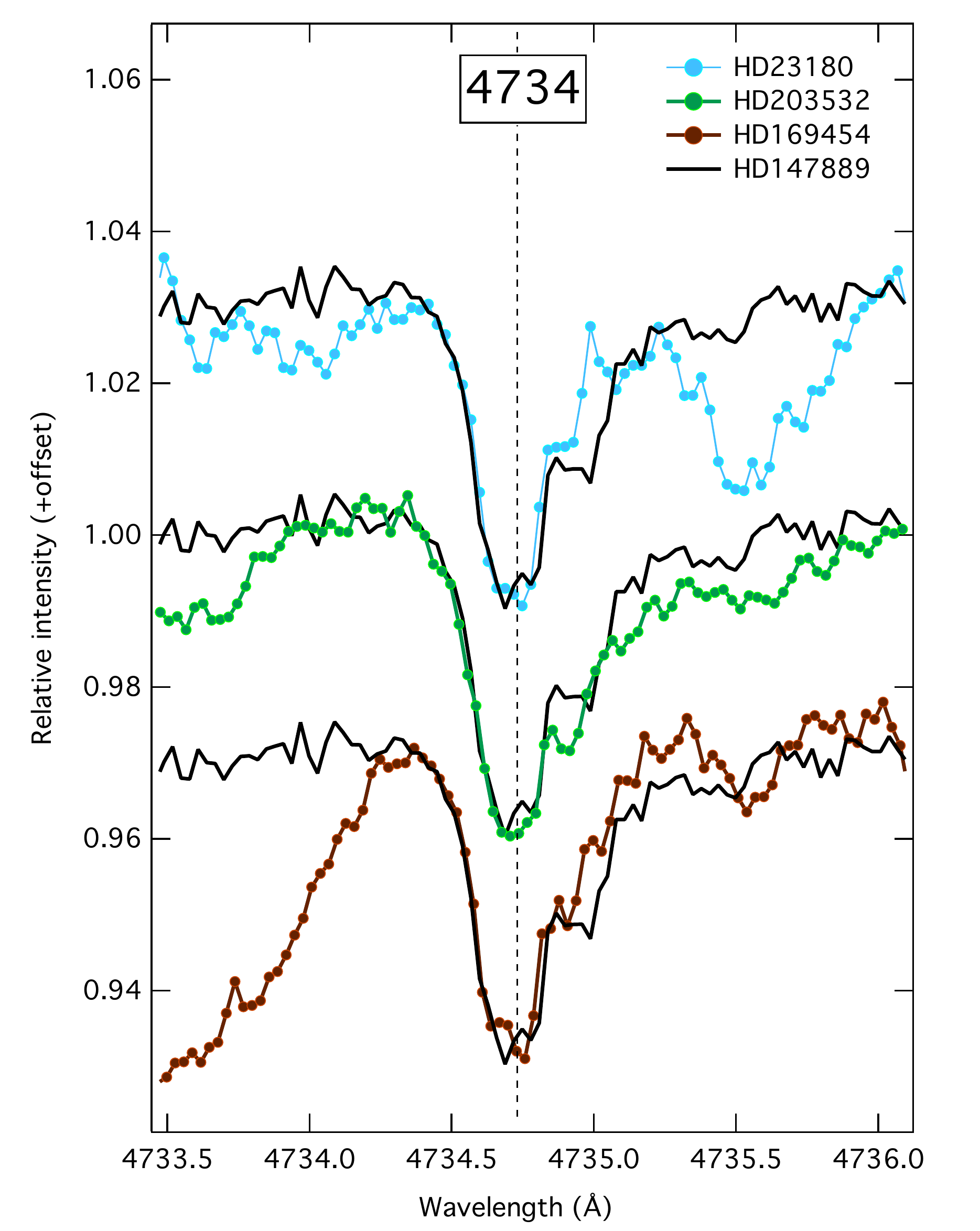}
\includegraphics[width=.3\textwidth]{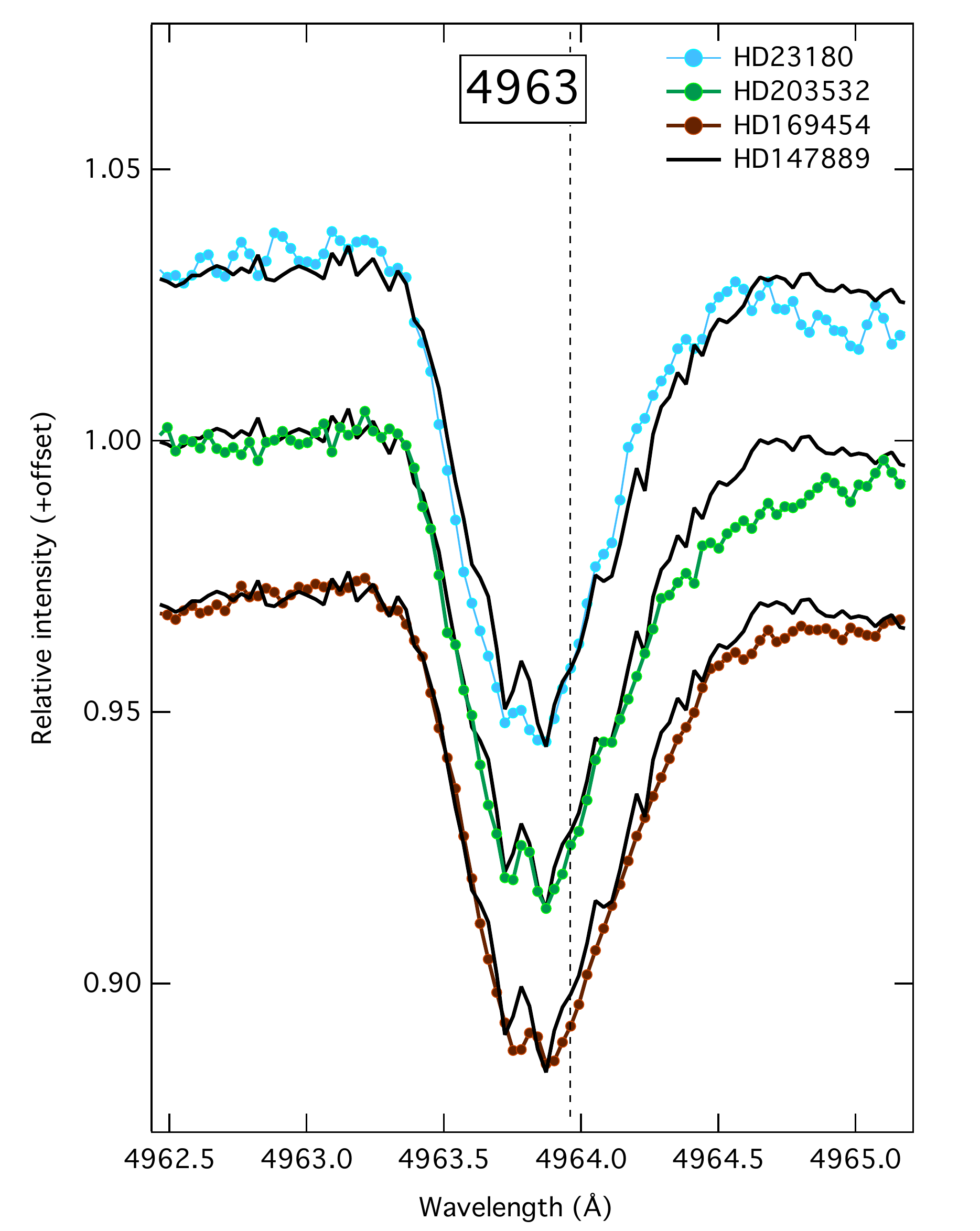}
\includegraphics[width=.3\textwidth]{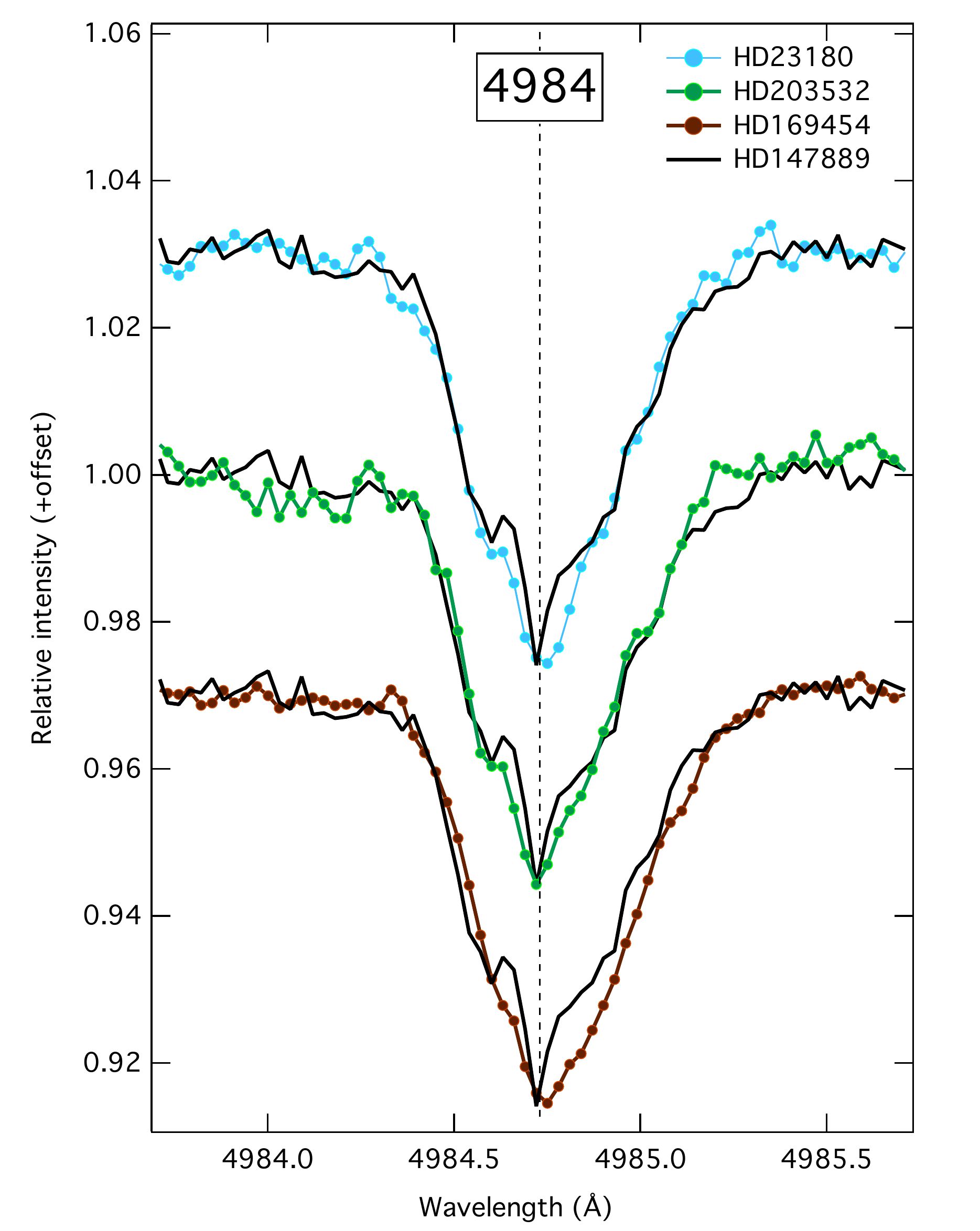}
\includegraphics[width=.3\textwidth]{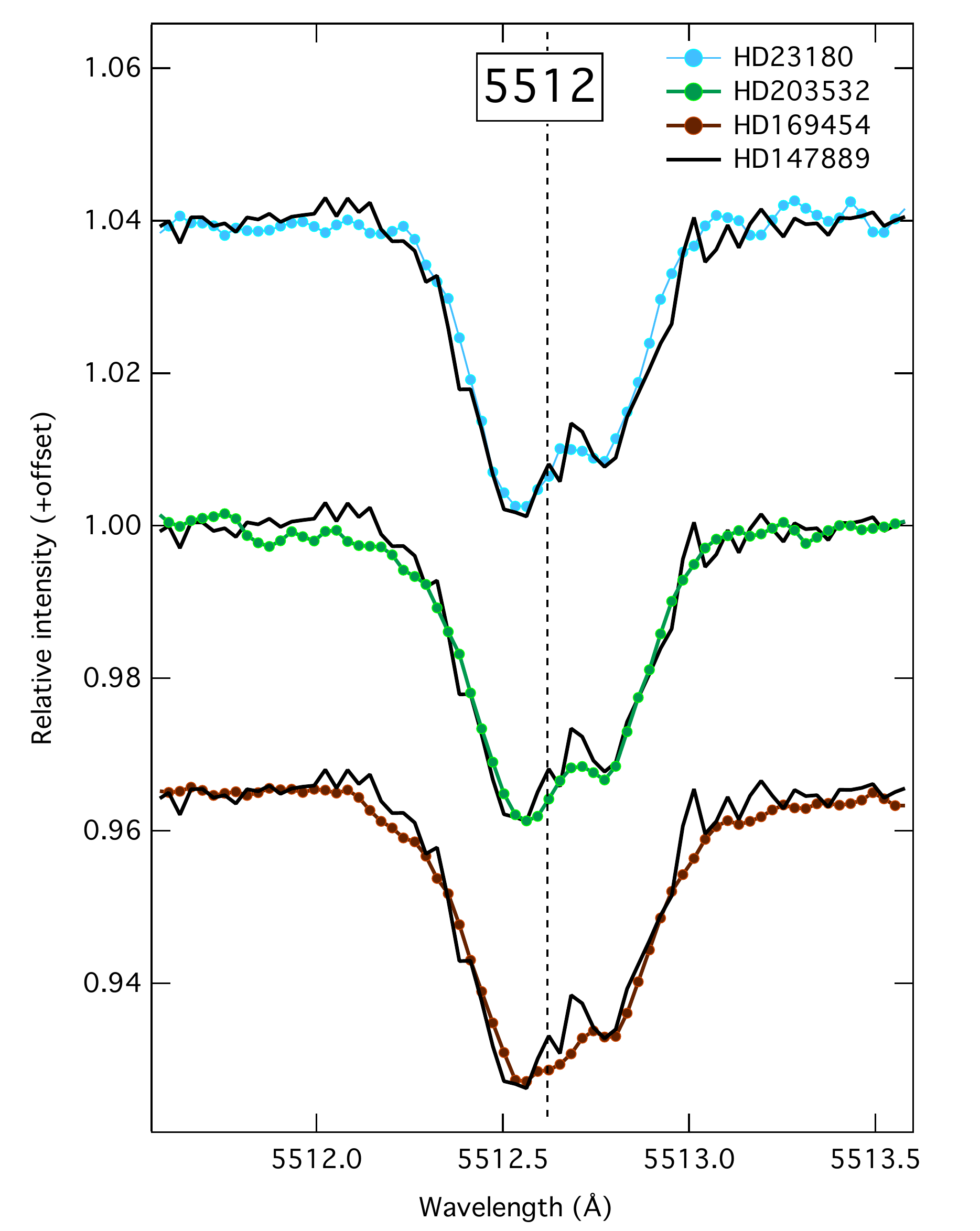}
\includegraphics[width=.3\textwidth]{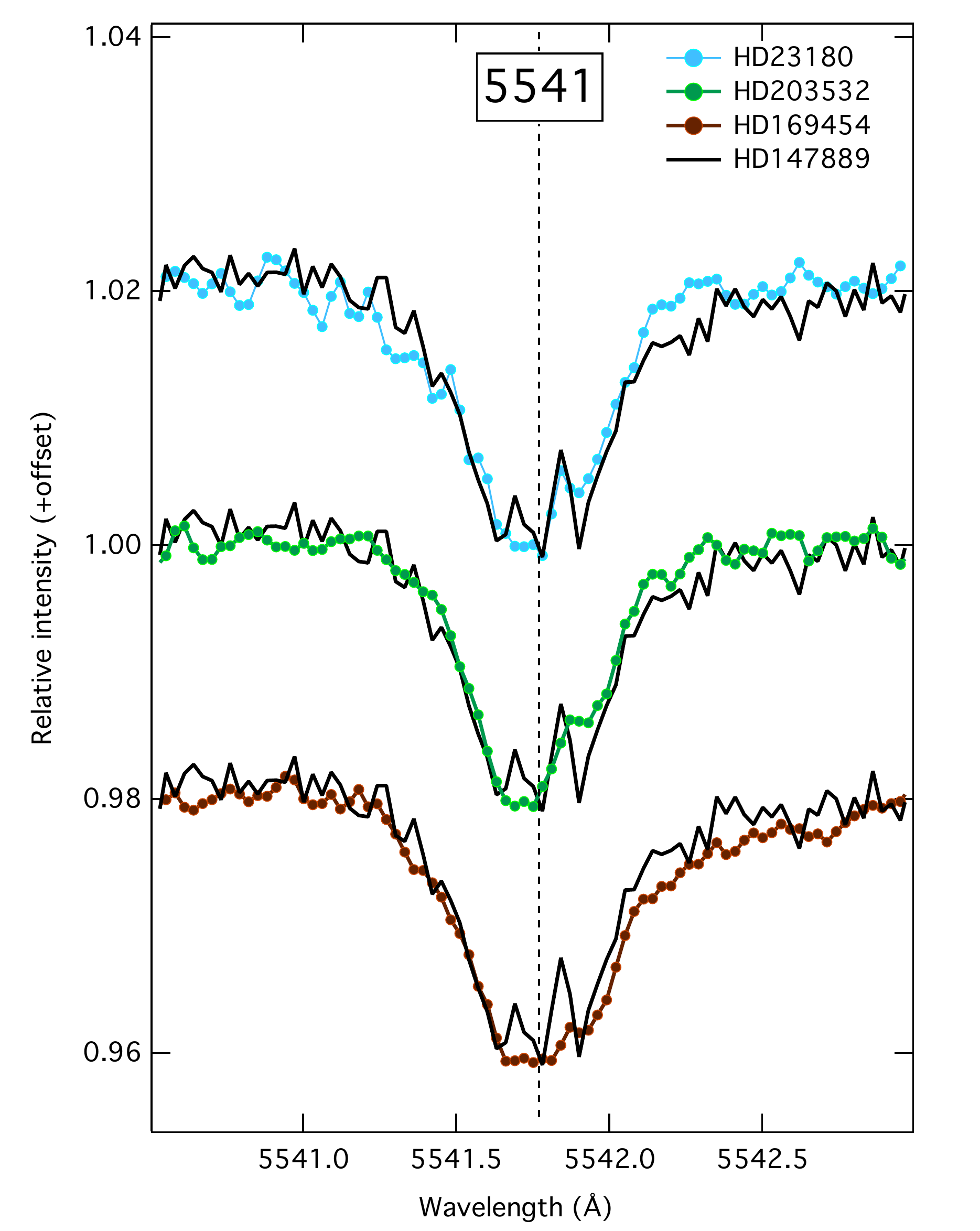}
\includegraphics[width=.3\textwidth]{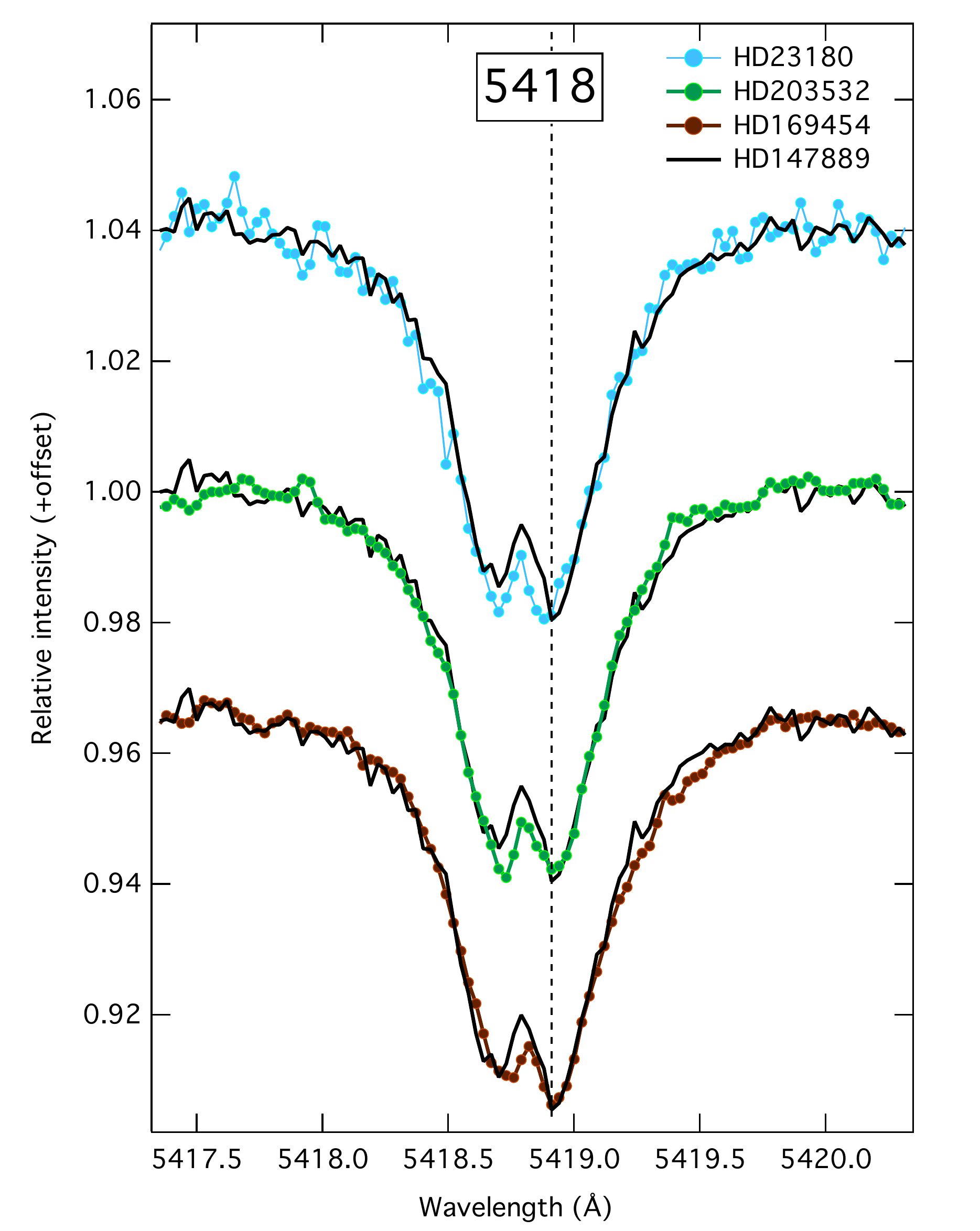}
\caption{Comparison between C$_{2}$ DIB profiles normalized to their central depths toward HD\,23180 (blue), HD\,203532 (green), HD\,147889 (black) and HD\,169454 (brown). The first three targets are those chosen for profile co-additions, for which results are shown in previous figures. The C$_{2}$ excitation temperatures are 20, 23 and 49K for HD\,23180, HD\,169454 and HD\,147889 respectively (see Table~\ref{table:c2_properties}).}
\label{category1}
\end{figure*}

\begin{figure*}
\centering
\includegraphics[width=.3\textwidth]{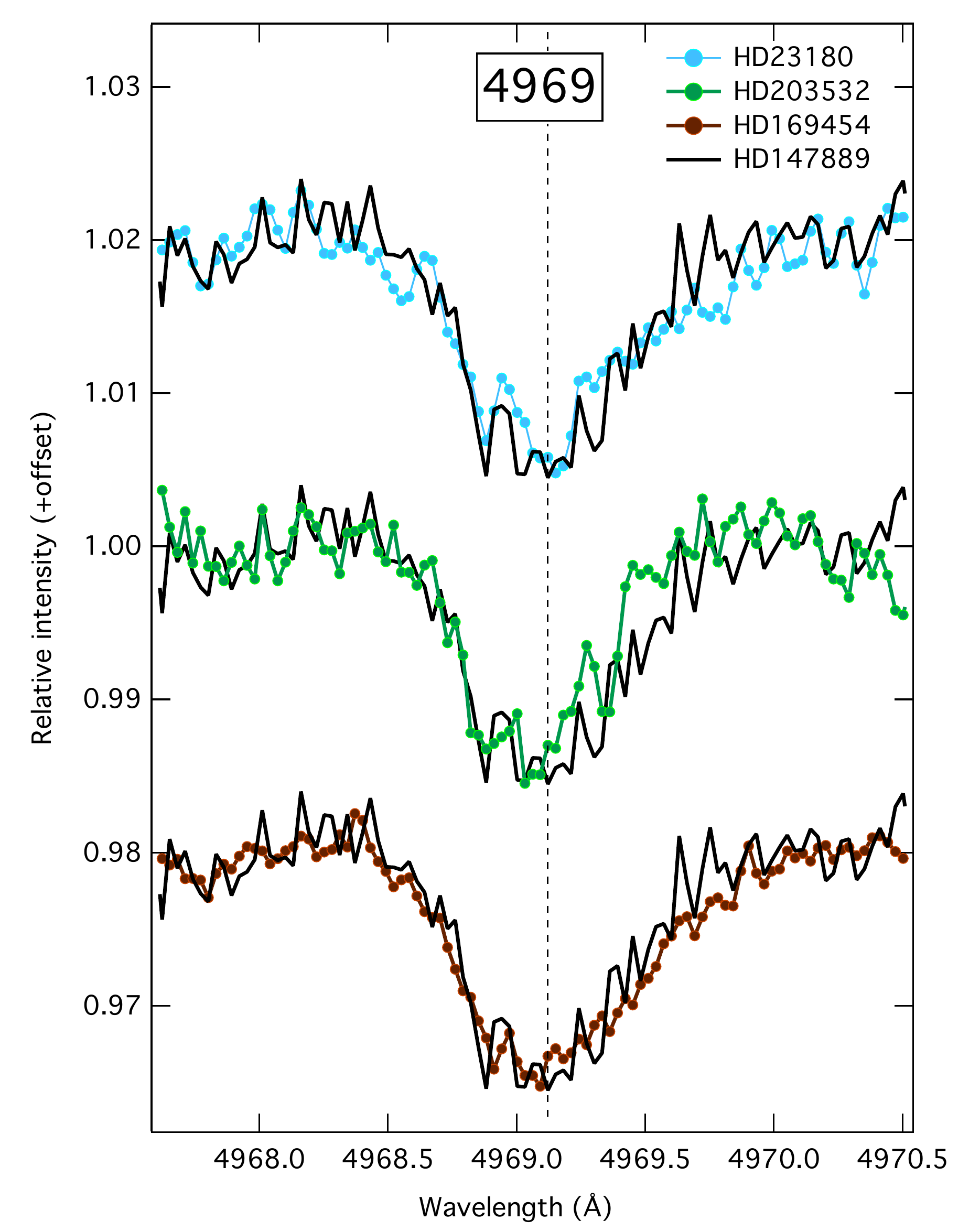}
\includegraphics[width=.3\textwidth]{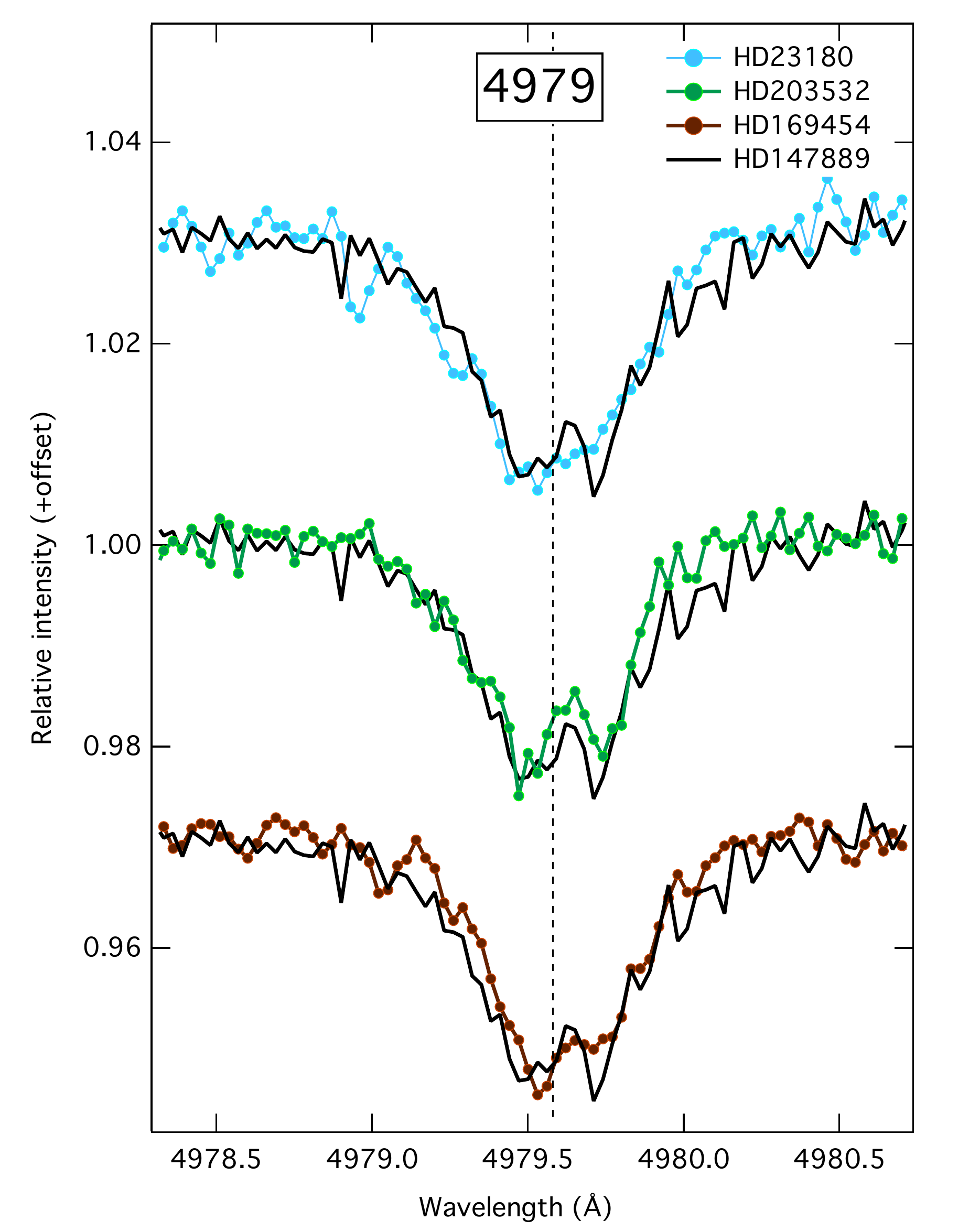}
\includegraphics[width=.3\textwidth]{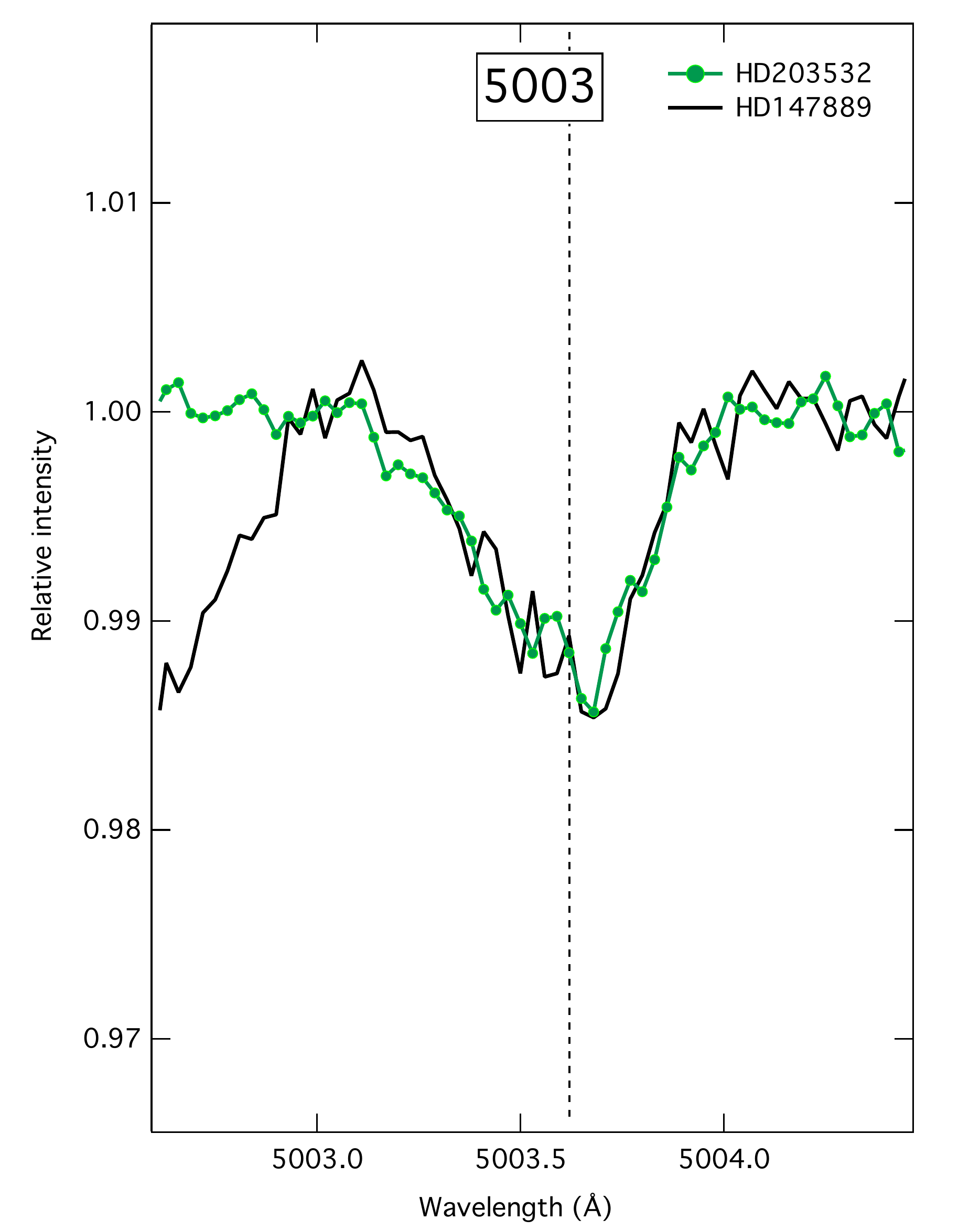}
\includegraphics[width=.3\textwidth]{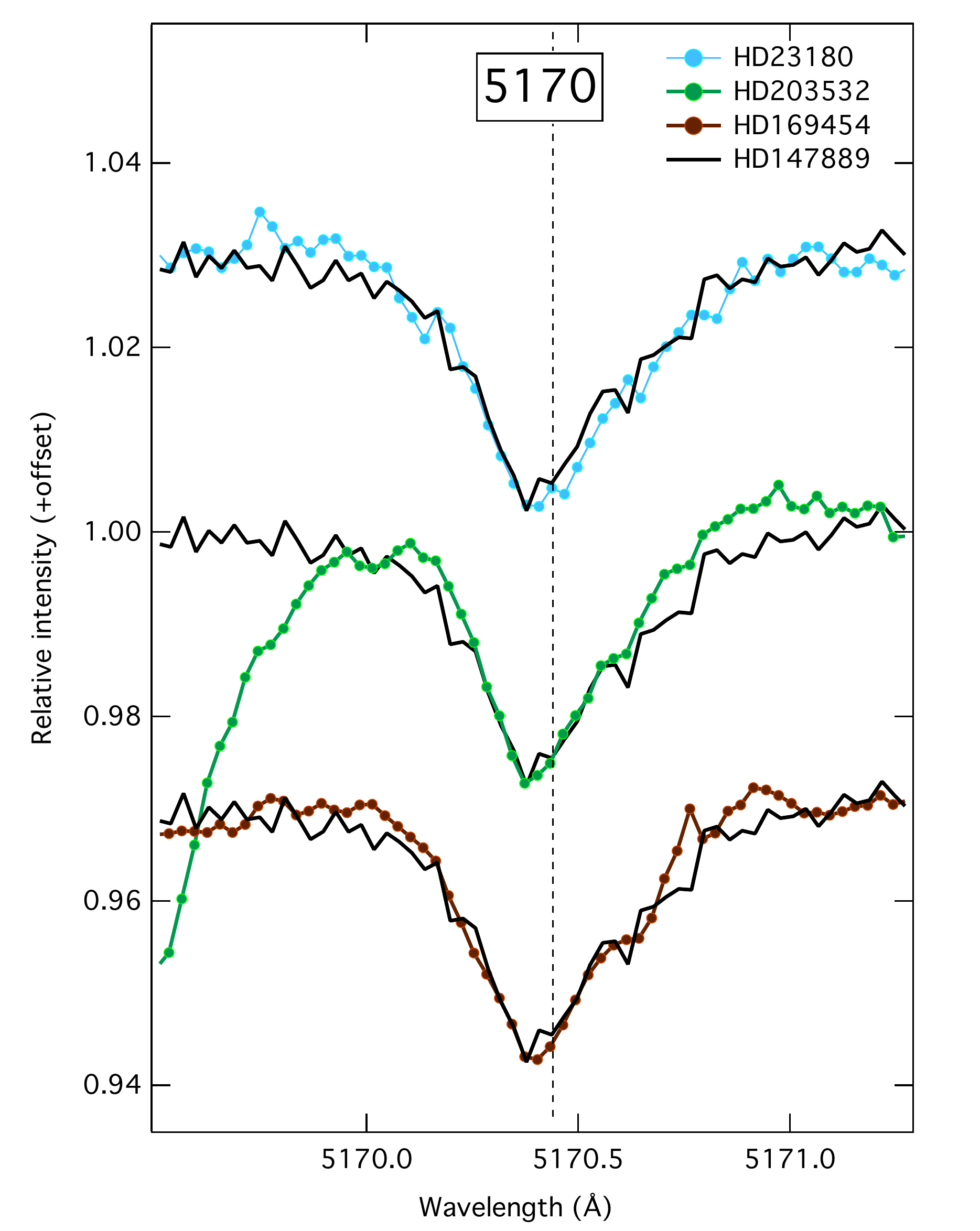}
\includegraphics[width=.3\textwidth]{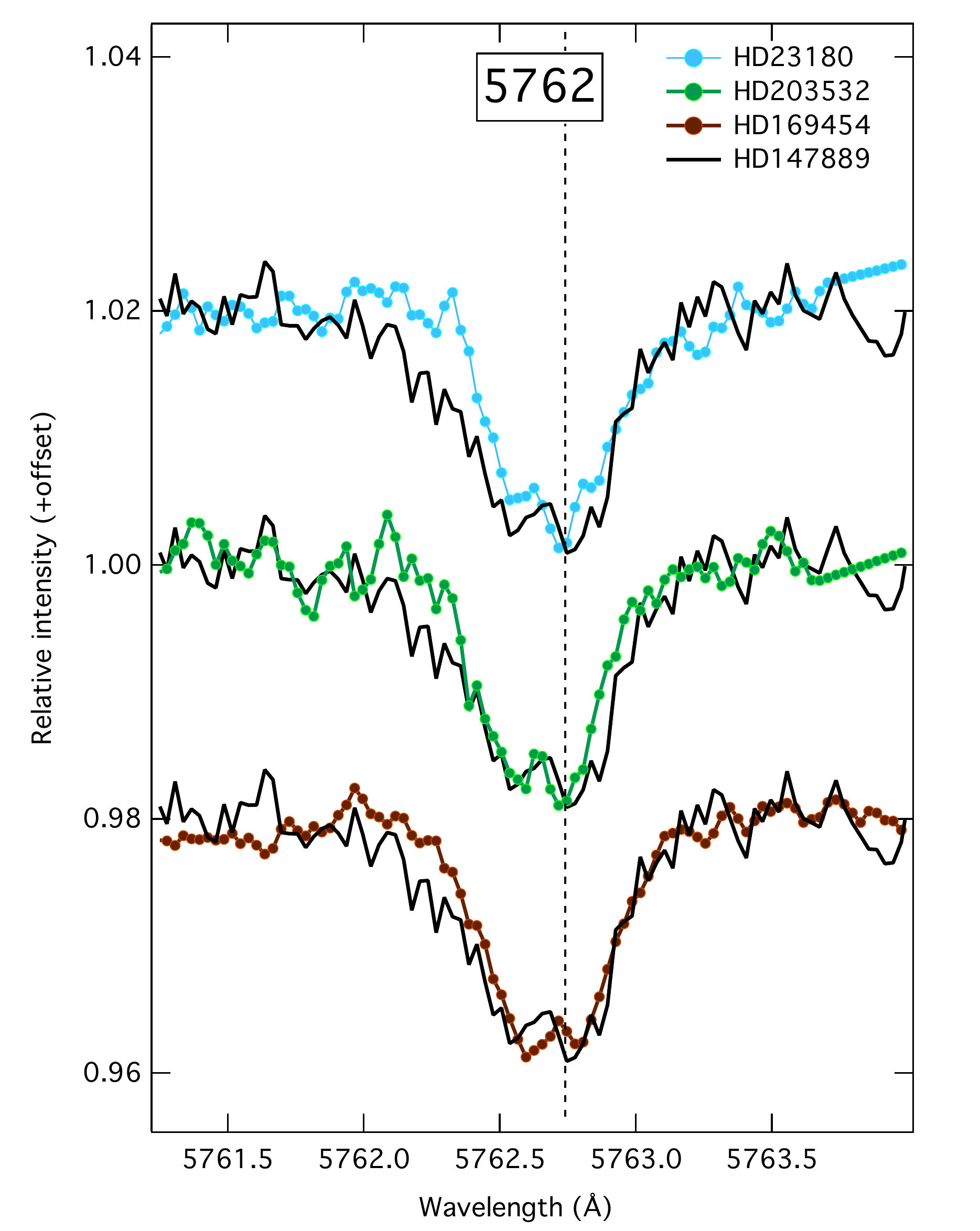}
\includegraphics[width=.3\textwidth]{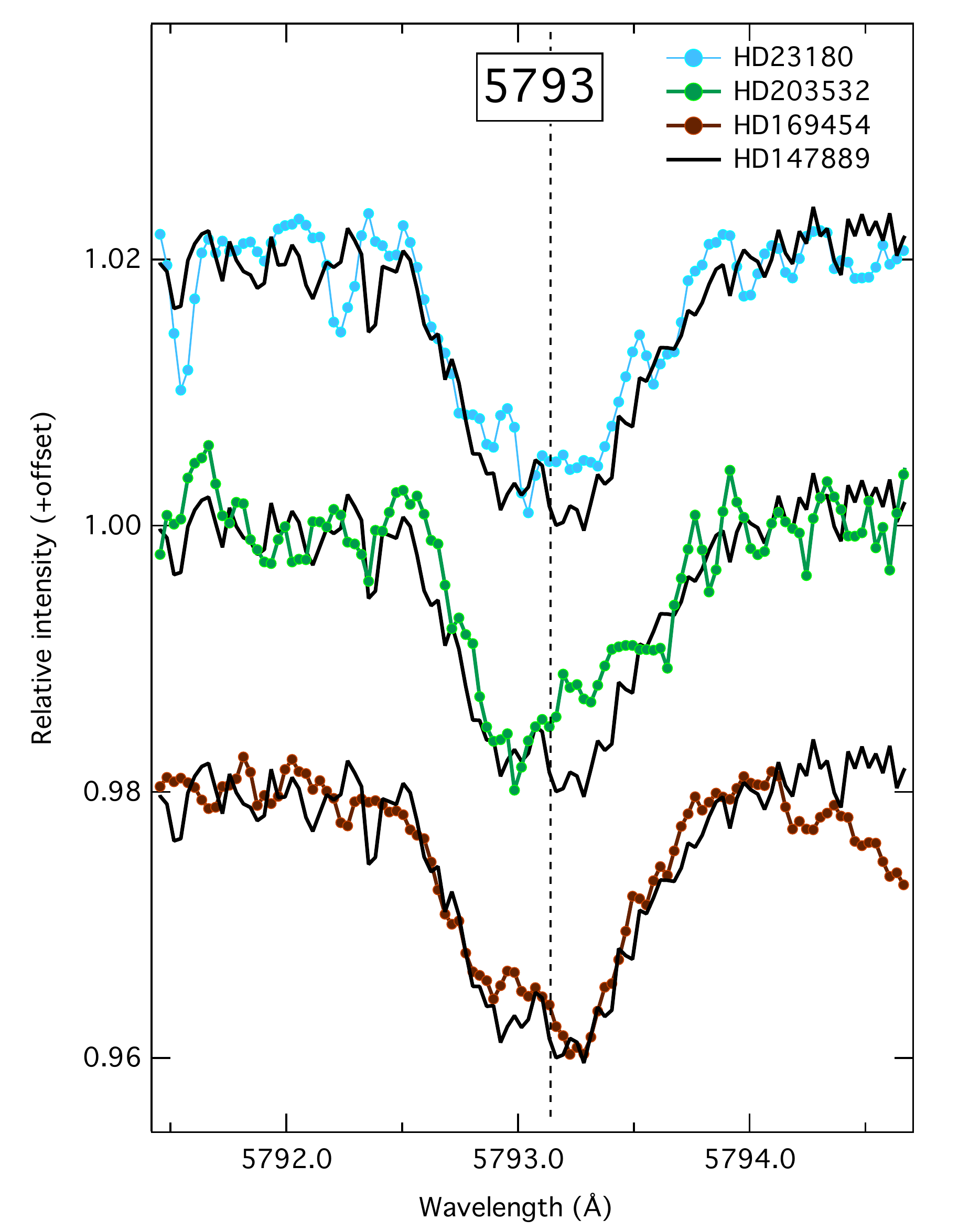}

\caption{Same as Fig.~\ref{category1} but for 4969, 4979, 5003, 5170, 5762, and 5793~\AA.}
\label{category2}
\end{figure*}

\begin{figure*}
\centering
\includegraphics[width=.3\textwidth]{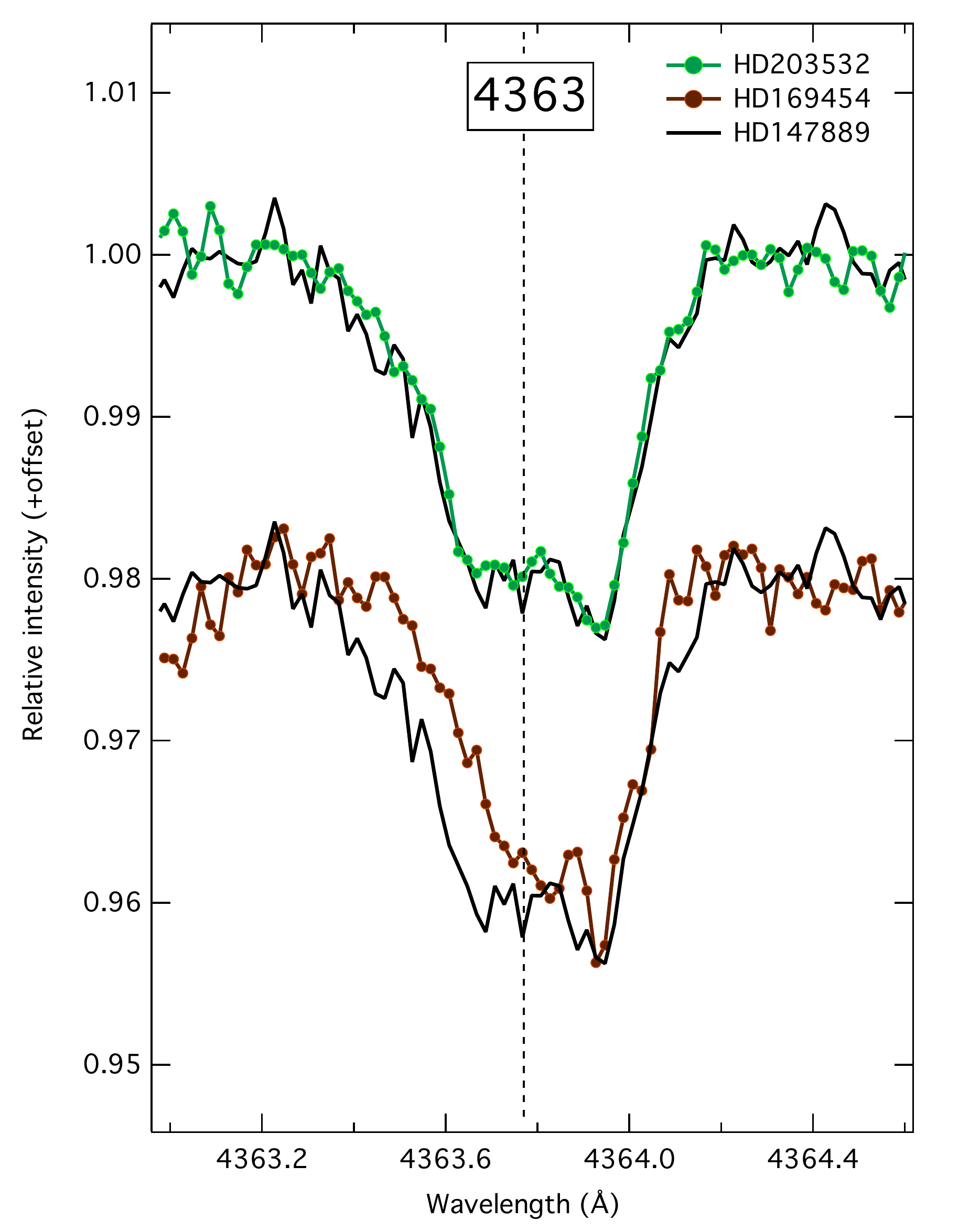}
\includegraphics[width=.3\textwidth]{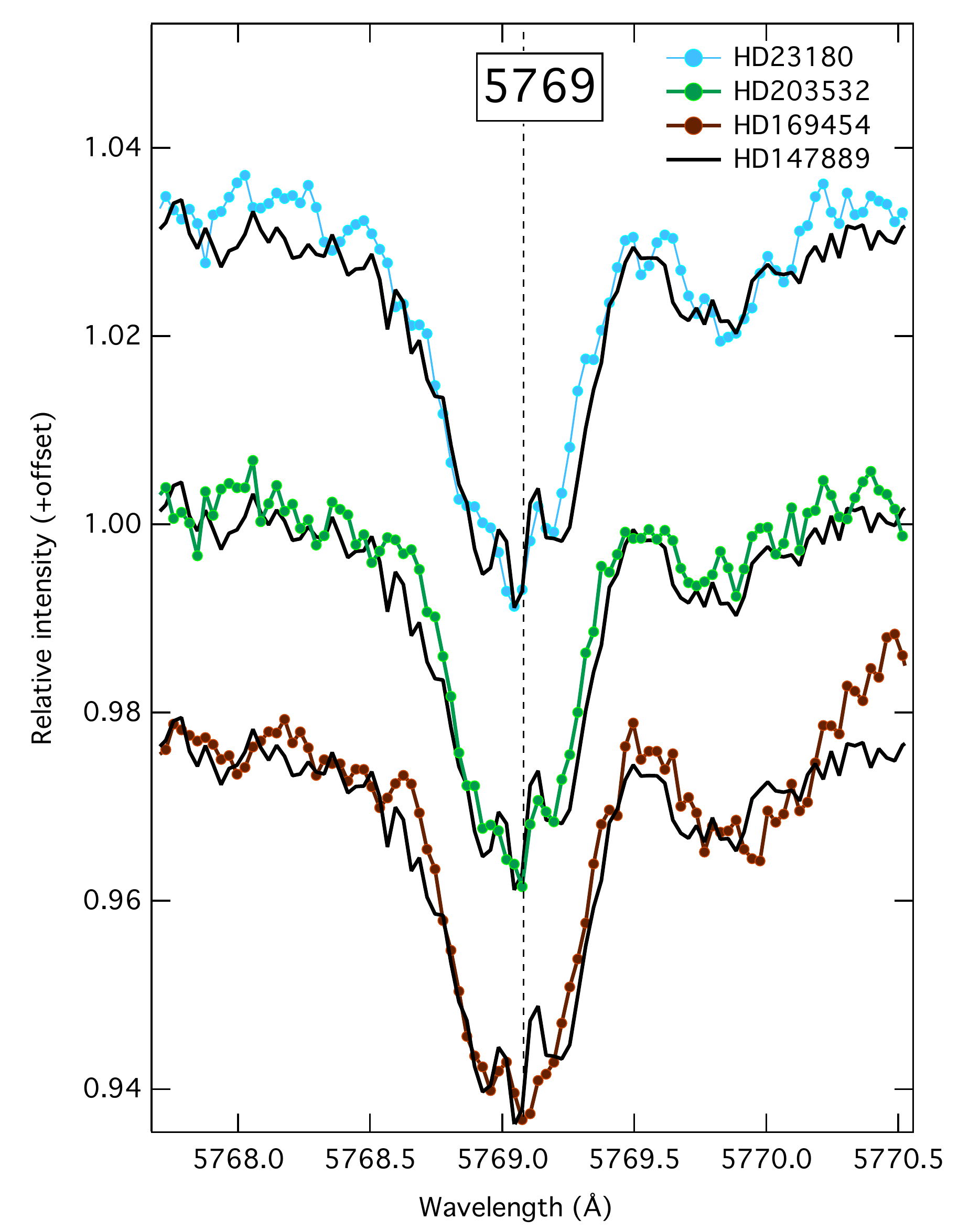}
\includegraphics[width=.3\textwidth]{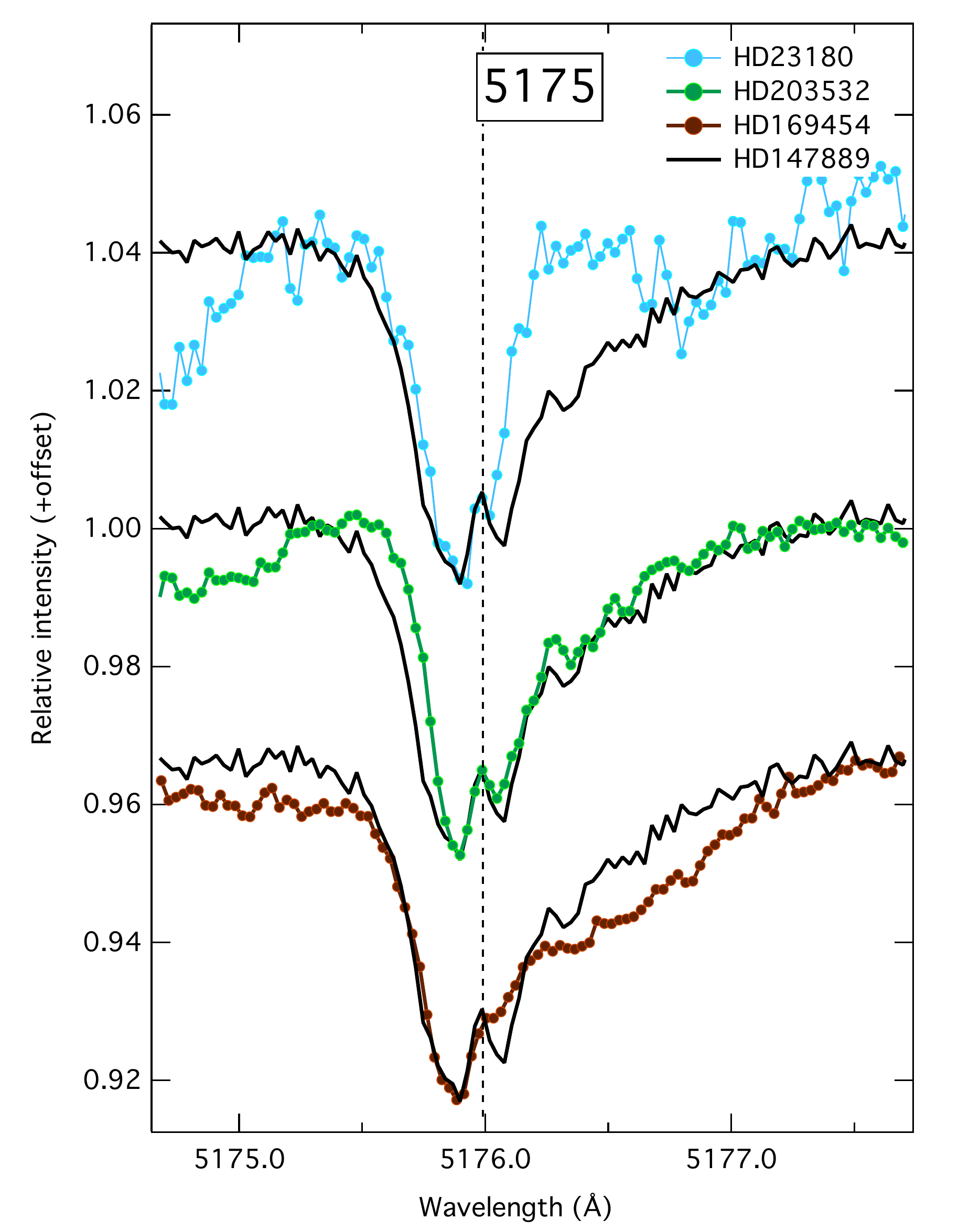}
\includegraphics[width=.3\textwidth]{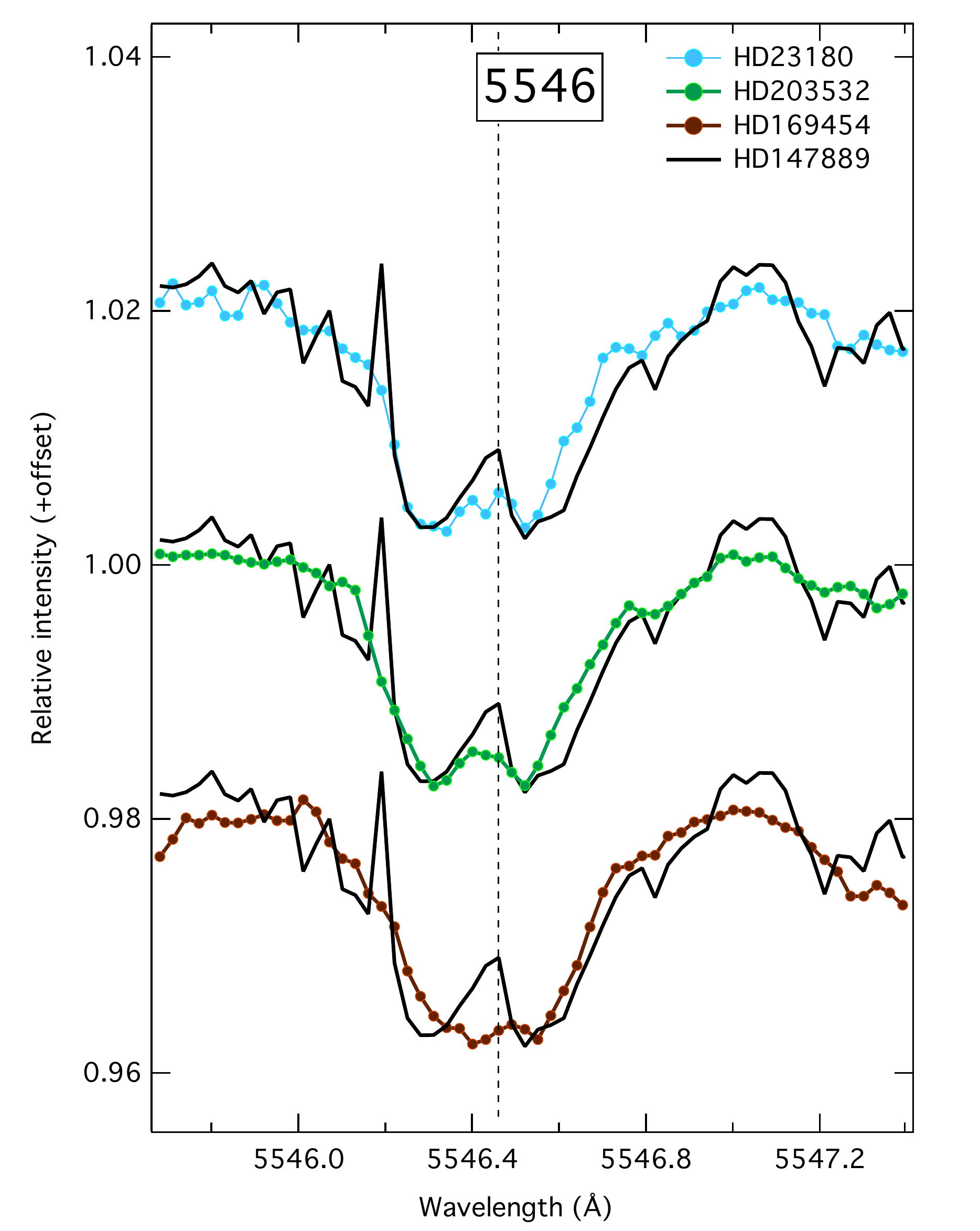}
\includegraphics[width=.3\textwidth]{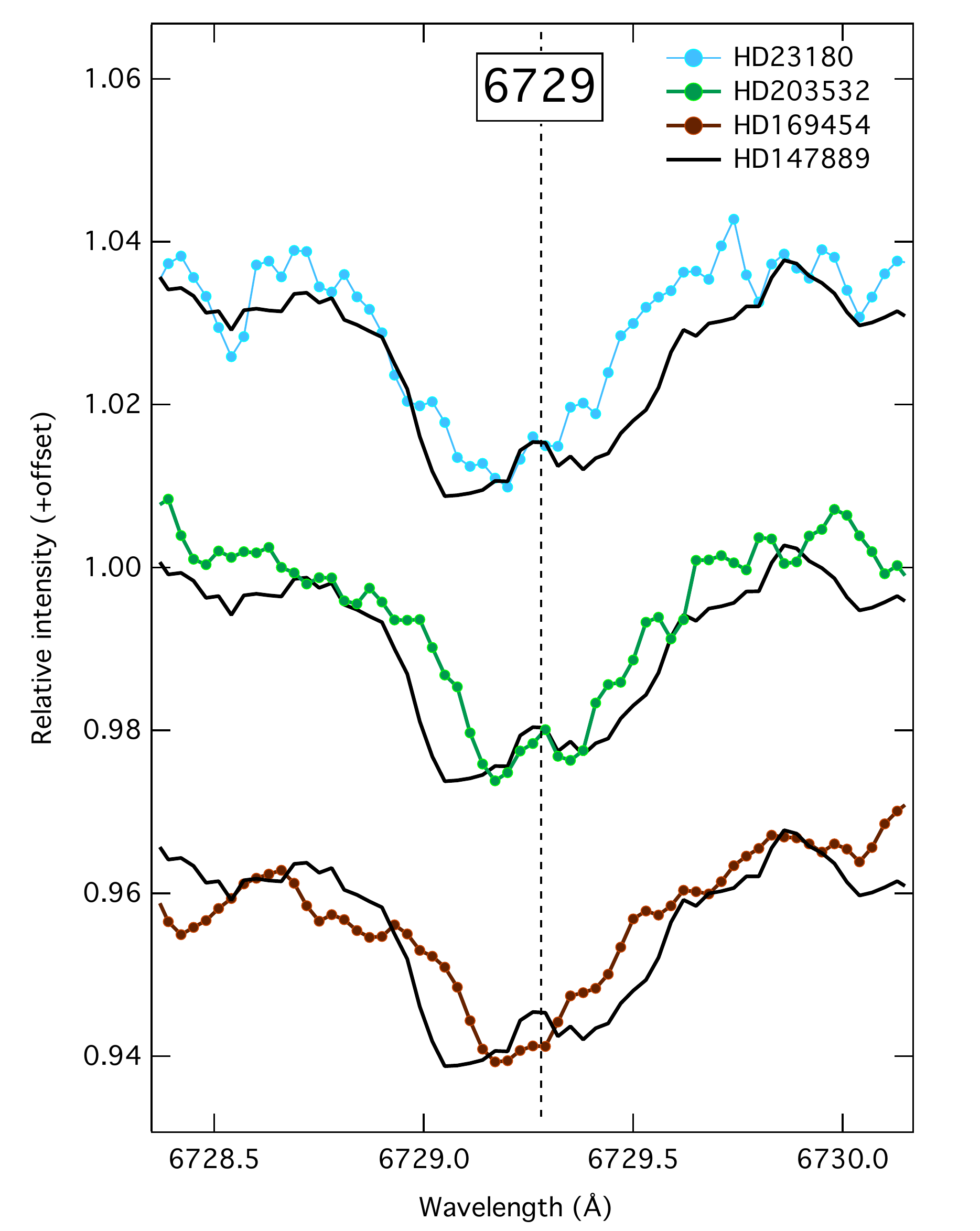}
\caption{Same as Fig.~\ref{category1} but for 4363, 5769, 5175, 5546 and 6729~\AA.}
\label{category3}
\end{figure*}

Figs.~\ref{category1}, \ref{category2} and~\ref{category3} show the 18 C$_{2}$-DIB profiles for the three EDIBLES targets after normalization to reach the same maximum depth. We also show the normalized profiles for the multi-cloud target HD\,169454. It is interesting to disentangle the effects of Doppler broadening due to cloud multiplicity and broadening due to increase of the excitation temperature. Coincidentally, $T_{02}$ = 23~K for HD\,169454, is very similar to the value measured for HD\,23180. Moreover, profiles of the 6196~\AA\: DIB have been investigated for this star as well as for HD\,147889 and HD\,23180 by \cite{Kazmierczak2009}. In order to better visualize differences, we show the profiles that correspond to HD\,147889 superimposed on the three others because for this sightline the sub-structures are the most prominent. After careful inspection of all comparison figures, we identified three categories of DIBs.

The first category corresponds to the existence of sub-structures for the four targets, and no detected variation in the total width of the band nor in the locations of the sub-peaks. Falling in this category are DIBs 4727, 4734, 4963, 4984, 5418, 5512 and 5541~\AA\: (Fig.~\ref{category1}) confirming an earlier result by \citet{Gala2008} for the 4963, 5418 and 5512~\AA\: DIBs. Therefore, for those bands the stacking is justified as a means to enhance the profile structures. It is worth noting that for 4984, 5512 and 5541~\AA\: DIBs the sub-structures are smoother in the case of lower temperatures.

The second category corresponds to the weakest bands for which individual profiles do not clearly reveal sub-structures well above the noise and no apparent variations of the total width. In those cases, the co-addition is justified since it simply improves the S/N. This is the case for the 4969, 4979, 5003, 5170  DIBs (Fig.~\ref{category2}). In the case of the 5762 and 5793~\AA\: DIBs the co-addition does not bring any improvement. We believe this may be due to the effect of telluric residuals (see the remark on the HITRAN database in Sect.~\ref{data}) that become very important for targets with the weakest absorbing column. Note that these are the two weakest of all \element[][][][2]{C}-bands. Further measurements and improved telluric corrections are needed to confirm their sub-peaks. 

The third DIB category corresponds to the detection of profile variations among the four targets, and this may potentially reveal a response to the excitation temperature. Indeed, the full DIB width appears smaller for HD\,23180 ($T_{02}=20$~K) and, to a lesser extent, for HD\,203532 (unmeasured $T_{02}$) compared to HD\,147889 ($T_{02}=49$~K). Interestingly, these DIBs are also narrower for the multi-cloud star HD\,169454 ($T_{02}=23$~K) than for HD\,147889, an effect opposite to the expected broadening due to cloud multiplicity. Moreover, the amplitudes of the sub-peaks appear smaller for HD\,23180, HD\,203532 and HD\,169454 compared to HD\,147889. Falling in this category are the 4363, 5541, and 5769~\AA\: DIBs. Identical, but less clearly measured effects can also be seen for DIBs 5175, 5546, and 6729~\AA\:. If these simultaneous variations in width and peak amplitudes are actually linked to the variation of the rotational excitation, then for these DIBs the average profile does not correspond to any intrinsic profile. 
However, we have kept these average profiles in Fig.~\ref{all}.
Interestingly, the DIBs from this third category are all very tightly correlated with the \element[][][][2]{C} column, as discussed in the next section.

\section{Correlational studies}\label{sec:correl}
\subsection{Correlations with the \element[][][][2]{C} column}
\begin{figure*}
\centering
\includegraphics[width=.33\textwidth]{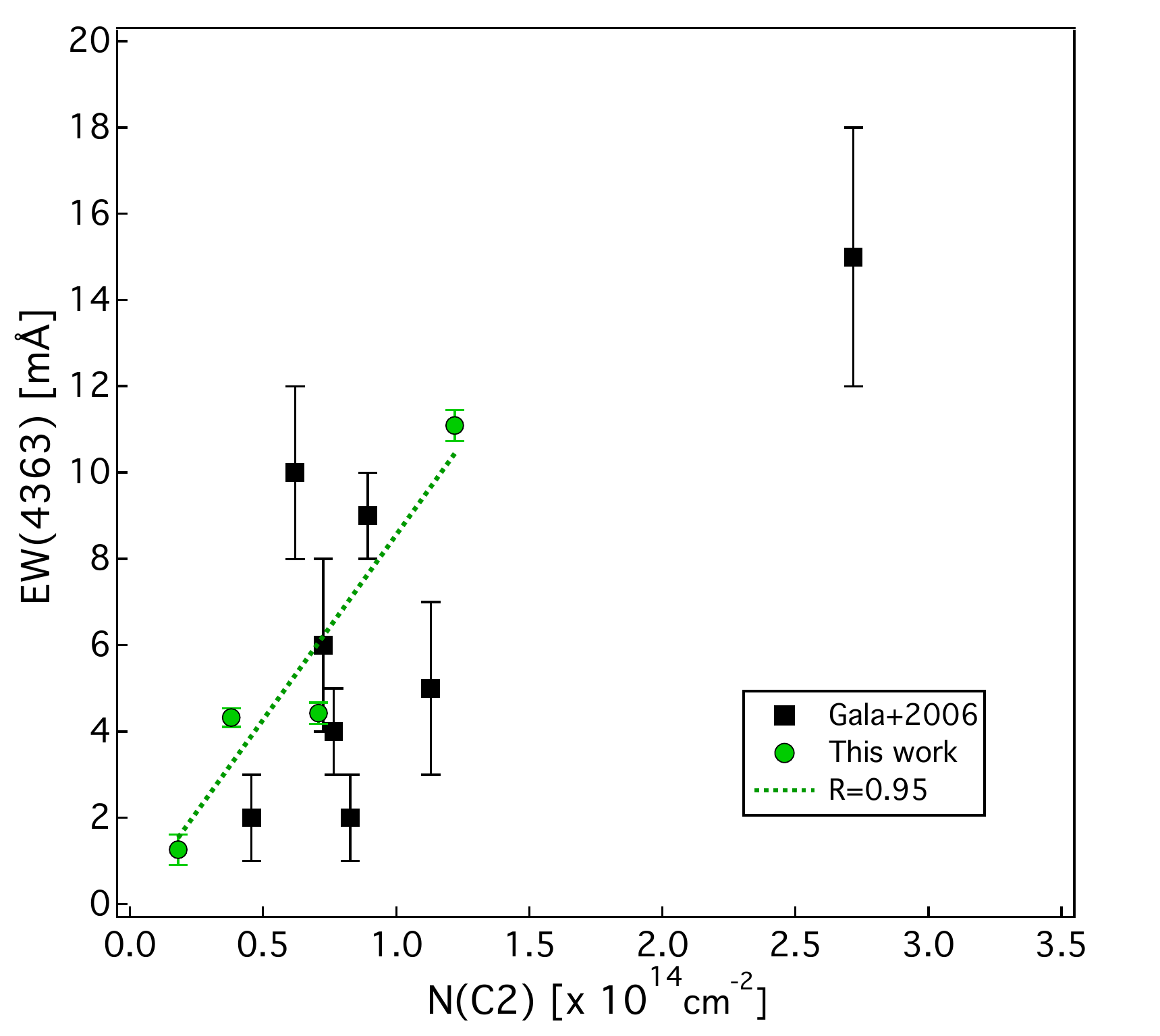}
\includegraphics[width=.33\textwidth]{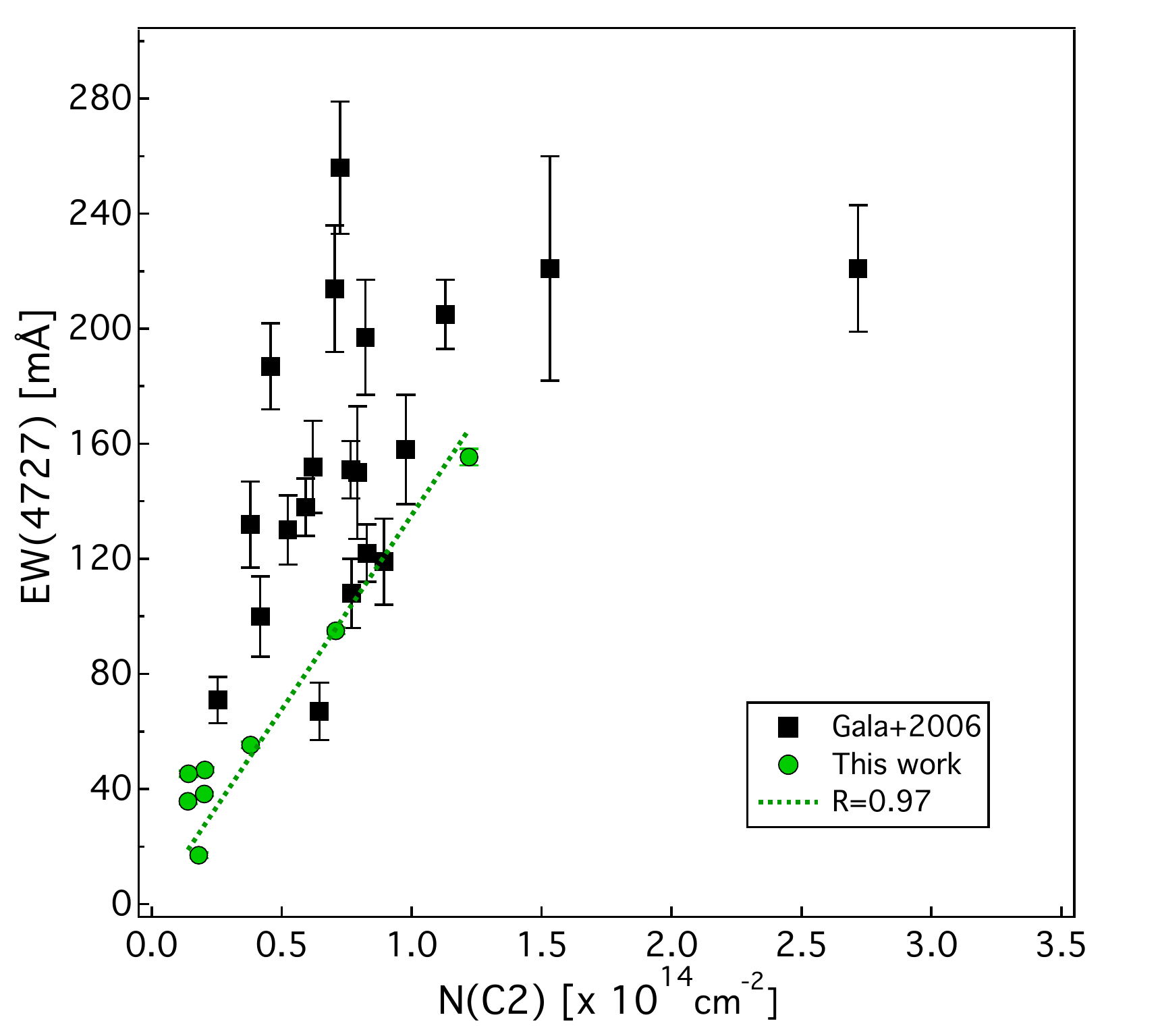}
\includegraphics[width=.33\textwidth]{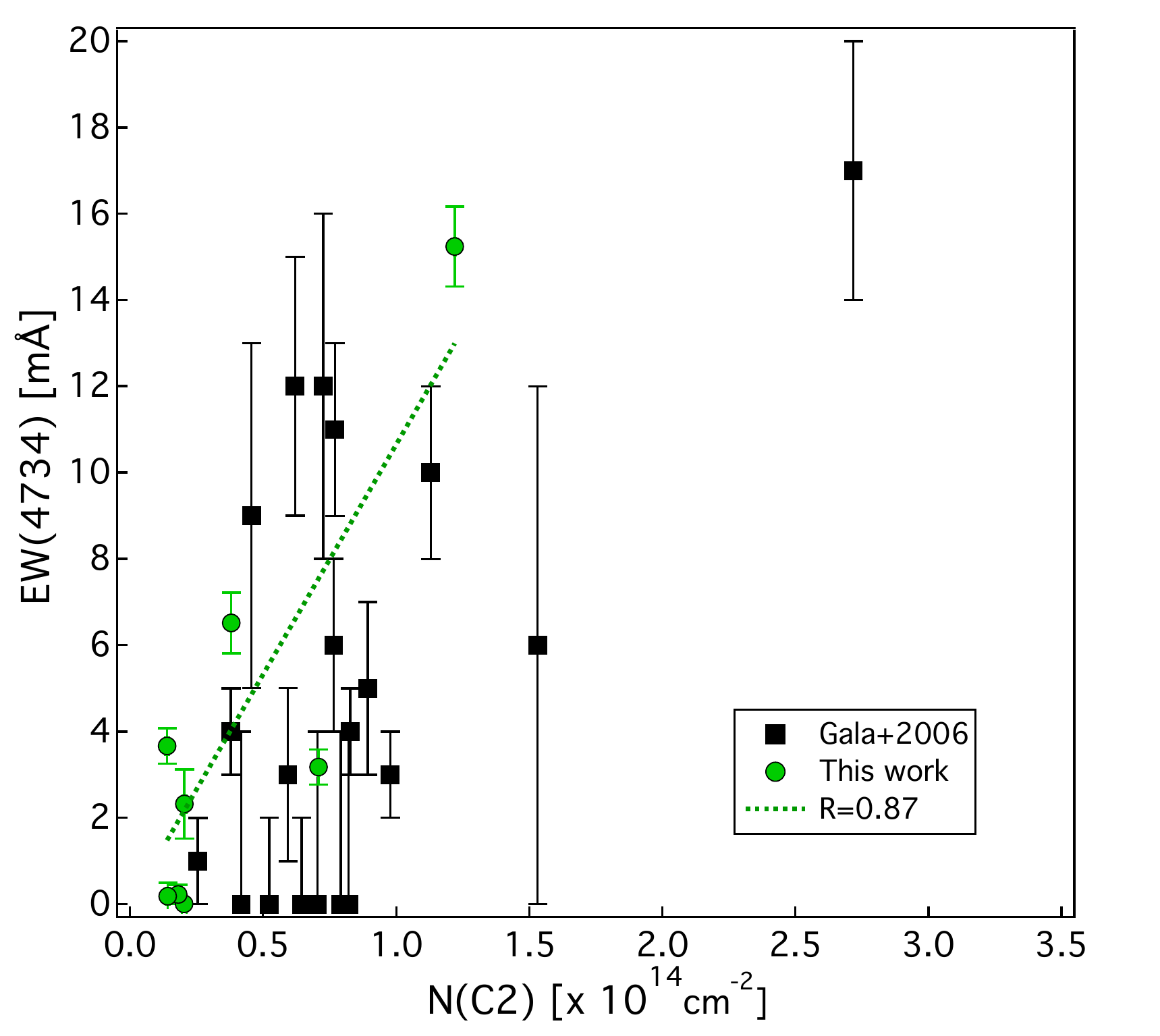}
\includegraphics[width=.33\textwidth]{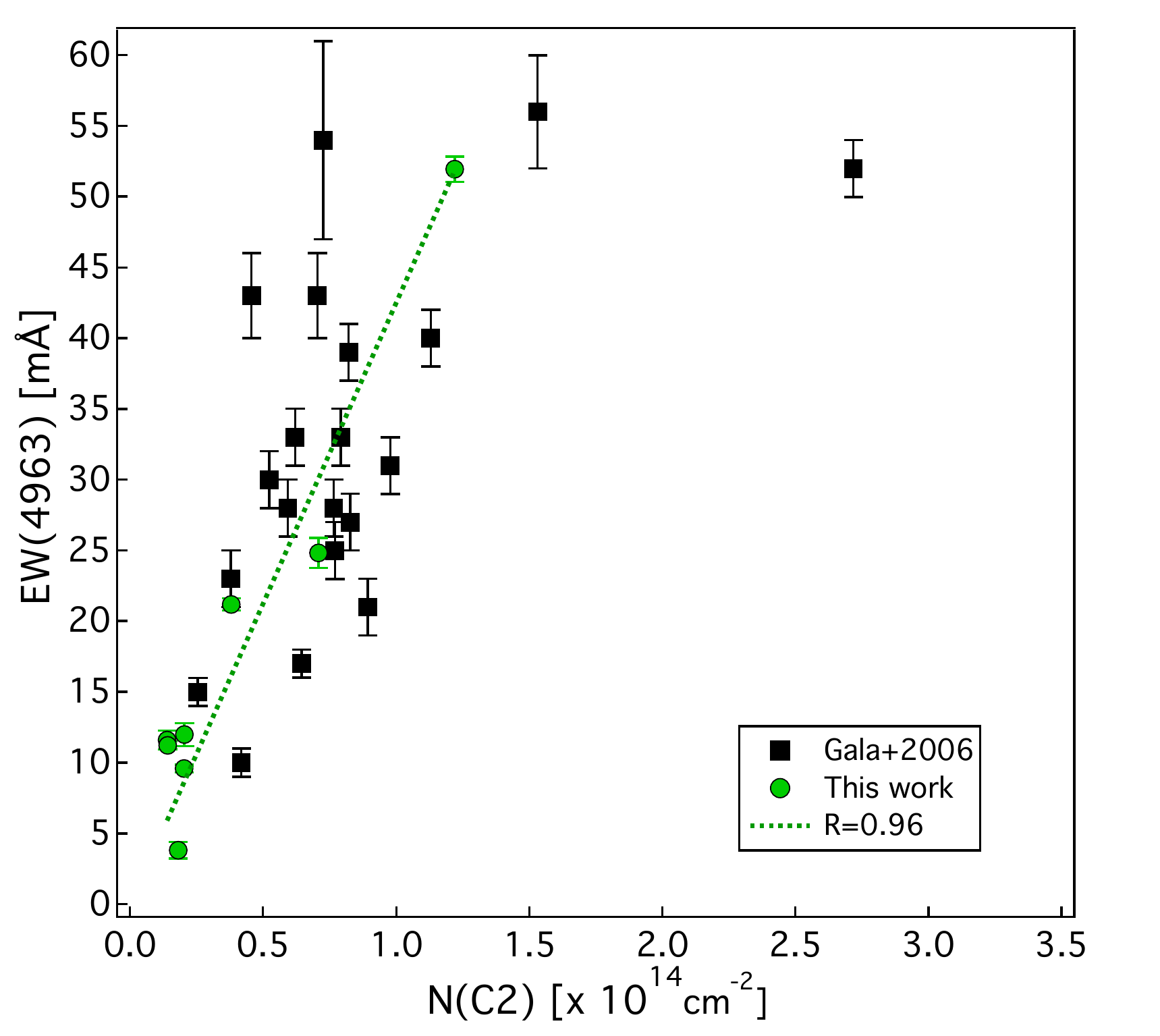}
\includegraphics[width=.33\textwidth]{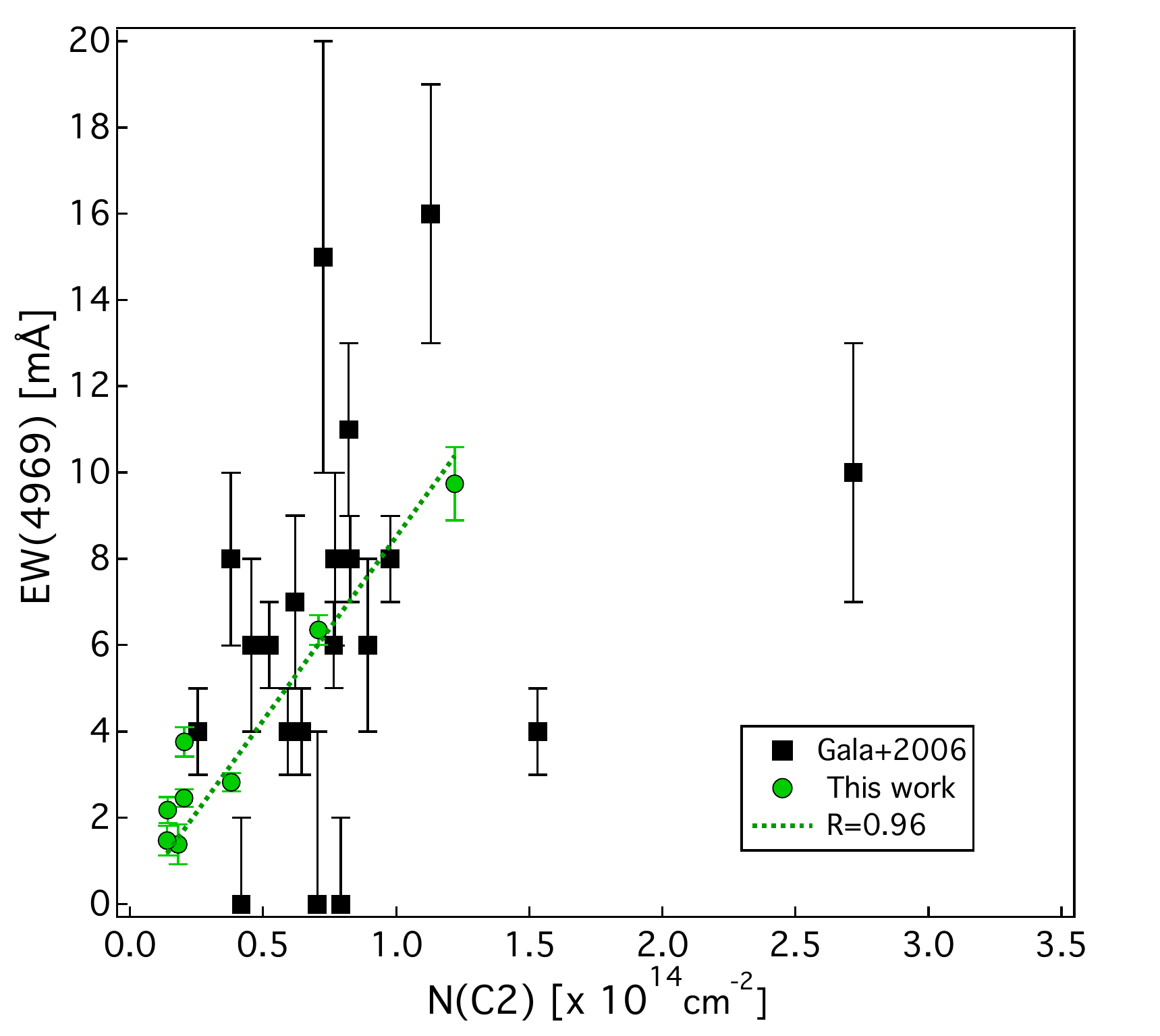}
\includegraphics[width=.33\textwidth]{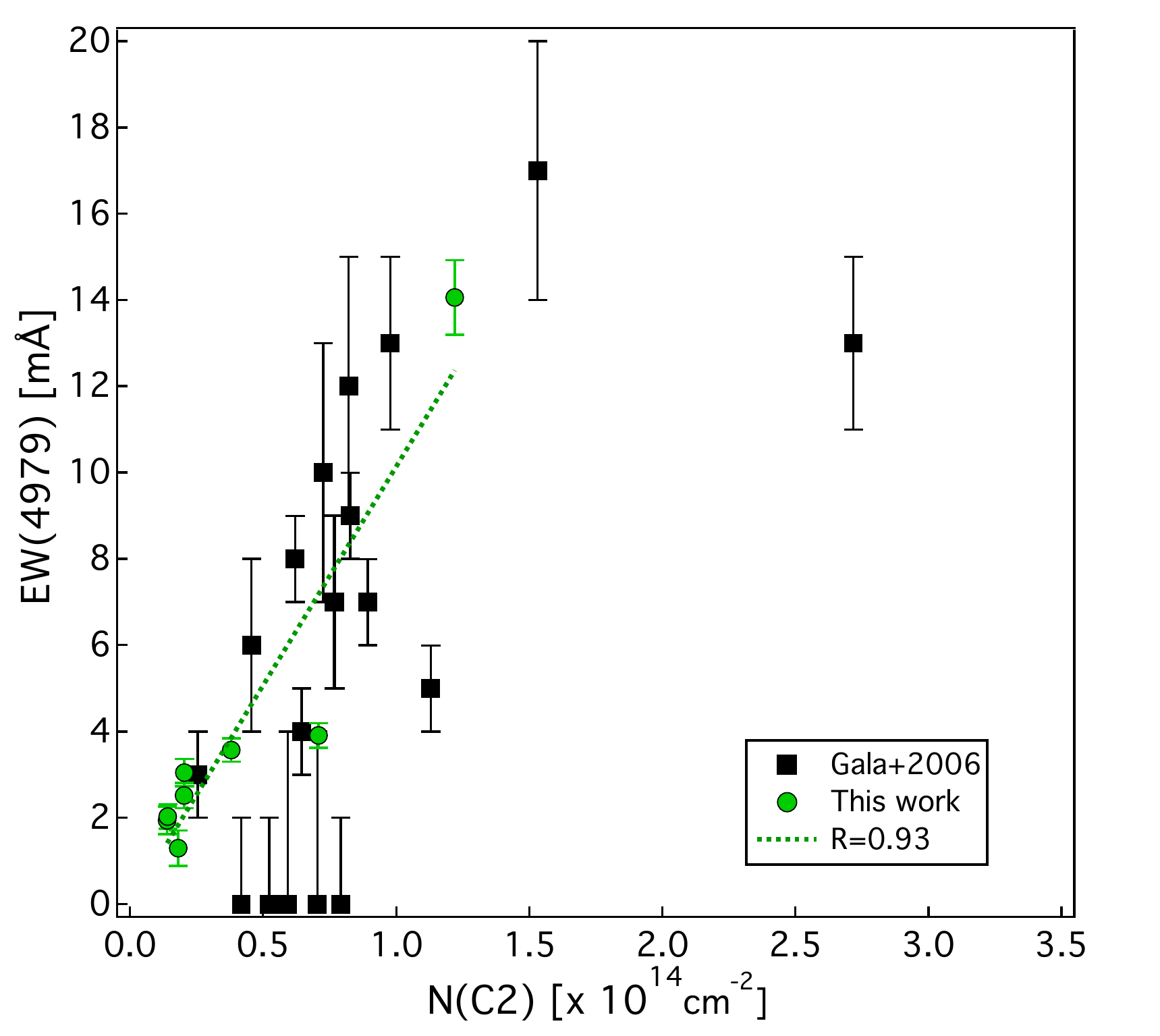}
\includegraphics[width=.33\textwidth]{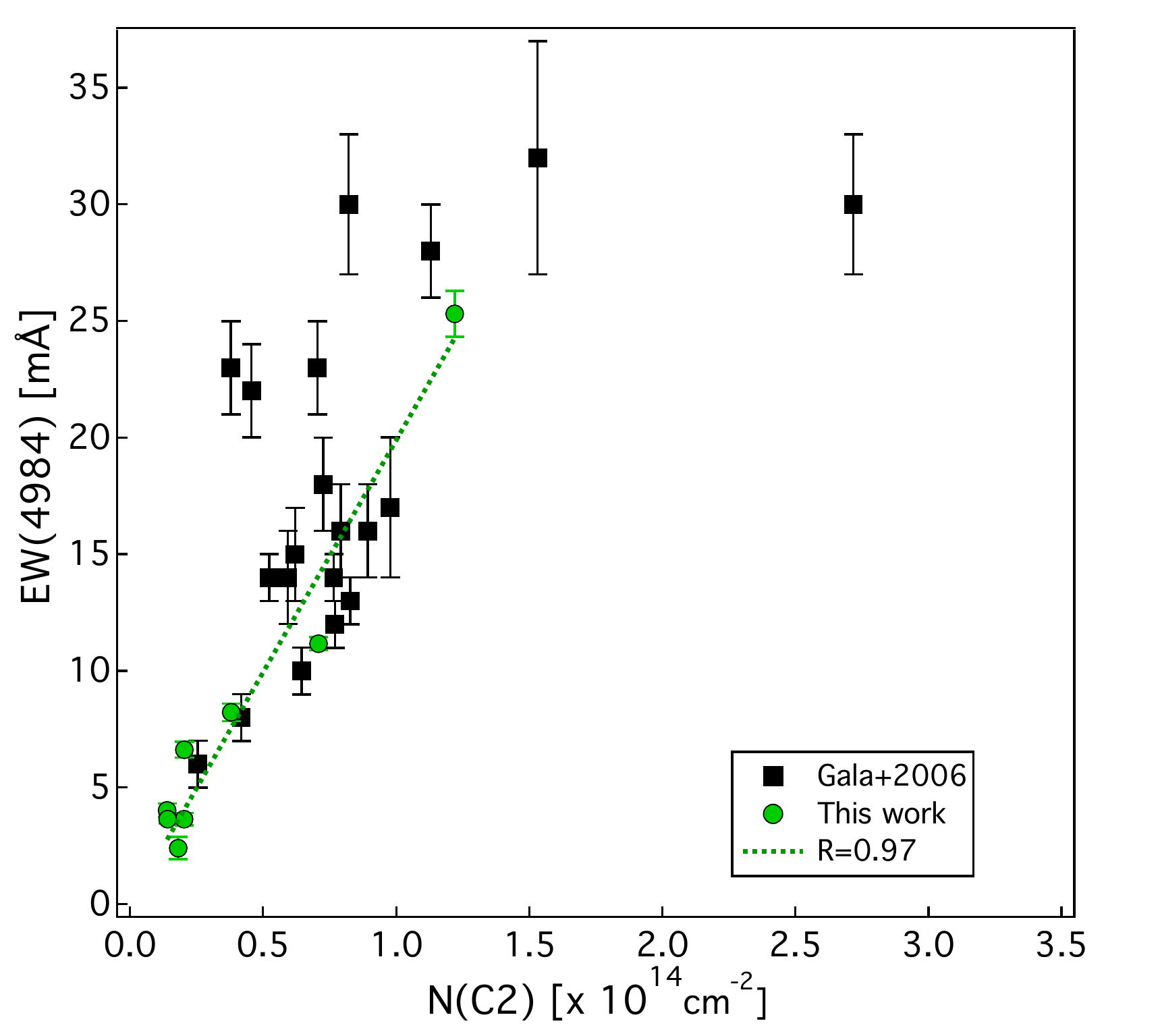}
\includegraphics[width=.33\textwidth]{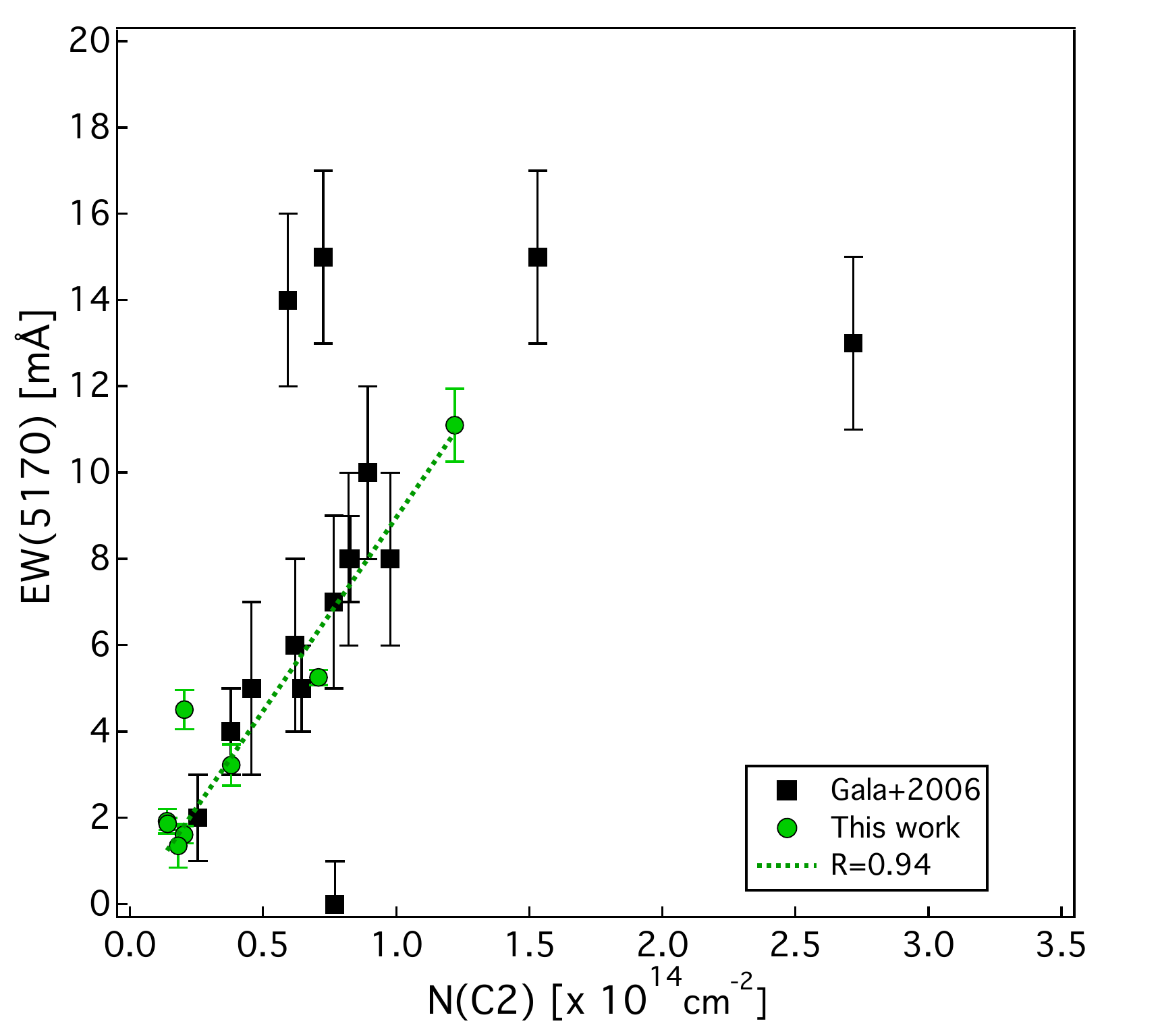}
\includegraphics[width=.33\textwidth]{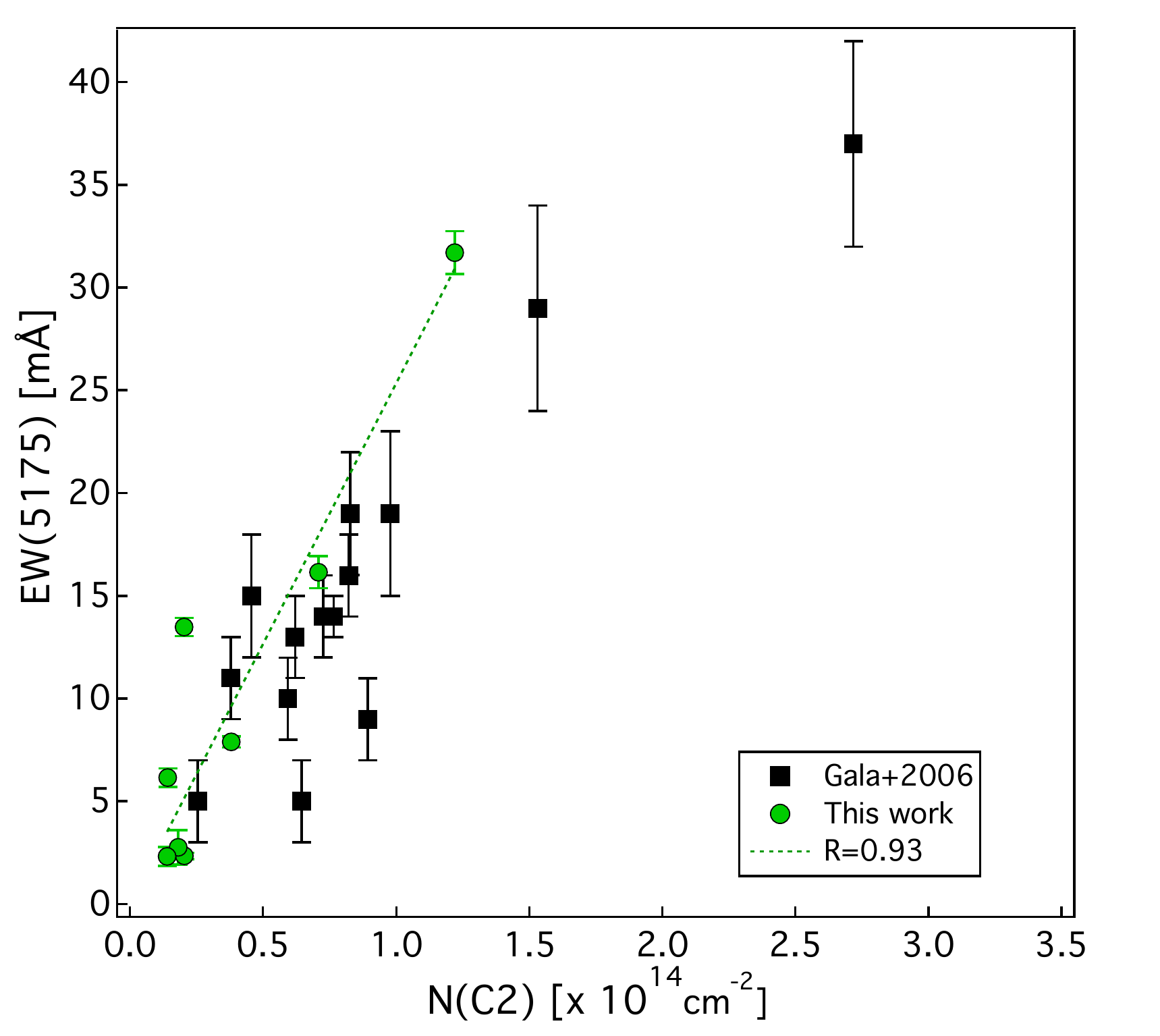}
\includegraphics[width=.33\textwidth]{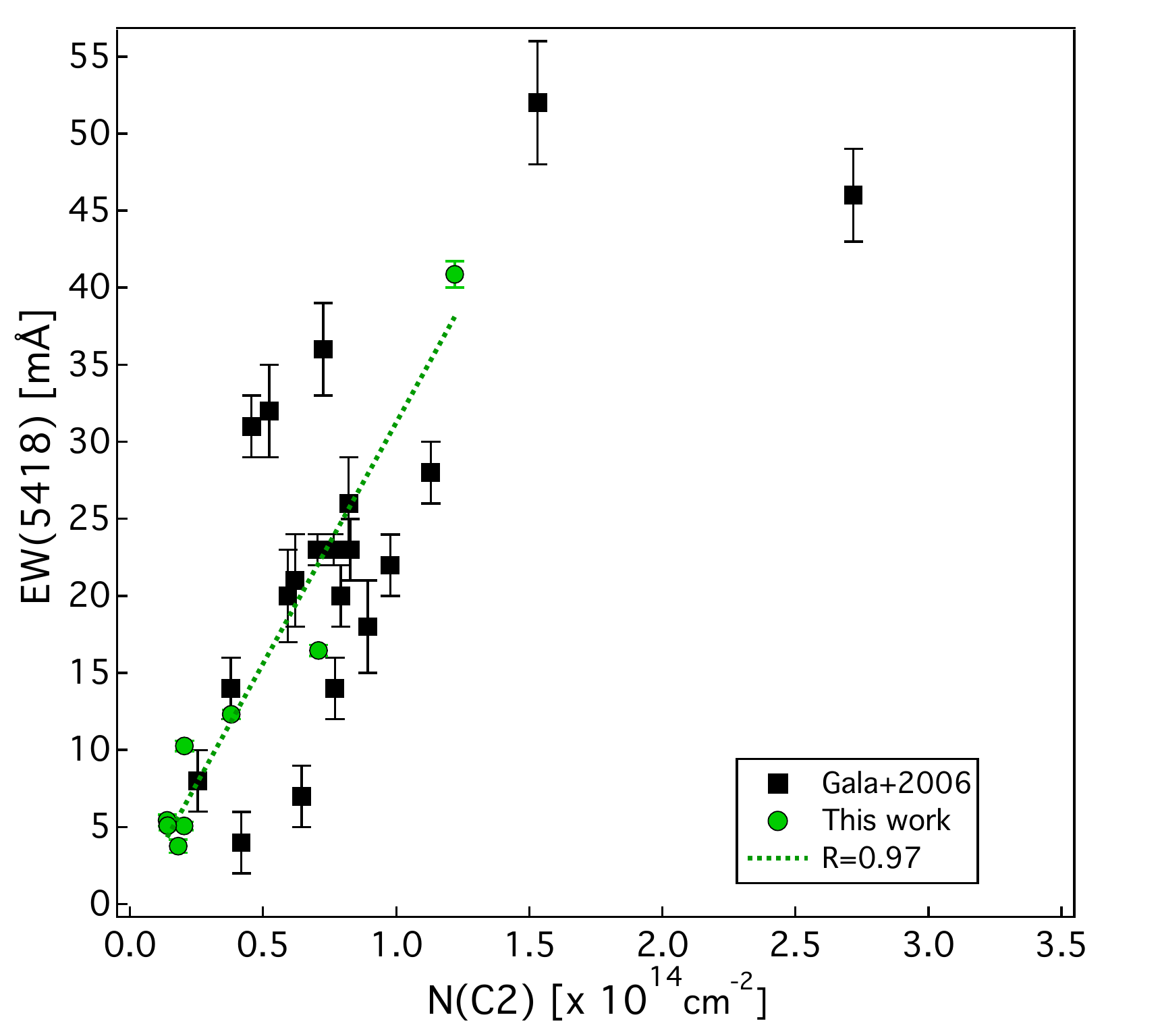}
\includegraphics[width=.33\textwidth]{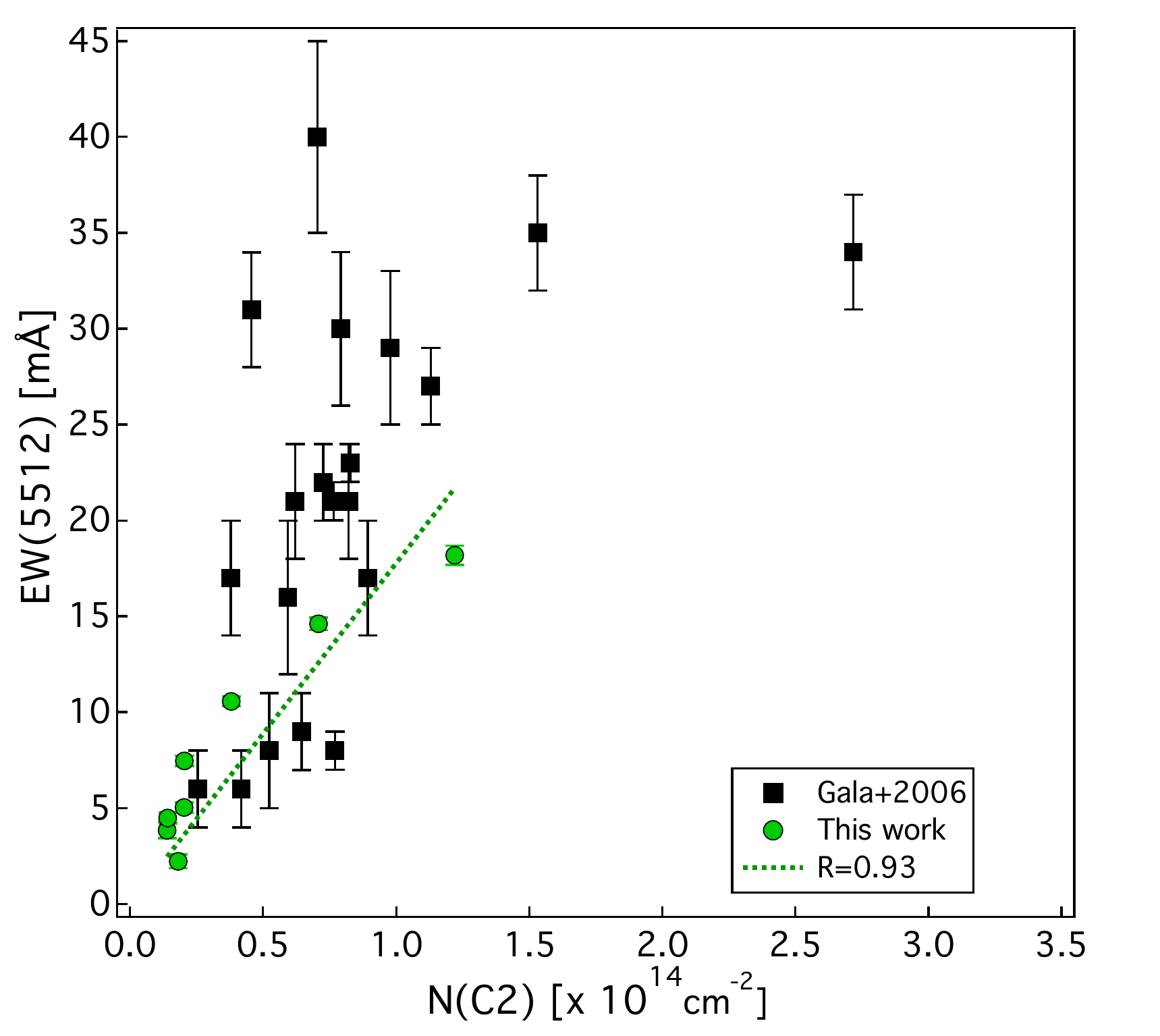}
\includegraphics[width=.33\textwidth]{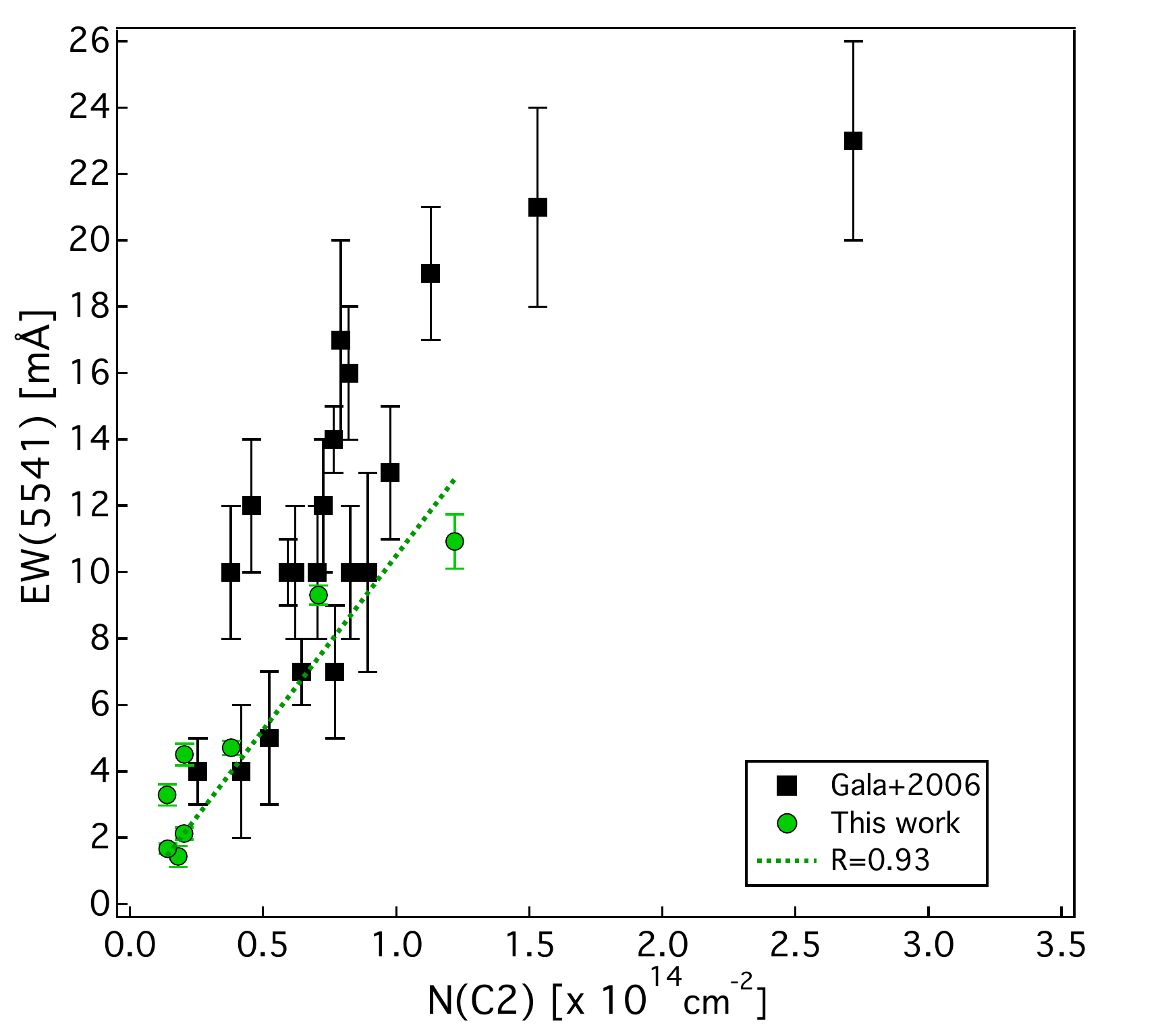}
\caption{Correlations between the DIB equivalent widths and the C$_{2}$ column (green filled circles). Superimposed as black squares are the results of \cite{Gala06}. R is the Pearson coefficient for our data only. The multi-cloud target HD\,169454 value corresponds to N(C$_{2}$)= 0.71 $\times$ 10$^{14}$~cm$^{-2}$\label{fig:correlations} 
}
\end{figure*}

Our combined measurements of \element[][][][2]{C} columns and weak DIB equivalent widths towards single-cloud targets constitute an ideal dataset for the study of any link between the \element[][][][2]{C} molecule and each DIB carrier. Figs.~\ref{fig:correlations} and~\ref{fig:correlations2} display the equivalent widths of the recorded \element[][][][2]{C}-DIBs as a function of the \element[][][][2]{C} column density $N(C_2)$. We have superimposed on each figure the corresponding previous measurements from \cite{Gala06}. We did not include the 5003~\AA\ band which was not studied by \cite{Gala06} and for which we have only three valid data points due to heavy stellar contamination. It can be seen that all our data points fall within the set of points from these authors, revealing good agreement with all these measurements. On the other hand, it can be  seen that the correlation with $N(C_2)$ is significantly tighter in the case of our targets. We interpret this as a consequence of our accurate determination of $N(C_2)$, the excellent signal-to-noise of EDIBLES spectra, our careful selection of mono-cloud sightlines and correction for telluric and stellar features. Note that we have also included the multi-cloud sightline HD\,169454 in these figures. The corresponding data point always falls within the \cite{Gala06} set of points. However, for many DIBs it is the most discrepant point if one considers our results only. This is a strong indication that the \element[][][][2]{C}-DIBs are very well correlated with the \element[][][][2]{C} molecule, and that a large fraction of the dispersion around the linear relationship is due to cloud multiplicity.
\renewcommand{\thefigure}{\arabic{figure} }
\addtocounter{figure}{-1}
\begin{figure*}
\centering
\includegraphics[width=.33\textwidth]{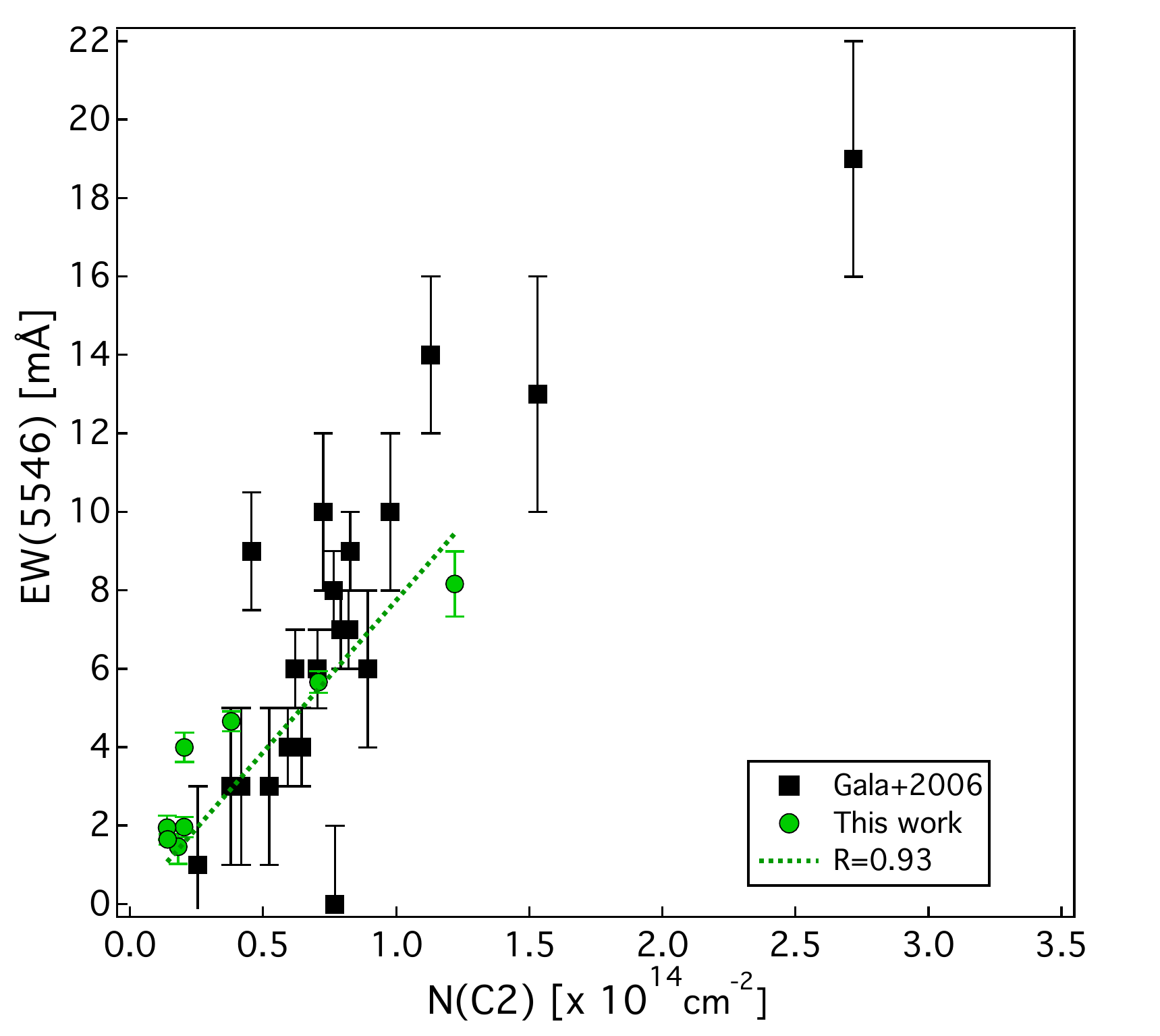}
\includegraphics[width=.33\textwidth]{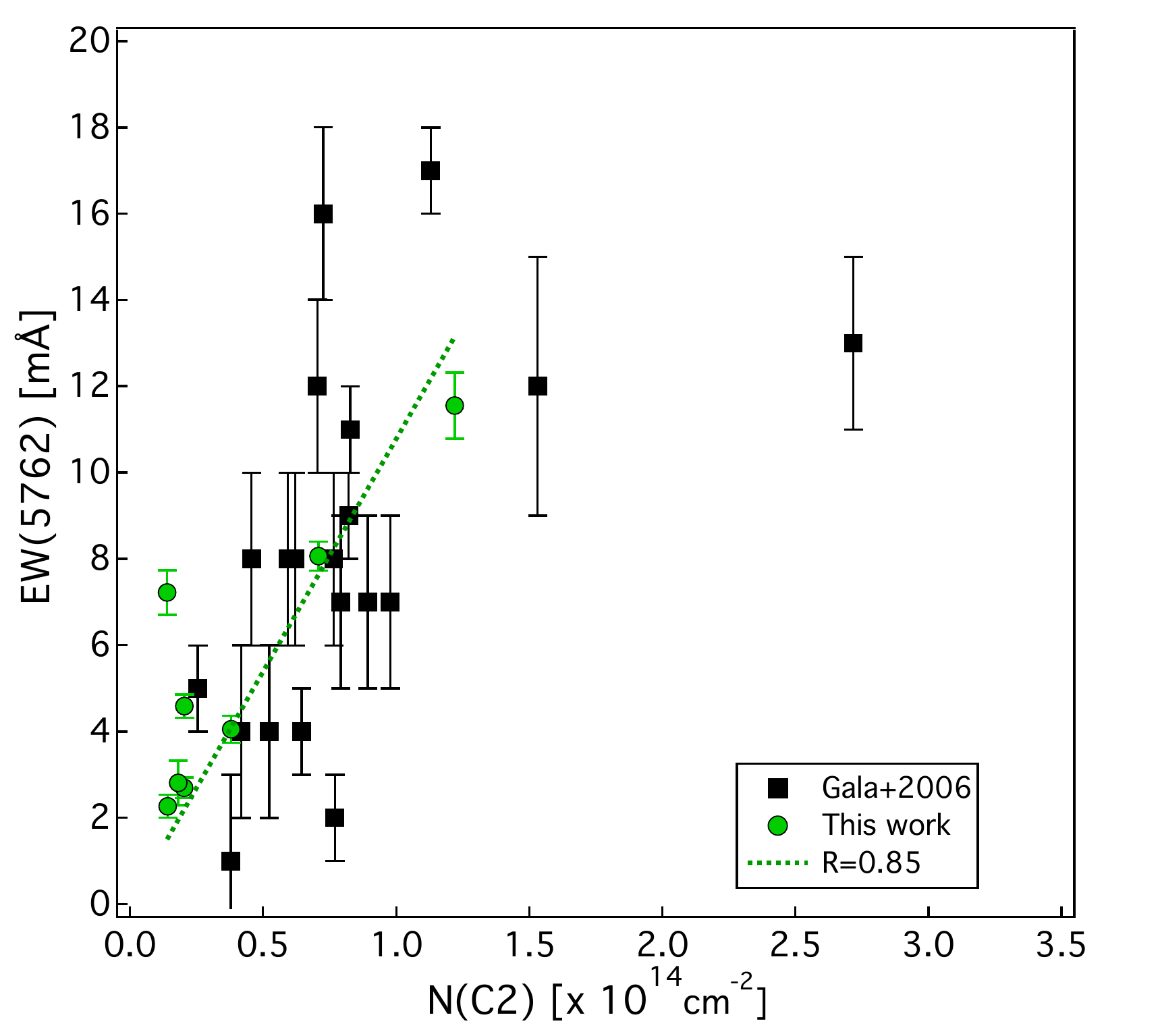}
\includegraphics[width=.33\textwidth]{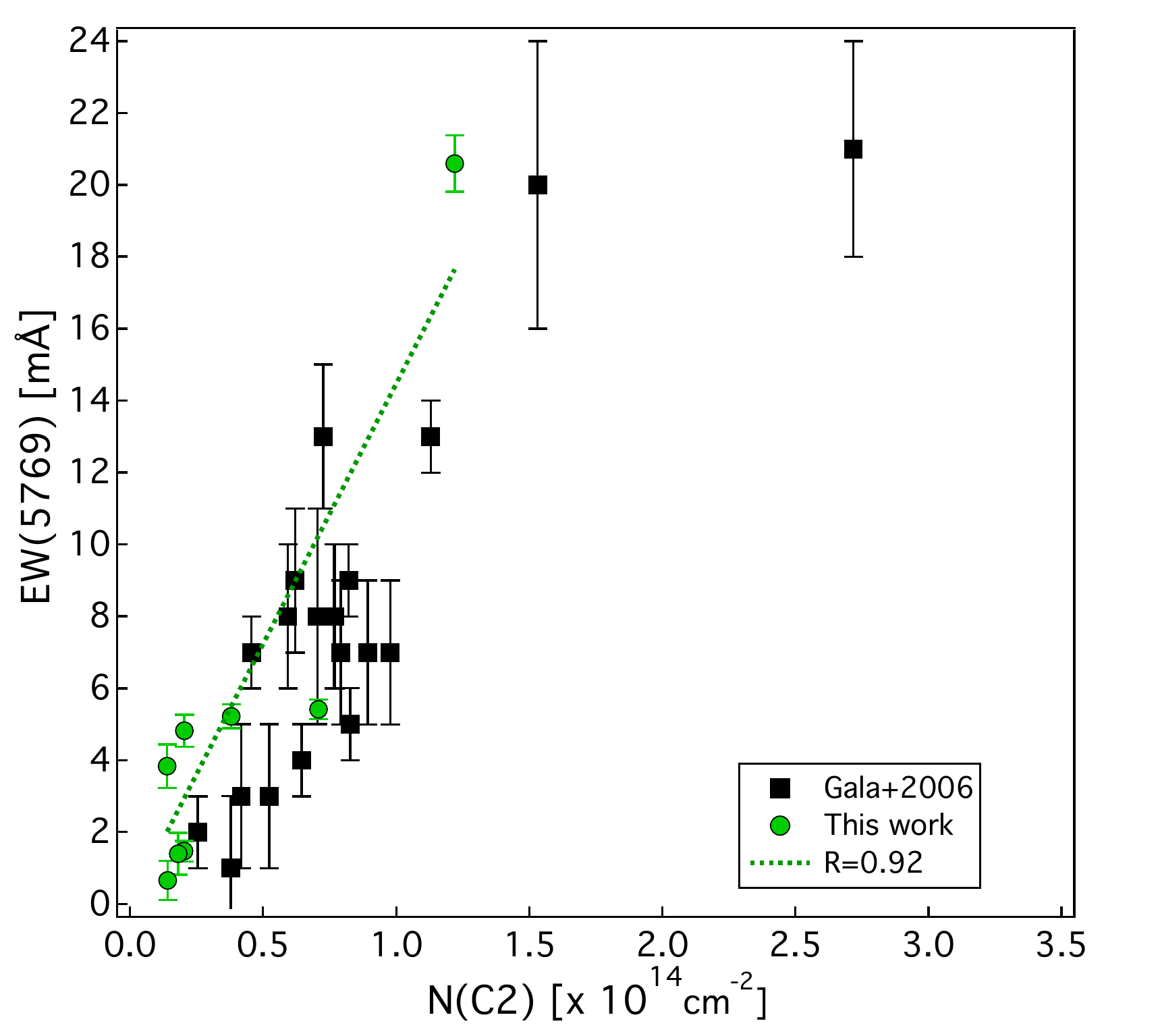}
\includegraphics[width=.33\textwidth]{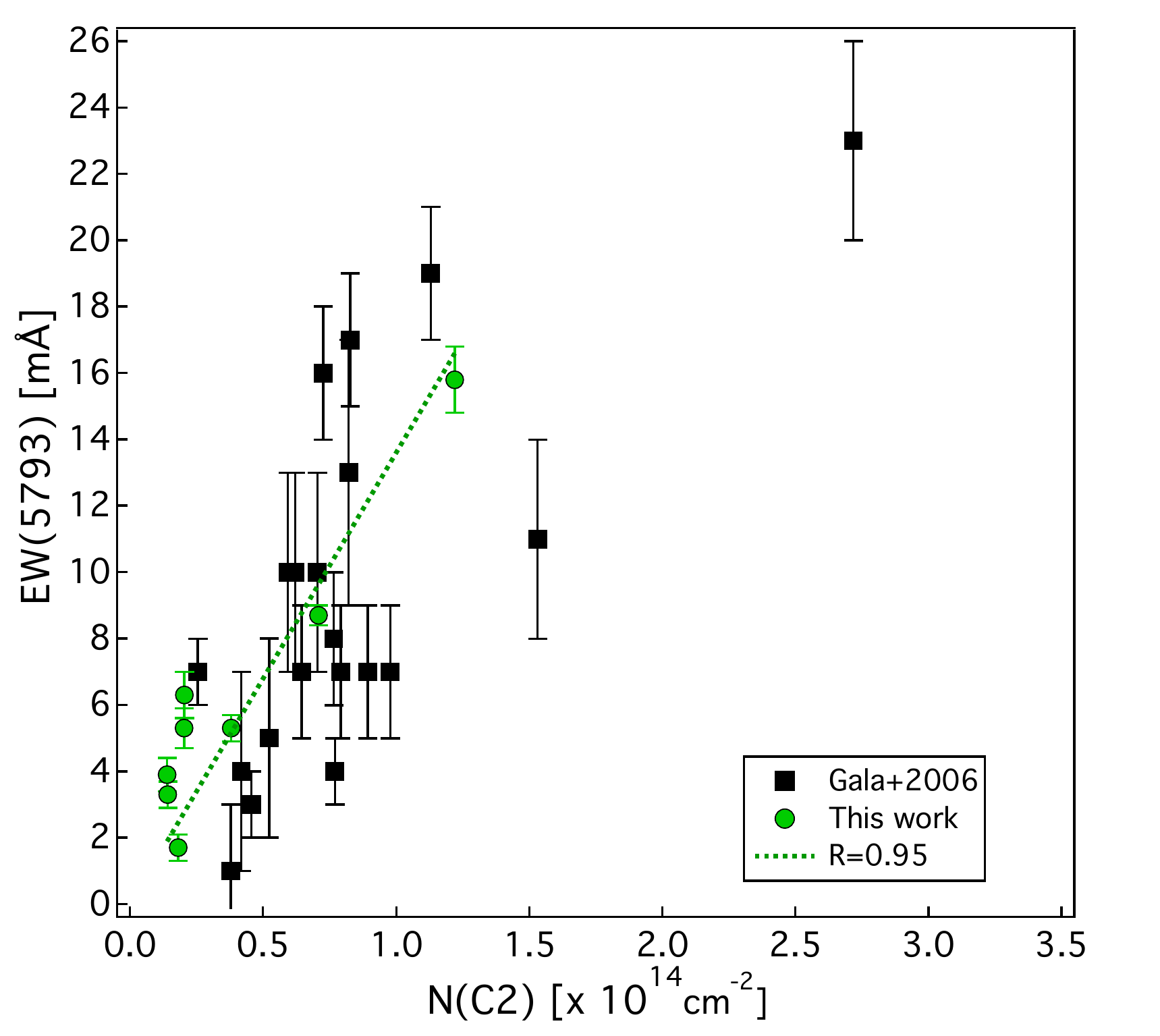}
\includegraphics[width=.33\textwidth]{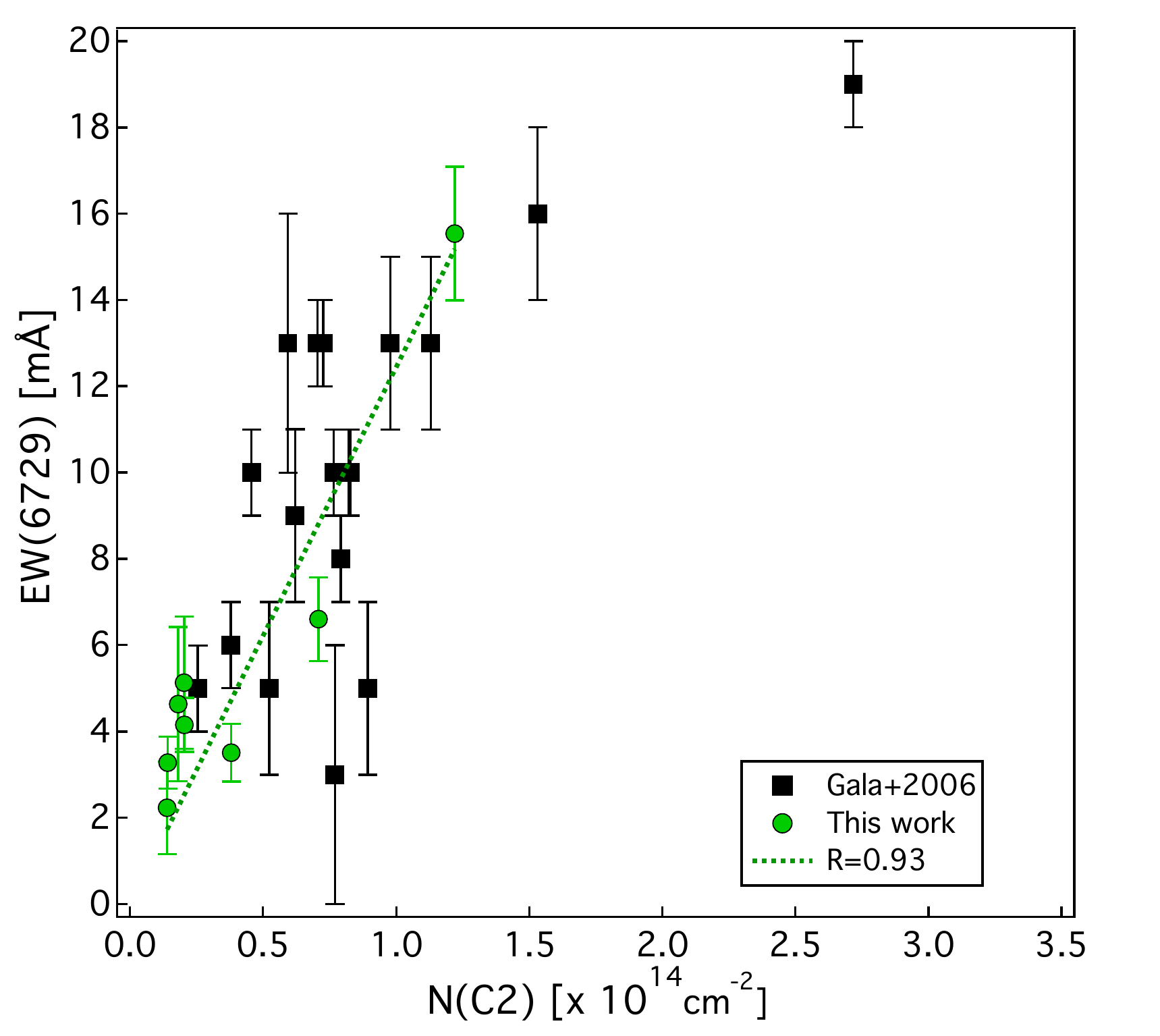}
\caption{ \label{fig:correlations2} Continued}
\end{figure*}
\subsection{Respective influence of \element[][][][2]{C} column and reddening}
\renewcommand{\thefigure}{\arabic{figure}}

The existence of a good correlation with $N(C_2)$ is not a sufficient condition for qualification as \element[][][][2]{C}-DIB and this has been discussed in detail by \citet{Gala06} and \citet{Kazmierczak2014}. One way to establish this qualification is to compare the dependencies of the measured EWs on both the \element[][][][2]{C} column and another parameter representing the quantity of interstellar matter, e.g. the reddening or the gas column. A \element[][][][2]{C}-DIB is one that is more tightly correlated with $N(C_2)$ than with these other parameters. Inspired by the study of \cite{Ensor2017}, we have used our new EDIBLES dataset of \element[][][][2]{C} and non-\element[][][][2]{C} DIBs to perform a simplified parametric analysis restricted to the reddening and $N(C_2)$.

For each DIB we performed a linear regression between the DIB EWs and both E(B-V) and the column $N(C_2)$:
\begin{equation}
EW = a \times N(C_2) + b \times E(B-V), 
\end{equation}
We took into account uncertainties on EWs but assumed E(B-V) and $N(C_2)$ are known with accuracy.
We then computed $A = a/\sigma(a)$ and $B = b/\sigma(b)$, $\sigma(a)$ and $\sigma(b)$ being the standard deviations found from the adjustment for the partial derivatives a and b. Since E(B-V) and $N(C_{2})$ are assumed to be accurately defined, the dimensionless quantities A and B represent the tightness of the links between the DIB and each of the two variables E(B-V) and $N(C_{2})$. We then normalized the vector A,B in the system of reference $N(C_{2})$ and E(B-V) by dividing A and B by $\sqrt{(A^2+B^2)}$.

\begin{figure*}
    \centering
    \includegraphics[width=0.8\textwidth]{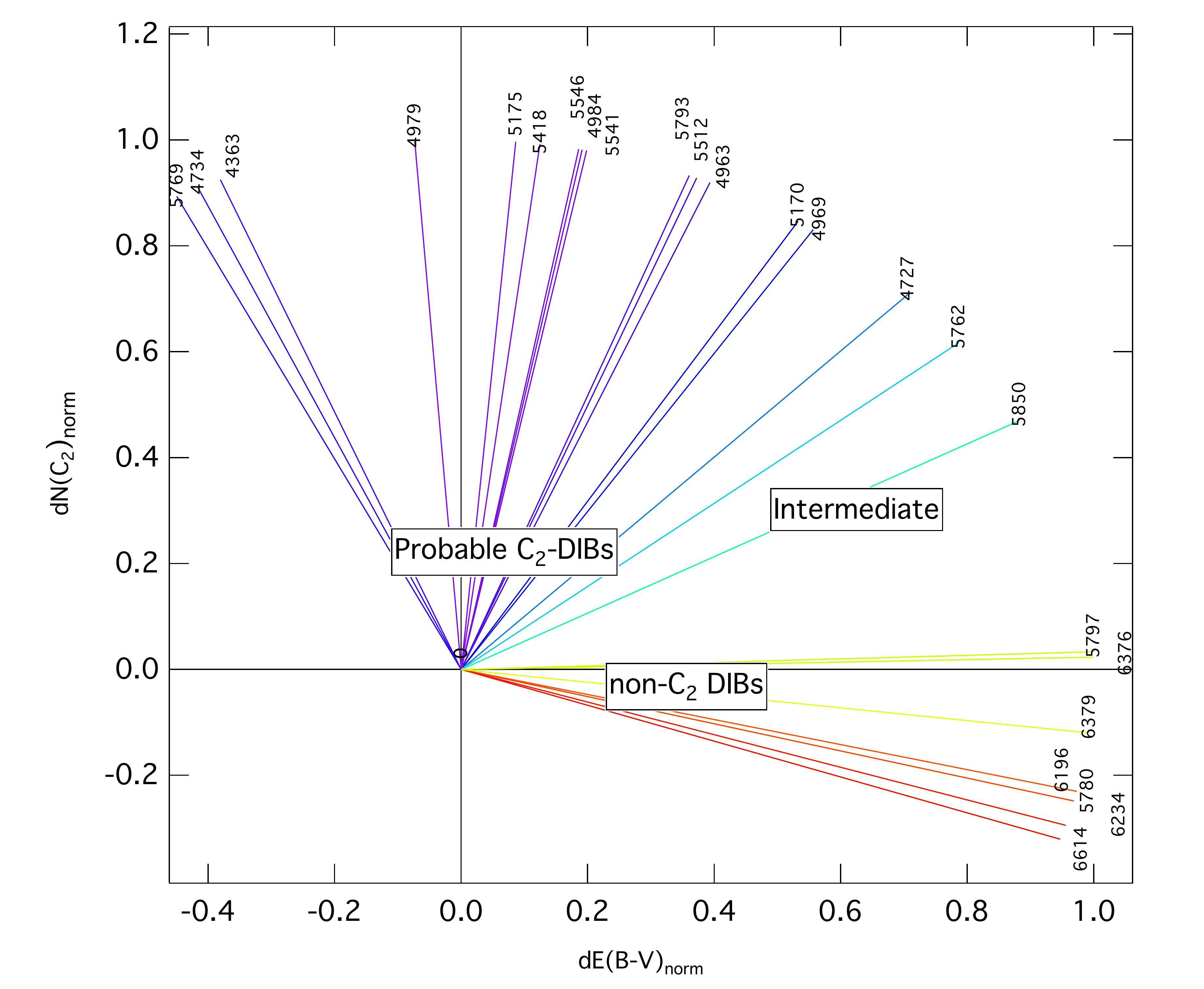}
    \caption{Decomposition of the C$_{2}$ and reddening influence.}
    \label{fig:my_PCA}
\end{figure*}

Fig.~\ref{fig:my_PCA} is a representation of the corresponding "vectors". The orientation of each vector points to the parameter that governs the DIB strength (for vectors oriented vertically or horizontally) or, in intermediate cases, represents their respective influence. 
Note that we performed the regression with and without inclusion of HD\,169454 (the multi-cloud target) and of the highly reddened HD\,147889 to test their influence. We find that the regression results do not change significantly except for the weak 6729~\AA\ band that has clearly the most contaminated spectrum (see Fig.~\ref{dib4}). We removed this DIB from the regression and show the results that were obtained using all targets.

Fig.~\ref{fig:my_PCA} reveals a gap between two series of DIBs. The first series corresponds to strong positive correlation with $N(C_{2})$ and weak or even negative correlation with E(B-V). The list of bands from this series is in total agreement with the 17 C$_{2}$-DIBs  reported by \citet{Thorburn03} in their Table 4.a as their main C$_{2}$-DIBs. In terms of relative correlation with C$_{2}$, the rankings of all six C$_{2}$-DIBs which are common to the two studies indeed agree to within the stated uncertainties. Moreover, our most tightly correlated 4734 and 5769~\AA\ bands are also those noticed by these authors as their main C$_{2}$-DIBs. Remarkably, the E(B-V) correlation coefficient for those two bands is found to be negative. The second series corresponds to the strong DIBs that follow the reddening E(B-V) rather than C$_{2}$. At the margin of the C$_{2}$-DIBs group is the particular 5850~\AA\:DIB. Interestingly, it has been classified as non-C$_{2}$ DIB \citep{Kazmierczak10b} and as C$_{2}$-DIB \citep{Kazmierczak2014} in subsequent analyses. This band is definitely peculiar and deserves further study. Finally, we note that 4727 and 5762~\AA\:DIBs seem to be similarly correlated to E(B-V) and $N(C_{2})$.
\begin{table*}
\caption{\element[][][][2]{C} DIB sub-structure peak separations (cm$^{-1}$).} \label{table:3} 
\centering                         
\begin{tabular}{l l l l l l l l l l l l l l}        
\hline                 
 HD
 &\multicolumn{3}{|c|}{4984} & 5418\tablefootmark{a} &\multicolumn{3}{|c|}{5512} &\multicolumn{3}{|c|}{5541} &\multicolumn{3}{|c}{5546}\\
 &$\Delta_{\nu_{PQ}}$&$\Delta_{\nu_{QR}}$&$\Delta_{\nu_{PR}}$&$\Delta_{\nu_{PR}}$&$\Delta_{\nu_{PQ}}$&  $\Delta_{\nu_{QR}}$&$\Delta_{\nu_{PR}}$&$\Delta_{\nu_{PQ}}$&$\Delta_{\nu_{QR}}$&$\Delta_{\nu_{PR}}$&$\Delta_{\nu_{PQ}}$& $\Delta_{\nu_{QR}}$&$\Delta_{\nu_{PR}}$\\
\hline 
23180 & 0.60&0.60	&1.20 &0.58 &-   &     -    & 0.71 &0.49	&0.39&	0.88 &0.32&	0.29&	0.61\\
24398  &0.60&	0.36	&0.96 &0.79 &- &       -  & 	0.73 &0.49&	0.49&	0.98& 0.29&	0.58&	0.87\\
203532& 0.42&	0.60&	1.02 &0.79&  -	&      - &    0.68& 0.29&	0.58&	0.88 &0.36&	0.32&	0.68\\
149757& 0.83&	0.81&	1.64& 0.62 &-&	      -  &   0.74& 0.39&	0.65&	1.04 & -	&   -  &  0.39\\
185418& 0.73&	0.35&	1.08 &0.83& 0.46&	0.27&	0.73& 0.38&	0.59&	0.97 &0.39&	0.29&	0.68\\
170740& 0.45&	0.39	&0.84& 0.41& 0.49	&0.49	&0.98& 0.29&	0.49&	0.78& -	&   -    &0.49\\
147889 &0.48&	0.85&	1.33 &0.73& 0.35&	0.42&	0.77& 0.48&	0.39&	0.87& -	 &  -  &  0.72\\  
169454 &0.48&	0.36&	0.84 &0.52& 0.31&	0.49&	0.80& 0.49&	0.29	&0.78& -	&   -   & 0.49\\
\hline
\end{tabular}
\tablefoot{
\tablefoottext{a}{The 5418 DIB shows a two sub-peak profile structure.}
}
\end{table*}

\section{Summary and conclusion}\label{sec:discussion}
We have used the high-quality spectra of the EDIBLES survey to select those sightlines that are characterized by both a high \element[][][][2]{C} column and a single Doppler component of interstellar absorption (the single-cloud criterion). To achieve the selection, EDIBLES spectra were corrected for atmospheric lines, exposures of the same target were combined and the spectral intervals co-added to form a unique spectrum. 
The \element[][][][2]{C} selection was based on the detection of the \element[][][][2]{C} (2-0) (8750–8849~\AA) Phillips system and the single-cloud selection was based on profile-fitting of the \ion{Na}{i} UV and optical doublets. Seven targets were selected and studied together with one multi-cloud sightline.

The \element[][][][2]{C} Phillips bands were fitted to extract the \element[][][][2]{C} column, radial velocity and kinetic temperature as well as the gas volume density for each sightline. The absorption bands of the 18 \citep{Thorburn03} \element[][][][2]{C}-DIBs were extracted. Their equivalent widths as well as the EWs of eight strong DIBs were measured and are provided with their measurement uncertainties. Three additional \element[][][][2]{C}-DIB candidates were detected. The \element[][][][2]{C}-DIB EWs were compared with the reddening and the \element[][][][2]{C} column by means of a linear regression on both quantities. Two categories of DIBs were derived, with a significant gap between the two. The first category corresponds to those classified as \element[][][][2]{C}-DIBs by \cite{Thorburn03} and the second to the strong DIBs more tightly correlated with the reddening. A particular case is the 5850~\AA\: band which seems to belong to a different family. In the case of the \element[][][][2]{C}-DIBs, their EWs are found to be very tightly correlated with \element[][][][2]{C} column densities. We interpret this as a consequence of accurate N(\element[][][][2]{C}), high signal-to-noise,  a careful selection of mono-cloud sightlines and correction for telluric and stellar features.

Individual profiles are presented for all targets and DIBs. We also co-added the DIB profiles of three different sightlines to enhance the sub-structures. The main result of this study is the presence of sub-structures with two or three sub-peaks in at least 14 DIBs, the remaining two being too weak to draw a definite conclusion. The two and three sub-peak profile structures are consistent with unresolved rotational branch structures of a molecular carrier. The two sub-peak profiles are very similar to those of the strong 6196 or 5797~\AA\: DIBs, while the three sub-peak profiles are very similar to the one of the strong 6614 ~\AA\: DIB. 
The profiles of the 18 \element[][][][2]{C} DIBs were studied as a function of the transition and as a function of the sightline: 

\begin{itemize}
\item Profile variability among the \element[][][][2]{C}-DIBs for the same sightline:

The differences in peak relative amplitudes and peak sub-structure separations found among the 18 \element[][][][2]{C} bands suggest that they correspond to different molecular carriers or electronic transition symmetries, and factors governing the level populations. 

\item Profile variability among the targets for the same \element[][][][2]{C}-DIB:

Three \element[][][][2]{C}-DIB categories were derived. Several \element[][][][2]{C}-DIBs appear to keep identical profiles among targets characterized by different \element[][][][2]{C} rotational temperatures, however, for others, there is a potential detection of a response to the temperature, namely a global broadening of the absorption band and slightly larger separations between the internal sub-structures at higher temperature. Such effects are similar to those detected for the 6614~\AA\: band by \cite{Cami04} and for the 6196~\AA\: band by \cite{Kazmierczak2009}, and are likely associated with shifting Boltzmann maxima of rotational branches of the molecular carrier. Finally, for a fraction of the DIBs, the variability (or absence of variability) could not be assessed due to their weakness.
\end{itemize}

\begin{acknowledgements}
ME acknowledges funding from the "Region Ile-de-France" through the DIM-ACAV project. RL acknowledges support from  Agence Nationale de la Recherche through the STILISM project (ANR-12-BS05-0016-02) and the CNRS PCMI national program. JC and AF acknowledge support from an NSERC Discovery Grant and a SERB Accelerator Award from Western University. PJS thanks the Leverhulme Trust for the award of a Leverhulme Emeritus Fellowship.
This work is based on observations collected at the European Southern Observatory, 
programme ID 194.C-0833.

\end{acknowledgements}

\bibliographystyle{aa}
\bibliography{mybib.bib}
\end{document}